\documentclass[aps,prd,twocolumn,superscriptaddress,showpacs]{revtex4-1}	
\usepackage[10pt]{type1ec}  
\usepackage[T1]{fontenc}
\usepackage{CJKutf8}
\usepackage[overlap, CJK]{ruby}
\usepackage{CJKulem}
\newenvironment{SChinese}{%
  \CJKfamily{gbsn}%
  \CJKtilde
  \CJKnospace}{}

\usepackage{dcolumn}
\usepackage{amsmath,amssymb,amsthm,paralist}
\usepackage{graphicx}
\usepackage{appendix}
\usepackage{subfigure}
\usepackage{longtable}
\usepackage{url}
\usepackage{verbatim}
\usepackage{slashed}
\usepackage{hyperref} 

\usepackage[usenames]{xcolor}
\usepackage{booktabs}
\usepackage{multirow}
\usepackage{pifont}
\newcommand{\xmark}{\text{\ding{55}}}

\newcommand{\be}{\begin{equation}}
\newcommand{\ee}{\end{equation}}
\newcommand{\ba}{\begin{eqnarray}}
\newcommand{\ea}{\end{eqnarray}}
\newcommand{\bc}{\begin{center}}
\newcommand{\ec}{\end{center}}

\newcommand{\Ga}{\Gamma}

\newcommand{\De}{\Delta}
\newcommand{\ep}{\varepsilon}

\newcommand{\la}{\lambda}
\newcommand{\La}{\Lambda}

\newcommand{\ptrans}{P_{\rm trans}}
\newcommand{\ntrans}{n_{\rm trans}}
\newcommand{\nb}{n_{\rm B}}
\newcommand{\etrans}{\varepsilon_{\rm trans}}
\newcommand{\Mtrans}{M_{\rm trans}}

\newcommand{\decrit}{\De\ep_{\rm crit}}
\newcommand{\Mchirp}{{\mathcal M}}

\newcommand{\Msolar}{{\rm M}_{\odot}}
\newcommand{\cQMsq}{c^2_{\rm QM}}

\newcommand{\Mmax}{M_{\rm max}}
\newcommand{\Rtyp}{R_{1.4}}

\newcommand\Eqn[1]{Eq.~(\ref{#1})}  

\begin{document}
\title{Treating quarks within neutron stars}

\author{Sophia Han (\begin{CJK}{UTF8}{}\begin{SChinese}韩 君\end{SChinese}\end{CJK})}
\email{sjhan@berkeley.edu}
\affiliation{Department of Physics and Astronomy, Ohio University,
Athens, OH 45701, USA}
\affiliation{Department of Physics, University of California Berkeley, Berkeley, CA~94720, USA}
\author{M.~A.~A.~Mamun}
\email{ma676013@ohio.edu}
\affiliation{Department of Physics and Astronomy, Ohio University,
Athens, OH 45701, USA}
\author{S.~Lalit}
\email{sl897812@ohio.edu}
\affiliation{Department of Physics and Astronomy, Ohio University,
Athens, OH 45701, USA}
\author{C.~Constantinou}
\email{cconsta5@kent.edu}
\affiliation{Department of Physics, Kent State University, Kent, OH 44242, USA }
\author{M.~Prakash}
\email{prakash@ohio.edu}\affiliation{Department of Physics and Astronomy, Ohio University,
Athens, OH 45701, USA}

\date{October 31, 2019} 
 
\begin{abstract}

Neutron star interiors provide the opportunity to probe properties of cold dense matter in the QCD phase diagram. Utilizing models of dense matter in accord with nuclear systematics at nuclear densities, we investigate the compatibility of deconfined quark cores with current observational constraints on the maximum mass and tidal deformability of neutron stars. We explore various methods of implementing the hadron-to-quark phase transition, specifically, first-order transitions with sharp (Maxwell construction) and soft (Gibbs construction) interfaces, and smooth crossover transitions. 
We find that within the models we apply, hadronic matter has to be stiff for a first-order phase transition and soft for a crossover transition. In both scenarios and for the equations of state we employed, quarks appear at the center of pre-merger neutron stars in the mass range $\approx 1.0-1.6\,\Msolar$, with a squared speed of sound $\cQMsq \gtrsim 0.4$ characteristic of strong repulsive interactions required to support the recently discovered neutron star masses $\geq 2\,\Msolar$. We also identify equations of state and phase transition scenarios that are consistent with the bounds placed on tidal deformations of neutron stars in the recent binary merger event GW170817. We emphasize that distinguishing hybrid stars with quark cores from normal hadronic stars is  very difficult from the knowledge of masses and radii alone, unless drastic sharp transitions induce distinctive disconnected hybrid branches in the mass-radius relation.

\label{abs}
\end{abstract}

\maketitle

\section{introduction}
\label{sec:intro}

The observation that the dense matter inside neutron stars might consist of weakly interacting quark matter owing to the asymptotic freedom of Quantum Chromodynamics (QCD) was first made by Collins and Perry \cite{CP75}. Since then, numerous explorative studies have been conducted to isolate neutron star observables that can establish the presence of quarks deconfined from hadrons. Starting from the QCD Lagrangian, lattice gauge simulations at finite temperature $T$ and net baryon number $\nb=0$ naturally realize hadronic and quark degrees of freedom in a smooth crossover transition. However, lattice simulations for finite $\nb$ at $T=0$, of relevance to neutron stars, have been thwarted due to the unsolved fermion sign problem and untenable imaginary probabilities. As a result, the possible phases of dense matter at $T=0$ have been generally explored by constructing equation of state (EoS) models of hadrons and quarks that are independent of each other although a few exceptions do exist.  

Extensive studies of nucleonic matter in neutron stars for $\nb \lesssim 0.5 \,n_0$, where $n_0 \simeq 0.16~{\rm fm}^{-3}$ is the isospin symmetric nuclear matter equilibrium density, have predicted the presence of a solid crust. Observations of the surface temperatures of accreting neutron stars in their quiescent periods have indeed confirmed the presence of a crust (see Ref.~\cite{Meisel:2018ufs}, and references therein). This region is characterized by a Coulomb lattice of neutron-rich nuclei surrounded by dripped neutrons with admixtures of light nuclei and a uniform background of electrons in chemical potential and pressure equilibrium in a charge-neutral state. Differences among different equations of state \cite{Baym71tg,Negele73ns,Douchin:2001sv,Sharma:2015bna} are small and are of minor importance to the structure of stars more massive than $1\,\Msolar$. In this work, we use the EoSs of Ref.~\cite {Negele73ns} (for $0.001 < \nb < 0.08~{\rm fm}^{-3}$) and Ref.~\cite{Baym71tg} (for $\nb < 0.001~{\rm fm}^{-3}$) to determine the structural properties of the star.

Models of the hadronic EoS for $\nb > 0.08~{\rm fm}^{-3}$ can be grouped into three broad categories: non-relativistic potential models, Dirac-Brueckner-Hartree-Fock models, and relativistic field-theoretical models. Microscopic many-body calculations in the first two of these categories (e.g., Brueckner-Hartree-Fock, variational, Greens' function Monte Carlo, chiral effective field theory, as well as Dirac-Brueckner-Hartree-Fock) employ free-space two-nucleon interactions  supplemented by three-nucleon interactions required to describe the properties of light nuclei as input. In contrast, coupling strengths of the two- and higher-body nucleon interactions mediated by meson exchanges are calibrated at $n_0$ in the relativistic field-theoretical models. Several schematic potential models based on zero- and finite-range forces also exist that take recourse in the Hohenberg-Kohn-Sham theorem \cite{HK64,KS65} which assures that the ground state energy of a many-body system can be expressed in terms of local densities alone. Refinements in all of these approaches are guided by laboratory data on the bulk properties of isospin symmetric and asymmetric matter, such as the binding energy $\rm{BE}= -16\pm 1$ MeV \cite{Myers:1966zz,Myers:1995wx} at the saturation density $n_0=0.16\pm 0.01~{\rm fm}^{-3}$ \cite{Myers:1966zz,Day:1978zz,Myers:1995wx}, compression modulus $K_{\rm nm}=240\pm 20$ MeV \cite{Garg:2004fsg,Colo:2004mj,Shlomo06}, nucleon's Landau effective mass $m^*/M=0.75\pm 0.1$ \cite{Bohigas:1978qu,Krivine:1980kzz,Margueron:2017eqc}, symmetry energy $S_2=28-35$ MeV \cite{Lattimer:2012xj,Tsang:2012se} and the symmetry energy slope parameter $L=60 \pm 20$ MeV \cite{Lattimer:2012xj,Tsang:2012se} at saturation, etc. Low-to-intermediate energy (0.5-2 GeV) heavy-ion collisions have been used to determine the EoS for densities up to 2-3 $n_0$ through studies of matter, momentum, and energy flow of nucleons \cite{Gale87, Prakash88b, Welke88, Gale90, Danielewicz00,Danielewicz:2002pu}. The consensus has been that as long as	momentum-dependent forces are employed in models that use Boltzmann-type kinetic equations, use of $K_{\rm nm} \sim 240\pm 20$ MeV, suggested by the analysis of the giant monopole resonance data \cite{Youngblood99, Garg04, Colo04}, fits the heavy-ion data as well \cite{Danielewicz:2002pu}.

The lack of Lorentz invariance in non-relativistic models leads to an acausal behavior at some high density particularly if contributions from three- and higher-body interactions to the energy are not screened in medium \cite{Bludman:1980xn,Prakash:1988md}. The general practice has been to enforce causality from thermodynamic considerations \cite{Nauenberg73,Lattimer:1990zz}. In some cases, the reliability of non-relativistic models is severely restricted, sometimes only up to $2\,n_0$ as in the case of chiral effective field-theoretical (EFT) models owing to the perturbative scheme and the momentum cut-off procedure employed there \cite{Hebeler:2010jx,Tews:2012fj}.

To explore consequences of the many predictions of these models at supra-nuclear densities, piecewise polytropic EoSs that are causal have also been extensively used to map out  the range of pressure vs density relations (EoSs) that are consistent with neutron star phenomenology \cite{Lattimer:2015nhk,Steiner:2015aea,Tews:2018kmu,Zhao:2018nyf}. The viability of these EoSs at supra-nuclear densities necessarily depends on the growing neutron star data to be detailed below.  

The possibility of non-nucleonic degrees of freedom such as strangeness-bearing hyperons, pion and kaon condensates, and deconfined quarks above $n_0$ has also been examined in many of these models \cite{Lattimer:2015nhk,Chatterjee:2015pua,Vidana:2018bdi}. At some $\nb \gtrsim (2-4) \,n_0$, the presence of quark degrees of freedom has been invoked on the physical basis that the constituents of hadrons could be liberated as the compression in density progressively increases. First-principle calculations \cite{Baym:1976yu,Baym:1975va,Freedman:1977gz,Farhi:1984qu,Kurkela:2009gj,Kurkela:2016was,Gorda:2018gpy,Alford:2007xm} of the EoS of quark matter have thus far been limited to the perturbative region of QCD valid at asymptotically high baryon densities. The Nambu--Jona-Lasinio (NJL) model \cite{Nambu:1961tp}, which shares many symmetries with QCD - but not confinement - has been used to mimic chiral restoration in quark matter \cite{Kunihiro:1989my,Buballa:1998pr,Buballa:2003qv}. Also in common use are variations \cite{Klahn:2015mfa,Gomes:2018eiv} of the MIT bag model \cite{Baym:1976yu}.  

Lacking knowledge about the nature of the phase transition, it has been common to posit a first-order phase transition in many recent studies~\cite{Nandi:2017rhy,Paschalidis:2017qmb,Alvarez-Castillo:2018pve,Wei:2018mxy,Gomes:2018eiv}. Even in this case, the magnitude of the hadron-quark interface tension is uncertain \cite{Alford:2001zr,Mintz:2009ay,Lugones:2013ema,Fraga:2018cvr}. If the interface tension is regarded as being infinite, a Maxwell construction can be employed to determine the range of density for which chemical potential and pressure equality between the hadronic and quark phases exists \cite{Glendenning:1992vb}. The other extreme case corresponds to a vanishing interface tension when a Gibbs construction is considered more appropriate. The Gibbs construction also corresponds to global charge neutrality instead of local charge neutrality, appropriate for matter with two conserved charges (baryon number and charge) \cite{Glendenning:2001pe}. 

Depending on the models used to calculate the EoSs of the hadron and quark phases, chemical potential and pressure equilibrium between the two phases may not be realized~\cite{Baym:2017whm}. In such cases, several interpolatory procedures have been used to connect the two phases on the premise that at $\nb >> n_0$, a purely hadronic phase is physically unjustifiable \cite{Masuda:2012ed,Fukushima:2015bda,Kojo:2014rca,Baym:2017whm}. As a result, the hadron-quark transition becomes one of a smooth crossover with the proportion of each phase depending on the specific interpolation procedure used. This is in contrast to the Gibbs construction (which also renders the transition into a mixed phase to be smooth) in which the fraction of each phase is determined self-consistently.  

Although differing in details, other examples of a smooth crossover transition are the chiral model of Ref.~\cite{Dexheimer:2014pea} and the quarkyonic model of Ref.~\cite{McLerran:2018hbz}. A quark phase with additional hadronic admixtures such as hyperons and meson condensates has also been explored \cite{Prakash:1996xs}. The precise manner in which the hadron-quark transition is treated influences the magnitudes of the mass and radius of the star. In addition, the behavior of the speed of sound with density affects the magnitude of tidal deformations. It is worth mentioning however, that stars with purely hadronic matter (HM) can sometimes masquerade as stars with quark matter (QM)~\cite{Alford:2004pf}. 

The objectives of this work are to seek answers to probing questions such as (a) What is the minimum neutron star (NS) mass consistent with the observational lower limit on the maximum mass ($\Mmax$) that is likely to contain quarks? (b) What is the minimum physically reasonable density at which a hadron-quark transition of any sort can occur? (c) Which astronomical observations have the best potential to attest to the presence of quarks?

Toward providing answers to the above questions, we have undertaken a detailed study of the hadron-to-quark matter transition in neutron stars. Our focus is to study the sensitivity of outcomes on neutron star structure, principally mass-radius relations, in the different treatments of the phase transition. Results so obtained are then subjected to the constraints provided by precise measurements of heavy neutron stars \cite{Demorest:2010bx,Antoniadis:2013pzd,Fonseca:2016tux,Cromartie:2019kug}, bounds on the tidal deformability of neutron stars in the binary merger event GW170817 \cite{LIGO:2017qsa,LIGO:2018exr,De:2018uhw,LIGO:2018wiz}, and radius estimates of $1.4\,\Msolar$ available from x-ray observations of neutron stars \cite{Steiner:2015aea,Lattimer:2015nhk,Ozel:2016oaf}. 

Earlier studies in this regard have generally chosen one favored EoS in the hadronic sector and one approach to the quark matter EoS \cite{Steiner:2000bi,Hanauske:2001nc,Bhattacharyya:2009fg,Klahn:2015mfa,Gomes:2018eiv,Wei:2018mxy}. Contrasts between the Maxwell and Gibbs constructions have also been made in some of these works, but with the result that $\Rtyp$ are typically larger than 14 km or more (characteristic of the use of mean-field theoretical (MFT) models) which is at odds with most of the available estimates. This work differs in that variations in the EoSs of both the hadronic and quark sectors are considered as well as a global view of the outcomes of different treatments of the transition is taken. By including terms involving scalar-vector and scalar-isovector interactions in MFT models, we  show that values of $\Rtyp$ more in consonance with data can be achieved.  Additionally, we present an  extension of the quarkyonic matter model of Ref.~\cite{McLerran:2018hbz} to isospin asymmetric matter with the inclusion of interactions between quarks (not considered there) to enable calculations of beta-equilibrated neutron stars. This extension will be useful in applications involving compositional and thermal gradients in quarkyonic stars, such as their long-term cooling as well as  quiescent cooling following accretion on them from a companion star and in investigating $f$-, $p$- and $g$- mode oscillations. Our in-depth study of the thermodynamics of quarkyonic matter sheds additional physical insight into the role that the nucleon shell plays in stiffening the EoS. 

Our findings in this work reveal that several aspects of neutron star properties deduced from observations may have to be brought to bear in finding answers to the questions posed above. These properties include the masses $M$, radii $R$, periods $P$ and their time derivatives $\dot P$ and $\ddot P$, surface temperatures $T_s$ of isolated neutron stars and of those that undergo periodic accretion from companions, tidal deformations $\La$ from the detection of gravitational waves during the inspiraling phase of neutron star mergers, etc. Currently, the accurately measured neutron star masses around and above $2\,\Msolar$ \cite{Demorest:2010bx,Antoniadis:2013pzd,Fonseca:2016tux,Cromartie:2019kug} pose stringent restrictions on the EoS. Even so, the EoS would be better restricted with knowledge of radii of stars for which the masses are also known, although this would not reveal the constituents of dense matter as the structure equations depend only on the pressure vs density relation $\ep(P)$, and not on how it was obtained. In contrast, the surface temperatures of both isolated neutron stars and of quiescent cooling of accreting neutron stars are sensitive to the composition, but simultaneous knowledge of their masses and radii are yet unknown. The anomalous behavior of the braking indices $n=\Omega \,\ddot \Omega/\dot \Omega^2$, where $\Omega=2\pi/P$ is the spin rate, of several known pulsars \cite {Magalhaes:2012za,Hamil:2015hqa,Johnston:2017wgm} can also be put to good use in this connection.  

The organization of this paper is as follows. In Sec.~\ref{sec:Models}, we present the models in the hadronic and quark sectors chosen for our study. The rationale for our choice and basic features of these models are highlighted here for orientation. We stress that our choices are representative, but not exhaustive. Results of neutron star properties for different treatments of the hadron-quark transition introduced in Sec.~\ref{sec:treat} are shown and discussed in Sec.~\ref{sec:Results}. Our conclusions and outlook are contained in Sec.~\ref{sec:con}. Appendix~\ref{sec:TI} contains details about the thermodynamics of nucleons in the shell of quarkyonic matter. 

We use units in which $\hbar=c=1$.

\section{Equation of State Models}
\label{sec:Models}

\subsection*{Nucleonic EoSs}
\label{sec:NEoSs}

To explore sensitivity to the hadronic part of the EoS, we use representative examples from both potential and relativistic mean field-theoretical (RMFT) models. In the former category, the EoS of Akmal, Pandharipande and Ravehall (APR)~\cite{APR98}, which is a parametrization of the microscopic variational calculations of Akmal and Pandharipande \cite{Akmal:1997ft}, is chosen as its energy vs baryon density up to $2\,n_0$ closely matches those of modern EFT calculations of pure neutron matter and symmetric nuclear matter \cite{Hebeler:2010jx,Tews:2012fj}. Moreover, it is compatible with current nuclear phenomenology from both structure (equilibrium density and energy, compression modulus, symmetry energy and its slope, etc.) and heavy-ion experiments~\cite{Danielewicz:2002pu} as well as with the latest constraints from astrophysical observations (largest known NS mass, upper limit on maximum NS mass, tidal deformability, NS radii, etc). Explicit expressions for the energy density $\ep$, pressure $P$, compression modulus $K_0$, Landau effective mass $m^*/M$, symmetry energy $S_2$, and the symmetry energy slope parameter $L$ along with the coupling strengths of the various terms therein can be found in Ref.~\cite{Constantinou:2014hha}. Recent fits of the APR calculations to the traditional Skyrme energy-density functional (EDF) can be found in Refs.~\cite{Steiner:2004fi,Schneider:2019vdm}. The latter also details the calculation of a complete tabular EoS based on the original APR parametric form.

To provide contrast, we have constructed three EoSs, MS-A, MS-B and MS-C using the RMFT model of M\"{u}ller and Serot \cite{Mueller:1996pm} employing terms that contain scalar-isovector and vector-isovector mixings as in Refs.~\cite{Horowitz:2001ya,Horowitz:2000xj}. The numerical results to be reported in this work are from these RMFT models; that is, we consider many-body effects at the Hartree level exclusive of quantum fluctuations in the meson fields. Fock (exchange) terms are beyond the scope of this paper. 
As demonstrated in Ref.~\cite{Chin:1977iz}, a simple re-parametrization of the couplings in the MFT Hartree models at $T=0$ yields very nearly the same $P$ vs $\ep$ relations (and, hence masses, radii and tidal deformabilities of neutron stars) as models with the inclusion of Fock terms. Note that Fock terms and additional many-body contributions do influence thermal effects in a way that is not reproducible by re-parametrization; see e.g. Refs.~\cite{Zhang:2016bem,Constantinou:2016hvf,Constantinou:2015mna}. In this work however, we do not consider hot matter.
Specifically, we have devised three new parametrizations for the coupling constants appearing in the MS Lagrangian such that consistency with contemporary experimental and observational data is achieved. Many other EoSs based on the MS model are currently in use; for an exhaustive list, see Ref.~\cite{Oertel:2016bki}. Explicitly, the Lagrangian density for this model is 
\ba
\label{eqn:Lag1}
\mathcal{L} &=& \overline{\Psi}[i \slashed{\partial} - g_\omega\slashed{\omega} - 
\frac{1}{2}g_\rho\slashed{\rho}.\tau - M + g_\sigma\sigma - \frac{1}{2}e(1+\tau_3)\slashed{A}]\Psi  \nonumber \\
 &+& \frac{1}{2}(\partial_\mu\sigma)^2 - V(\sigma) - \frac{1}{4}f_{\mu\nu}f^{\mu\nu} + 
 \frac{1}{2}m_\omega^2\omega^\mu\omega_\mu  \nonumber \\
 &-& \frac{1}{4} B_{\mu\nu}B^{\mu\nu} + \frac{1}{2}m_\rho^2\rho^\mu\rho_\mu - 
 \frac{1}{4}F_{\mu\nu}F^{\mu\nu}  \nonumber \\
 &+& \frac{\zeta}{24}g_\omega^4(\omega^{\mu}\omega_{\mu})^2 + 
 \frac{\xi}{24}g_\rho^4(\rho^\mu\rho_\mu)^2 + g_\rho^2f(\sigma,\omega_\mu\omega^\mu)\rho^\mu.\rho_\mu \nonumber \\ 
\ea
with
\ba
V(\sigma) &=& \frac{1}{2}m_\sigma^2\sigma^2 + \frac{\kappa}{6}(g_\sigma\sigma)^3 
+ \frac{\lambda}{24}(g_\sigma\sigma)^4 \nonumber \\
f(\sigma,\omega) &=& \Lambda_\sigma g_\sigma^2\sigma^2 + \Lambda_\omega g_\omega^2\omega^2 \,.
\ea
Expressions for the energy per particle $\ep/n$, $P$, $K_0$, the Dirac effective mass $M^*$ and hence the sigma field $\sigma_0 = (M-M^*)/g_\sigma$ in the mean-field approximation can be found in Ref.~\cite{Steiner:2004fi}. With input values of these quantities at $n_0$, the coupling strengths  $g_\sigma$, $g_\omega$, $\kappa$ and $\lambda$ are straightforwardly determined by numerically solving the system of nonlinear equations containing these quantities. 
The strengths $\zeta$ and $\xi$, $\Lambda_\sigma$ and $\Lambda_\omega$ of the quartic $\omega$ and $\rho$ fields, remain as adjustable input parameters to control the high-density behavior. The density-dependent symmetry energy in this model is~\cite{Horowitz:2000xj} 
\ba
S_2 &=& S_{2k} + S_{2d} \nonumber \\ 
&=& \frac {k_F^2}{6E_F^*} + \frac {1}{8} \frac {g_\rho^2~ n}{m_\rho^{*2}} \,, \quad E_F^* = {\sqrt{k_F^2+M^{*2}}} \nonumber \\ 
m_\rho^{*2} &=& 2g_\rho^2 \left(\Lambda_\sigma g_\sigma^2 \sigma_0^2 + \Lambda_\omega g_\omega^2 \omega_0^2\right) \,.
\ea
The first term on the right-hand side above contains effects of interaction through $\sigma$-meson exchange, whereas the second term includes those from the $\rho$-meson exchange along with $\rho$-$\sigma$ and $\rho$-$\omega$ mixing. The corresponding slope parameter at $n_0$ becomes 
\ba
\label{eqn:Ls}
L &=&  3 \, n_0 \left. \frac {dS_2}{dn}\right|_{n_0} = \left. L_{k} \right|_{n_0} + \left. L_{d}\right|_{n_0} \nonumber \\ 
L_{k} &=& 2 S_{2k}
 \left[ 1 - 18  \left( \frac {S_{2k}}{k_F} \right)^2 \left\{ 1 + 3 \left( \frac {M^*}{k_F} \right)^2 \frac {d\ln M^*}{d\ln n} \right\} \right] 
 \nonumber \\
 L_{d} &=& 3  S_{2d} \left[ 1 -  32 S_{2d} 
 \left\{ \Lambda_\sigma g_\sigma^2 \sigma_0 
   \frac {d\sigma_0}{dn}  \right. \right. 
 + \left. \left.  {\Lambda_\omega}  g_\omega^2 \omega_0    \frac {d\omega_0}{dn}  \right\} \right] \,. \nonumber \\
\ea
Analogous expressions but without the term involving $\Lambda_\sigma$ can be found in Ref.~\cite{Chen:2014sca}. The strength $g_\rho$ may be fixed with a prescribed value of $S_2$ at $n_0$, which leaves one or a combination of $\Lambda_\sigma$ and $\Lambda_\omega$ to obtain a desired value of $L$. The values of the various couplings used in this work are listed in Table~\ref{tab:Couplings1}. \\

\begin{table}[!htb]
\caption{RMFT coupling strengths. Values of the meson masses used are $m_\sigma=660$ MeV, $m_\omega=783$ MeV and $m_\rho=770$ MeV. }
\begin{center} 
\begin{tabular}{crrrrr}
\hline
\hline
Model & $g_\sigma$ & $g_\omega$ & $g_\rho$ & $\kappa$ & $\lambda$ \\ \hline
MS-A & 12.819 & 12.258 & 12.079 & 0.02544 & -0.02179 \\
MS-B & 11.369 & 10.143 & 9.446 & 0.05098 & -0.03396 \\
MS-C & 10.026 & 7.961 & 8.492 & 0.10841 & -0.00365 \\ \hline
Model & $\zeta$ & $\xi$ & $\Lambda_\sigma$ & $\Lambda_\omega$ &  \\ \hline
MS-A & 0.0001 & 1.0 & 0.001 & 0.05 &  \\
MS-B & 0.0001 & 1.0 & 0.001 & 0.05 &  \\
MS-C & 0.0001 & 1.0 & 0.001 & 0.05 &  \\
\hline \hline
\end{tabular}
\end{center}
\label{tab:Couplings1}
\end{table}

As noted in Refs.~\cite{Horowitz:2000xj,Steiner:2004fi}, the quartic and scalar-isovector and vector-isovector terms in \Eqn{eqn:Lag1} enable acceptable values~\cite{Lattimer:2012xj} of the symmetry energy slope parameter $L$ at $n_0$ to be obtained. The reduction in $L$ from its generally large value found  for RMFT models is made possible by the second term in $L_d$ of \Eqn{eqn:Ls}, the term in braces being positive definite. These density-dependent terms also influence the high-density behavior of these EoSs, leaving the near-nuclear-density behavior intact. Salient properties at $n_0$ for these nucleonic models are presented in Table~\ref{tab:Props1}. The values  of $L$ in Table~\ref{tab:Props1} are to be compared with those of the FSU models \cite{Chen:2014sca,Fattoyev:2017jql} in the literature; see e.g. Fig. 2 and Table IV in Ref.~\cite{Chen:2014sca}: $L=60.5$ MeV for FSU (but it does not achieve $2\,\Msolar$) and $L=112.8\pm16.1$ MeV for FSU2 with $\Mmax=2.07\pm0.02\,\Msolar$, $R_{\rm max} = 12.2$ km and $\Rtyp=14.42\pm 0.26$ km. In comparison to FSU2, the values of $L$ for the MS models of this work are significantly smaller, which result in smaller radii for the maximum mass and $1.4\,\Msolar$ neutron stars (see Table~\ref{tab:Struc1} below).

\begin{table}[!htb]
\caption{Properties at the nuclear equilibrium density $n_0$ for EoSs used in this work compared to that of the APR EoS \cite{Constantinou:2014hha}. Entries in this table are the Landau effective mass $m^*/M$, isospin symmetric matter compression modulus $K_0$, kinetic and interaction parts $S_{2k}$ and $S_{2d}$ of the total symmetry energy $S_2$, and the corresponding parts of the symmetry energy slope parameter $L$. In MS models, $m^*=E_F^*={\sqrt{k_F^2+M^{*2}}}$.} 
\begin{center} 
\begin{tabular}{crrrrc}
\hline
\hline
Property & APR   & MS-A  & MS-B  & MS-C  & Units \\ \hline
$n_0$    & 0.16  & 0.16  & 0.16  & 0.16  & ${\rm fm}^{-3}$ \\   
$m^*/M$  & 0.698 & 0.662  & 0.763  & 0.847  & \\
$K_0$    & 266   & 230   & 230   & 230   & MeV \\ 
\hline \hline
$S_{2k}$ & 9.79 & 18.55 & 16.09 & 14.49 & MeV \\
$S_{2d}$ & 22.80 & 11.45 & 13.91 & 15.51 & MeV \\
$S_2$    & 32.58 & 30.0  & 30.0  & 30.0  & MeV \\ 
\hline \hline 
$L_k$ & 12.69 & 61.74 & 44.35 & 34.52 & MeV \\
$L_d$ & 45.78 & -13.40 & 8.65 & 30.88 & MeV \\
$L$      & 58.47 & 48.34 & 53.00 & 65.40 & MeV \\
\hline \hline
\end{tabular}
\end{center}
\label{tab:Props1}
\end{table}

\subsection*{Properties of nucleonic neutron stars}

Structural properties of charge-neutral and  beta-equilibrated neutron stars resulting from the chosen EoSs are listed in Table~\ref{tab:Struc1}. Two of the  three MS EoSs satisfy the requirement of supporting a star with mass $\geq 2\,\Msolar$. The EoS of MS-C does not obey the $2\,\Msolar$ constraint, but we have retained it in our analysis because, in conjunction with crossover transitions involving quark matter, masses well in excess of this observational limit can be obtained (see Secs.~\ref{sec:treat} and \ref{sec:Results}). Although the RMFT models employ terms that contain scalar-isovector and vector-isovector mixings as in Refs.~\cite{Horowitz:2001ya,Horowitz:2000xj}  to yield acceptable values of the symmetry energy slope parameter $L$ at $n_0$, the radii of neutron stars stemming from these models are somewhat larger than that of the APR model, but lie within the range of those extracted from data \cite{Lattimer:2012xj}. The largest differences between the APR and RMFT models are in the central pressures of the maximum-mass stars. The proton fractions, $y_{c,{1.4}}$ and $y_{c,{\rm max}}$, are such that stars close to the maximum-mass stars allow the direct Urca processes with  electrons and muons to occur~\cite{Lattimer:1991ib}.

\begin{table}[!htb]
\caption{Structural properties of nucleonic neutron stars with $M=1.4\,\Msolar$ and $\Mmax$ for the indicated EoSs. For each mass, the compactness parameter $\beta=(GM/c^2R) \simeq (1.475 \,R)(M/\Msolar)$; $n_c$, $P_c$ and $y_c$ are the central values of the density, pressure and proton fraction, respectively.}

\begin{center} 
\begin{tabular}{crrrrc}
\hline
\hline
Property            & APR    & MS-A   & MS-B   & MS-C   & Units \\ \hline
$\Rtyp$           & 11.74  & 13.21  & 12.41  & 11.85  & {\rm km} \\
$\beta_{1.4}$       & 0.176  & 0.157  & 0.167  & 0.174  & \\
$n_{c,1.4}/n_0$     & 3.35   & 2.05   & 2.80   & 3.72   & \\
$P_{c,1.4}$         & 89.33  & 41.78  & 64.43  & 94.24 & ${\rm MeV~fm^{-3}}$ \\
$y_{c,1.4}$         & 0.11   & 0.104  & 0.106  & 0.106  & \\ \hline
$R_{\rm max}$       & 10.26  & 12.44  & 10.91  & 9.94   & {\rm km} \\
$\Mmax$             & 2.185  & 2.63   & 2.21   & 1.83   & $\Msolar$ \\ 
$\beta_{\rm max}$   & 0.314  & 0.312  & 0.299  & 0.273  & \\
$n_{\rm c,max}/n_0$ & 6.97   & 4.71   & 6.38   & 8.30  & \\
$P_{\rm c,max}$     & 884.69 & 498.32 & 632.66 & 664.60 & ${\rm MeV~fm^{-3}}$ \\
$y_{\rm c,max}$     & 0.16  & 0.14 & 0.14 & 0.128 & \\
\hline \hline
\end{tabular}
\end{center}
\label{tab:Struc1}
\end{table}

\begin{figure*}[htb]
\parbox{0.45\hsize}{
\includegraphics[width=\hsize]{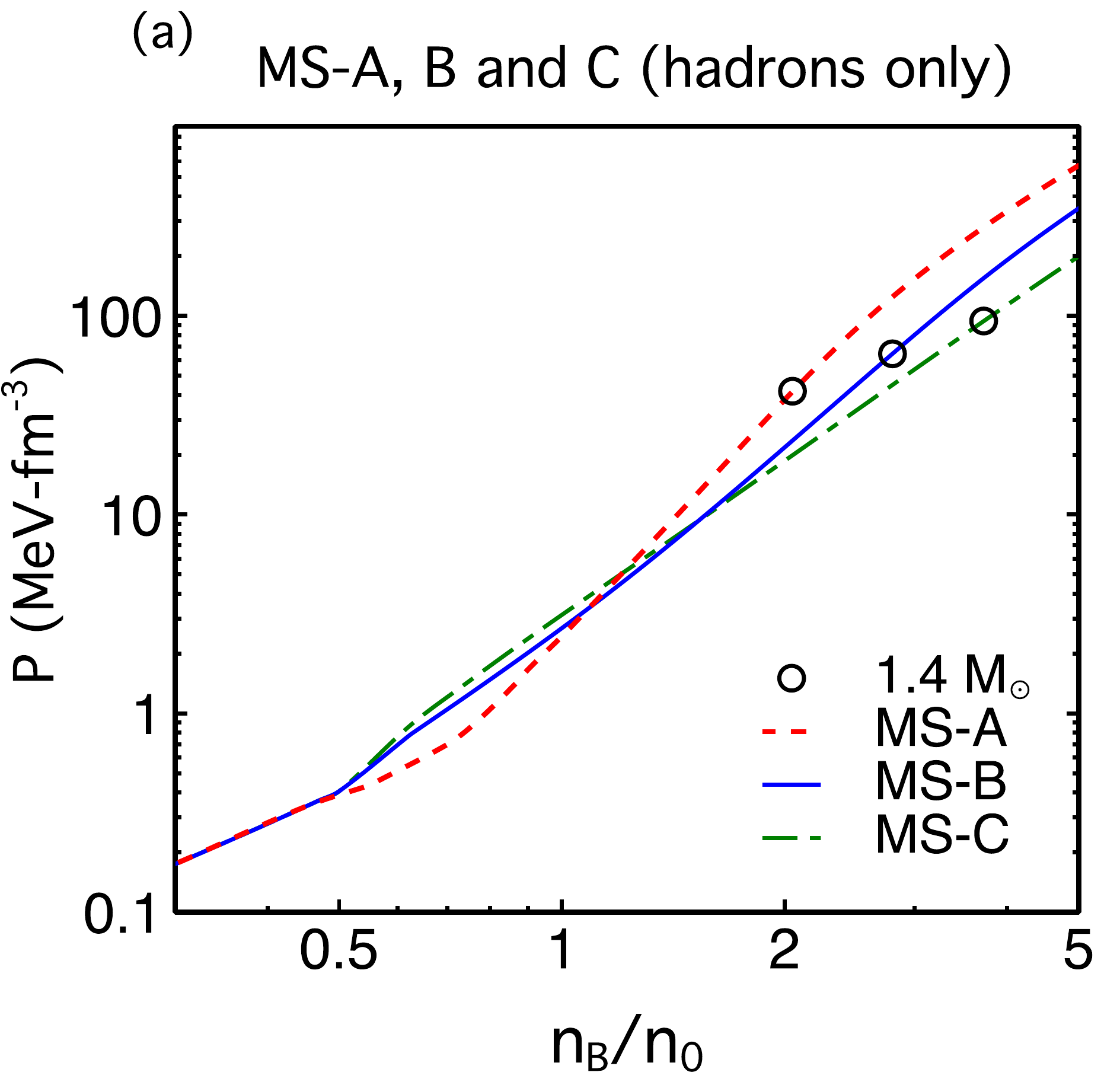}\\[-2ex]
}\parbox{0.45\hsize}{
\includegraphics[width=\hsize]{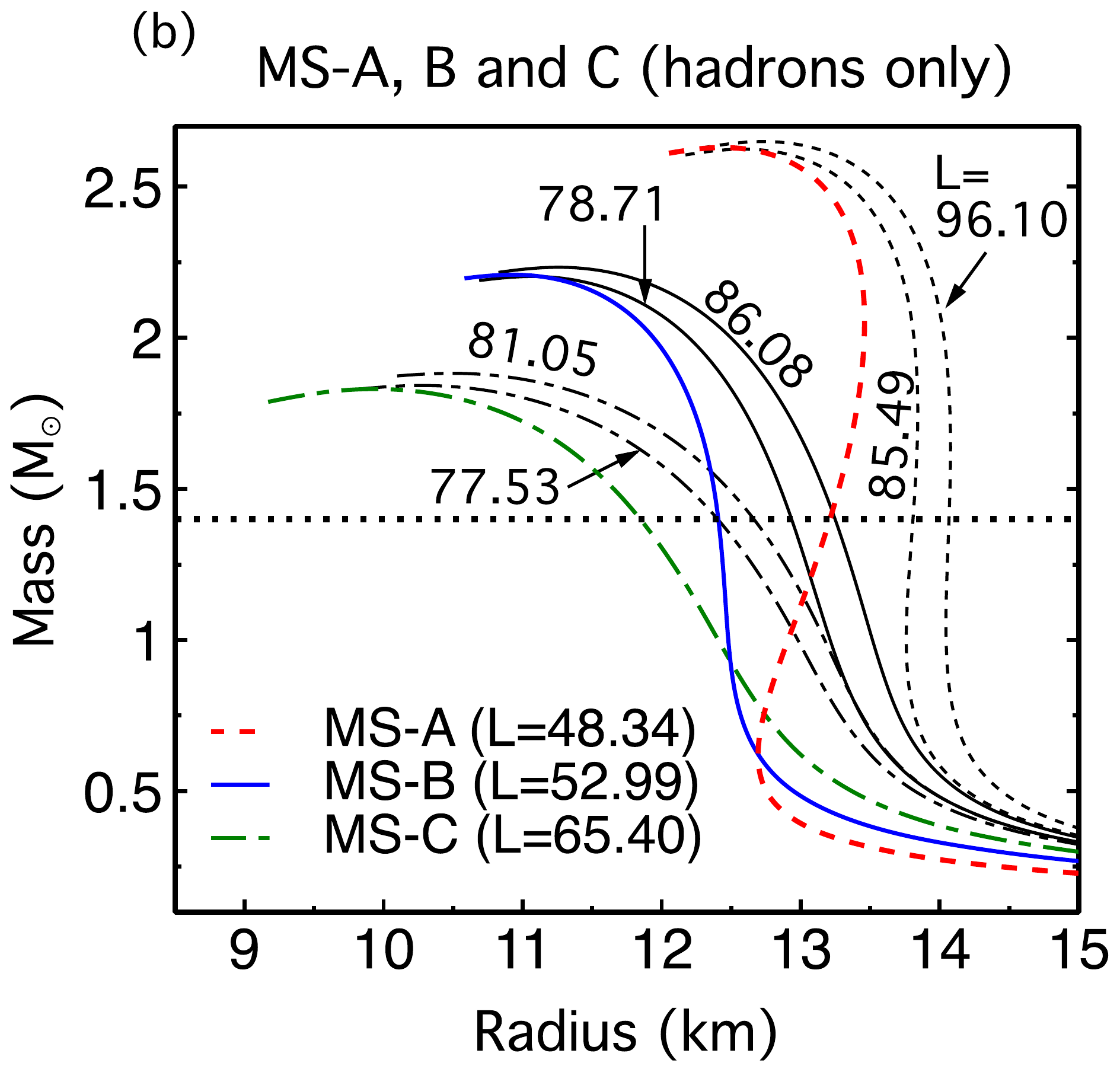}\\[-2ex]
}
\caption{(Color online) Pressure versus baryon density and $M$-$R$ curves for the MS models in Table~\ref{tab:Struc2}. The circles on the pressure curves in panel (a) shown indicate the densities of $1.4\,\Msolar$ stars for EoSs that yield the smallest radius $\Rtyp$ in each class of MS models.}
\label{fig:LR}
\end{figure*}


An examination of $L$ in Table~\ref{tab:Props1} and $\Rtyp$ and $R_{\rm max}$ in Table~\ref{tab:Struc1} would seem to imply an anti-correlation between these quantities for the MS models. That is, smaller values of $L$ appear to lead to larger values of $\Rtyp$ and $R_{\rm max}$, which is a trend opposite to that observed for many EoS models. The reason for this reversal becomes clear when $L$'s corresponding to different $m^*$'s within the same model are compared, see Fig.~\ref{fig:LR} (b) and Table~\ref{tab:Struc2}. In other words, the standard $L-R$ correlation holds within a class  of MS models with the same effective mass, whereas there exists an anti-correlation between $m^*-R$ which, if not taken into account explicitly, manifests itself as a turnabout in $L-R$. Similar trends of correlation with $L$ and anti-correlation with $m^*$ are also seen in Ref.~\cite{Hornick:2018kfi} which used the MS Lagrangian but without the term involving $\Lambda_{\sigma}$ in Eq. (2) as in Ref.~\cite{Chen:2014sca}. Fig. 7 of Ref.~\cite{Hornick:2018kfi} suggests that, when both $m^*$ and $L$ are varied, $L$ and $R$ can appear correlated, anti-correlated, or uncorrelated. The latter two possibilities are due to the competing effects of $m^*$ and $L$ on neutron star radii. We have verified that nonrelativistic potential models also yield similar trends, which are not shown here for brevity.

Moreover, further examination of the $P$ versus $\nb$ and $M$-$R$ relations for the MS models (Fig.~\ref{fig:LR}) shows that the central densities of $1.4\,\Msolar$ stars for the EoSs chosen are all $\geq 2\,n_0$ with that of the MS-C star being the largest. The symmetry energy slope parameter $L$ however, refers to that at $n_0$. The behaviors of the pressures (see panel (a) in this figure) at $\nb \geq 2\,n_0$ for all of these EoSs are distinctly different from their corresponding behaviors at $\nb \simeq n_0$. The $M$-$R$ curves in Fig.~\ref{fig:LR} (b) and Table~\ref{tab:Struc2} also clearly show how the value of $n_{c,1.4}$ differs in each of these cases. Evidently, the manner in which the size of a $1.4\,\Msolar$ is built depends sensitively on the behavior of the EoS well above $n_0$. These features deliver the alert that the standard $L-\Rtyp$ correlation involves more subtleties than generally thought.

\begin{table}[!htb]
\caption{Effective mass dependence of the $L$-$\Rtyp$ relation. Entries are as in Tables~\ref{tab:Props1} and \ref{tab:Struc1}, but organized differently.}

\begin{center} 
\begin{tabular}{crrrrrc}
\hline
\hline
Model &  $m^*/M$ &  $\La_\sigma$ &   $\La_\omega$ & $L \,({\rm MeV})$ & $\Rtyp \,({\rm km})$  & $n_{c,1.4}/n_0 $  \\ \hline
MS-A & 0.662 & 0.001 & 0.05 & 48.34 & 13.21 & 2.05 \\ 
          & 0.662 & 0.001 & 0.01 & 85.49 & 13.81 & 2.00 \\
          & 0.662 & 0.0 & 0.0 & 96.1 & 14.07 & 1.93 \\
           \hline \hline
MS-B  & 0.763 & 0.001 & 0.05 & 52.99 & 12.41 & 2.80 \\ 
           & 0.763 & 0.001 & 0.01 & 78.71& 12.93  & 2.68 \\
           & 0.763 & 0.0 & 0.0 & 86.08 & 13.25 & 2.53 \\
           \hline \hline
MS-C  & 0.847 & 0.001 & 0.05 & 65.40 & 11.85 & 3.72 \\ 
           & 0.847 & 0.001 & 0.01 & 77.53 & 12.41 & 3.35 \\
           & 0.847 & 0.0 & 0.0 & 81.05 & 12.67 & 3.12 \\
\hline \hline
\end{tabular}
\end{center}
\label{tab:Struc2}
\end{table}

\subsection*{Quark EoSs}
\label{sec:QEoSs}

For completeness, we briefly describe the quark matter EoSs considered in this work; details can be found in the references cited. Since the discovery of $2\,\Msolar$ neutron stars \cite{Demorest:2010bx,Antoniadis:2013pzd,Fonseca:2016tux,Cromartie:2019kug}, the traditional MIT bag~\cite{Baym:1976yu} and NJL \cite{Kunihiro:1989my} models have been supplemented with vector interactions \cite{Klahn:2015mfa} to achieve consistency with data. These models have been termed vMIT, vBag, vNJL, etc., and are outlined below. Common and different features of these models will be highlighted after a brief description of each model.

\subsubsection*{The bag model and its variations}
\label{sec:Bag}

The Lagrangian density of the MIT bag model is \cite{Baym:1976yu}
\ba
\mathcal{L} &=& \sum_i \big [~\overline{\psi}_i(i \slashed{\partial} - m_i - B) \psi_i + \mathcal{L}_{\rm int}~ \big] \Theta 
\ea
which describes quarks of mass $m_i$ confined within a bag as denoted by the $\Theta$ function. For three flavors $i=u,d,s$ and three colors $N_c=3$ of quarks, the number and baryon densities, energy density, pressure and chemical potentials in the bag model are \cite{Baym:1976yu}
\ba
n_i &=& 2 N_c  \int^{k_{Fi}} \frac {d^3k}{(2\pi)^3}\,, \quad \nb =  \frac 13 \sum_i n_i\nonumber \\
\ep_Q &=& 2 N_c \sum_i \int^{k_{Fi}} \frac {d^3k}{(2\pi)^3} \, \sqrt{k^2+m_i^2}  + \ep_{\rm pert} + B\nonumber \\
P_Q &=& \frac {2 N_c}{3} \sum_i \int^{k_{Fi}} \frac {d^3k}{(2\pi)^3} \, \frac {k^2}{\sqrt{k^2+m_i^2}}  + P_{\rm pert}  - B \nonumber \\
\mu_{i} &=& {\sqrt{k_{Fi}^2+m_i^2}} + \mu_{{\rm pert},i} \,.
\label{eqn:Bag}
\ea
The superscript $k_{Fi}$ in the integral signs is the Fermi wave number for each species $i$, at which the integration over $k$ is terminated at zero temperature. The first terms in $\ep_Q$ and $P_Q$ are free Fermi gas contributions, $\ep_{\rm FG}$ and $P_{\rm FG}$, respectively,  the second terms are QCD perturbative corrections due to gluon exchange corresponding to ${\cal L}_{\rm int}$, and $B$ is the so-called bag constant which accounts for the cost in confining the quarks into a bag. The quark masses $m_i$ are generally taken to be current quark masses. Often, the $u$ and $d$ quark masses are set to zero (as at high density, $k_{Fi}$ in these cases far exceed $m_i$), whereas that of the $s$ quark is taken at its Particle Data Group (PDG) value. Refs. \cite{Baym:1976yu,Baym:1975va,Freedman:1977gz,Farhi:1984qu,Kurkela:2009gj,Kurkela:2016was,Gorda:2018gpy} detail the QCD perturbative calculations of $\ep_{\rm pert}$ and $P_{\rm pert}$, and the ensuing results for the structure of neutron stars containing quarks within the cores as well as self-bound strange quark stars. At leading order of QCD corrections, the results are qualitatively similar to what is obtained by just using the Fermi gas results with an appropriately chosen value of $B$ \cite{Prakash:1990at}.

In recent years, variations of the bag model have been adopted \cite{Klahn:2015mfa,Gomes:2018eiv,Xia:2019xax} to calculate the structure of neutron stars with quarks cores to account for $\geq 2\,\Msolar$ maximum-mass stars. Termed as vMIT or vBag models, the QCD perturbative results are dropped and replaced by repulsive vector interactions between quarks in such works. We will provide some numerical examples of the vMIT model for contrast with other models as those of the vBag model turn out to be qualitatively similar. 

\subsubsection*{The vMIT model}

The form $\mathcal{L}_{\rm int} = - G_v\sum_i ~\overline{\psi}\gamma_\mu V^\mu \psi + (m_V^2/2) V_\mu V^\mu$, where interactions among the quarks occur via the exchange of a vector-isoscalar meson $V^\mu$ of mass $m_V$, is chosen in Ref.~\cite{Gomes:2018eiv}. Here, the quark masses are chosen close to their current quark masses. Explicitly,  
\ba
 \ep_Q &=& \sum_i \ep_{\rm FG,i} + \frac {1}{2}  \left(\frac {G_v} {m_V}\right)^2   n_Q^2 + B \nonumber \\
 P_Q &=&  \sum_i P_{\rm FG,i}  + \frac {1}{2} \left(\frac {G_v} {m_V}\right)^2 n_Q^2 - B \nonumber \\
\mu_i &=& \sqrt{k_{Fi}^2+m_i^2} + \left(\frac {G_v} {m_V}\right)^2 n_Q \,,
\label{eqn:vMIT}
\ea
where $n_Q=\sum_i n_i$, and the bag constant $B$ is chosen appropriately to enable a transition to matter containing quarks. Note that terms associated with the vector interaction above are similar to those in hadronic models. In the results reported below, we vary the model parameters in the range $B^{1/4}=(155-180)$ MeV and $a=(G_v/m_V)^2=(0.1-0.3)~{\rm fm^{2}}$. 

\subsubsection*{The vNJL model}
\label{sec:vNJL}

In its commonly used form, the Lagrangian density for the vNJL model in the mean field approximation is 
\ba
\mathcal{L} &=& \overline{q}(i \slashed{\partial} - \hat{m_0})q  + 
G_s\sum_{k=0}^8 \bigg[ (\overline{q}\lambda_k q)^2 + (\overline{q}i\gamma_5\lambda_kq)^2 \bigg] \nonumber \\ 
&-& K \bigg[ {\rm det}_f(\overline{q}(1+\gamma_5)q) + {\rm det}_f(\overline{q}(1-\gamma_5)q) \bigg]  \nonumber \\
            &+& G_v\sum_i ~(\overline{q}\gamma^\mu q)^2 \,.
\label{LvNJL}
\ea
Here, $q$ denotes a quark field with three flavors $u,d,s$ and three colors, $\hat{m_0}$ is the 3$\times$3 diagonal current quark mass matrix, $\la_k$ represents the 8 generators of SU(3), and $\la_0$ is proportional to the identity matrix. The four-fermion interactions are from the original formulation of this model \cite{Nambu:1961tp}, whereas the flavor mixing, determinental  interaction is added to break the $U_A(1)$ symmetry~\cite{tHooft:1986ooh}. The last term accounts for vector interactions~\cite{Buballa:2003qv}. As the constants $G_s$, $K$, and $G_v$ are dimensionful, the quantum theory is non-renormalizeable. Therefore, an ultraviolet cutoff $\La$ is imposed, and results are meaningful only for quark Fermi momenta well below this cutoff.  

The Lagrangian density in \Eqn{LvNJL} leads to the energy density 
\ba
\ep &=& \ep_{\rm FG} + \ep_{\rm int}   \nonumber \\
\label{eqn:epsInt}
\ep_{\rm int} &=& -2N_c \sum_i \int^\La  \frac {d^3k}{(2\pi)^3} \left( {\sqrt{k^2+m_i^2}} - {\sqrt{k^2+m_{o,i}^2}} \right) \nonumber \\
&+& 2\,G_s \sum_i  \langle\overline{q}_iq_i\rangle - 4\,K \prod_i \langle\overline{q}_iq_i\rangle 
+ 2\,G_v \sum_i n_i^2  \,,
\ea
where the sums above run over $u,d,s$. The subscript $``0"$ denotes current quark masses and the superscript $\La$ in the integral sign indicates that an ultraviolet cutoff $\La$ is imposed on the integration over $k$. In both $\ep_{\rm FG}$ [see \Eqn{eqn:Bag}] and $\ep_{\rm int}$, the quark masses $m_i$ are dynamically generated by requiring that $\ep$ be stationary with respect to variations in the quark condensate $\langle \overline{q}_i q_i\rangle$: 
\ba
m_i = m_{0,i} - 4\,G_s \langle \overline{q}_i q_i\rangle + 2K \langle \overline{q}_j q_j\rangle \langle \overline{q}_k q_k\rangle \,,
\label{dynmass}
\ea
$(q_i,q_j,q_k)$ representing any permutation of $(u,d,s)$. The quark condensate $\langle \overline{q}_i q_i\rangle$ 
is given by 
\ba
\langle \overline{q}_i q_i\rangle &=& - 2N_c  \int_{k_{Fi}}^\Lambda \frac {d^3k}{(2\pi)^3} \frac {m_i}{{\sqrt{k^2+m_i^2}}} \,,
\label{scaden}
\ea
and the quark number density $n_i = \langle q_i^\dagger q_i\rangle$ is as in \Eqn{eqn:Bag}. Note that the integrals appearing in Eqs. (\ref{eqn:epsInt})-(\ref{scaden}) can all be evaluated analytically. Eqs. (\ref{dynmass}) and (\ref{scaden}) render the dynamically generated masses $m_i$ density dependent, which tend to $m_{0,i}$ at high density mimicking the restoration of chiral symmetry in QCD.  

To facilitate a comparison between the vMIT and vNJL models, Ref.~\cite{Buballa:1998pr} recommends a constant energy density $B_0=\ep_{\rm int}|_{m_u=m_d=m_s=0}$  to be added to $\ep_{\rm int}$ which makes the vacuum energy density zero. With this addition, the energy density takes the form $\ep = \ep_{\rm FG} + B_{\rm eff}$, with $B_{\rm eff}=B_0 + \ep_{\rm int}$. 

The quark chemical potentials are
\ba
 \mu_i = \left. \frac {d\ep}{d n_i}\right|_{n_j,n_k}  = \sqrt{m_i^2 + k_{Fi}^2} + 4\,G_v n_i
\ea
using which the pressure is obtained from the thermodynamic identity
\ba
 P = \sum_i n_i \mu_i - \ep = P_{\rm FG} - B_{\rm eff}\,.
\ea
To mimic confinement absent in the vNJL model, often a constant term $B_{\rm dc}$ is used with the  replacement $B_{\rm eff} \rightarrow B_{\rm eff} - B_{\rm dc}$. 

For numerical calculations, we use the HK parameter set \cite{Hatsuda:1994pi}: $\La=631.4$ MeV, $G_s\La^2=1.835$, $K\La^5=0.29$, $m_{u,d}=5.5$ MeV, $m_s=135.7$ MeV and $B_{\rm dc}=0$.

\subsubsection*{The vBag model}

In Ref.~\cite{Klahn:2015mfa}, vector interactions are used in the form of flavor-independent four-fermion interactions as in the NJL models (described below): 
 $\mathcal{L}_{\rm int} = G_v\sum_i ~(\overline{\psi}\gamma^\mu \psi)^2$. In this case \cite{Klahn:2015mfa}, 
 \ba
 \ep_Q &=&  \sum_i \ep_{\rm FG,i} + \frac {G_v}{2} \sum_i n_i^2 + B_{\rm eff} \nonumber \\
 P_Q &=&  \sum_i P_{\rm FG,i}  + \frac {G_v}{2} \sum_i n_i^2 - B_{\rm eff} \nonumber \\
\mu_i &=& \sqrt{k_{Fi}^2+m_i^2} + G_v n_i \,,
\label{vBag1}
 \ea
 where the explicit forms of $\ep_{\rm FG,i}$ and $P_{\rm FG,i}$ can be read off from \Eqn{eqn:Bag}. The effective bag constant $B_{\rm eff}$ in this model is composed of two parts: $B_{\rm eff} = \sum_i B_\chi^i - B_{\rm dc}$, where the flavor-dependent  chiral bag constant
 \ba
 B_\chi^i &=& P(m_{0,i},k_{Fi}=0) - P(m_i,k_{Fi}=0) \nonumber \\
 &=& 2N_c \sum_i \int^\Lambda  \frac {d^3k}{(2\pi)^3} \left( {\sqrt{k^2+m_{o,i}^2}} - {\sqrt{k^2+m_i^2}} \right) \,, \nonumber \\
 \ea
where $m_i$ is the dynamically generated quark mass as in the NJL model, $m_{0,i}$ is the current quark mass, and $\La$ is an ultraviolet cut-off on the integration over $k$. The quantity $B_{\rm dc}$ is tuned to control the onset of quark deconfinement.

\subsubsection*{Charge neutrality and beta-equilibrium conditions}

Equilibrium with respect to weak-interaction processes $d \rightarrow u e^- \bar\nu_e$ and $s \rightarrow u e^- \bar\nu_e$ leads to the chemical potential equalities $\mu_d = \mu_u + \mu_e = \mu_s$ in neutrino-free matter. Charge neutrality requires that $2\,n_u - n_d- n_s- 3\,n_e=0$. Together with the baryon number relation $n_u + n_d + n_s = 3\,\nb$, the simultaneous solution of the equations
\ba
\label{Equil}
(1+x_e) + \left[ x_s^{2/3} + \frac {C_s-C_d}{r^{2/3}} \right]^{3/2} + x_s &=& 3 \nonumber \\
\left[ (1+x_e)^{2/3} + \frac {C_u}{r^{2/3}} \right]^{1/2} - \left[ x_s^{2/3} + \frac {C_s}{r^{2/3}} \right]^{1/2} + (3x_e)^{1/3} &=& 0 \nonumber \\
\ea
assures that quark matter with the three flavors $u, d, s$ is charge neutral and in beta equilibrium. In \Eqn{Equil}, $x_i=n_i/\nb$ denote the particle concentrations, $r=\nb/n_0$, and the factor $C_ i = {m_i^2}/({\pi^{4/3} n_0^{2/3} })$, $i=u,d,s$. Note that $C_i$ can depend on the density compression ratio $r$ through $m_ i \equiv m_i(r)$ as in the vNJL model. The concentrations of the $u$ and $d$ quarks are given by
\ba
x_u = 1+x_e \quad {\rm and} \quad x_d = \left[ x_s^{2/3} + \frac {C_s-C_d}{r^{2/3}}  \right]^{3/2} \,,
\ea
respectively. Owing to the charges carried by the quarks, the electron concentration in quark matter is generally very small with increasing $r$.

\subsubsection*{Distinguishing features of the quark EoSs}

The vMIT and vNJL models differ in important ways. Fashioned after the MIT bag for the nucleon, the vMIT model incorporates overall confinement of quarks within a giant bag \cite{Baym:1976yu,Baym:1975va} through its density-independent (non-perturbative) bag constant $B$. Repulsive vector interactions in this model are $\propto n_Q^2$ with $n_Q = \sum n_{u,d,s}$ in the pressure and energy density. The kinetic energy is calculated with current quark masses, although use of constituent masses $m_{u,d,s}=M_n/3$ can also be found in the literature. Effects of interactions are included from perturbative QCD calculations, but often they are set to zero in favor of an altered value of $B$ to simulate the same effect. 

The most important and distinguishing feature of the mean-field vNJL model is the chiral restoration of the quark masses present in the original QCD Lagrangian. Starting from the dynamically generated quark masses $m_u=m_d \simeq 350~{\rm MeV}$ and $m_s\simeq 525~{\rm MeV}$ in vacuum,  the masses decrease steadily toward their current quark values with increasing density. In our numerical calculations, we have used the current quark masses $m_{0,u} = m_{0,d} = 5$ MeV and $m_{0,s} = 140$ MeV to conform to the values used in Ref.~\cite{Buballa:1998pr} (use of the current PDG values $m_u\simeq 2~{\rm MeV}$, $m_d\simeq 5~{\rm MeV}$, and $m_s\simeq 100~{\rm MeV}$ does not significantly affect the results). In addition to generating the quark masses, the scalar field energies involving the couplings $G_s$ and $K$ also enter the energy density (and hence the pressure). Vector interactions in the vNJL model are $\propto \sum_i n_i^2$, $i=u,d,s$; this differs from the vMIT model in that cross terms such as $n_j n_k$ are absent in the former case. The vNJL model lacks confinement, although a constant $B_0$ is added to the energy density so that $B_{\rm eff} = B_0 + \ep_{\rm int}$ to facilitate comparison with the $B$ of the vMIT bag model. $B_{\rm eff}$ is however density dependent, unlike the $B$ of the vMIT  model.

In short, both the vMIT and vNJL models incorporate some aspects of the QCD Lagrangian, but only partially. Lacking a truly non-perturbative approach to QCD, we have explored both models as representative of the current status.

Note that in the vBag model, $B_\chi^i$ and $B_{\rm dc}$, and thus $B_{\rm eff}$, are independent of density. Unlike in the vNJL model in which all terms in the energy density and pressure are calculated with density-dependent dynamical masses $m_i$, the Fermi gas contributions in the vBag model are calculated with $m_i(k_{Fi} = 0)$. 

The striking similarity of the vMIT and vBag models is worthy of an explicit discussion. For the purpose of comparison, we can impose $a^{\rm vMIT} = G_v^{\rm vBag}$ (numerically), $B=B_{\rm eff}$ and set the quark masses the same. Then, the difference in the vector-interaction terms in $\ep_Q$ and $P_Q$ in Eqs. (7) and (14) becomes apparent. Specifically, those of vMIT are $\propto n_Q^2$ with $n_Q = \sum_{i}n_i$ whereas those of vBag are $\propto \sum_i n_i^2$. These differences are caused by the associated terms in the respective Lagrangians. The corresponding terms in the chemical potentials will be $ a^{\rm vMIT} n_Q$ for vMIT and $a^{\rm vMIT} n_i$ for vBag. When charge neutrality and beta equilibrium are imposed, even the Fermi gas parts in the two models will be different as the corresponding Fermi momenta will be different. Thus, although the two models look similar they are different because of the way interactions are treated.

Consequences of the vBag model on neutron star structure have been studied extensively in Refs.~\cite{Klahn:2015mfa,Wei:2018mxy} (and in this work), and will not be repeated here.

\section{Treatment of Phase Transitions}
\label{sec:treat}

\subsection* {First-order transitions}
\label{result:1st} 

The manner in which the hadron-quark transition occurs is unknown. Even if the phase transition is assumed to be of first-order, description of the transition depends on the knowledge of the surface tension $\sigma_s$ between the two phases~\cite{Alford:2001zr,Mintz:2009ay,Lugones:2013ema,Fraga:2018cvr}. In view of uncertainties in the magnitude of $\sigma_s$, two extreme cases have been studied in the literature. \\

\subsubsection*{Maxwell Construction} 

For very large values of $\sigma_s$, a Maxwell construction in which the pressure and chemical potential equalities, $P(H)=P(Q)$ and $\mu_n(H)=\mu_n(Q)$, are established between the two phases, hadronic (H) and quark (Q), has been deemed appropriate. In charge-neutral and beta-equilibrated matter, only one baryon chemical potential, often chosen to be $\mu_n$, is needed to conserve baryon number as local charge neutrality is implicit. The range of densities over which these equalities hold can be found using the methods detailed in Refs.~ \cite{Constantinou:2014hha,Lamb:1983djd}. \\

\subsubsection*{Gibbs Construction} 

For very low values of $\sigma_s$, a Gibbs construction in  which a mixed phase of hadrons and quarks is present is more appropriate \cite{Glendenning:1992vb,Glendenning:2001pe}. The description of the mixed phase is achieved by satisfying Gibbs' phase rules: $P(H)=P(Q)$ and $\mu_n=\mu_u+2\mu_d$. Further, the conditions of global charge neutrality and baryon number conservation are imposed through the relations 
\ba
Q &=& f \,Q(H) + (1-f) \, Q(Q) = 0 \nonumber \\
\nb &=& f \, \nb(H) + (1-f)\,  \nb(Q) \,, 
\ea
where $f$ represents the fractional volume occupied by hadrons and is solved for at each $\nb$. Unlike in the pure phases of the Maxwell construction, $Q(H)$ and $Q(Q)$ do not separately vanish in the Gibbs mixed phase. The total energy density is $\ep = f \,\ep(H) + (1-f) \,\ep(Q)$. Relative to the Maxwell construction, the behavior of pressure vs density is smooth in the case of Gibbs construction. Discontinuities in its derivatives with respect to density, reflected in the squared speed of sound $c_s^2=dP/d\ep$, will however be present at the densities where the mixed phase begins and ends. \\

The Maxwell and Gibbs constructions represent extreme cases of treating first-order phase transitions, and reality may lie in between these two cases. However, there are situations in which neither method can be applied as the required pressure and chemical potential equalities cannot be met for many hadronic and quark EoSs~\cite{Baym:2017whm}. In such cases, an interpolatory method which makes the transition a smooth crossover has been used \cite{Masuda:2012ed,Fukushima:2015bda,Kojo:2014rca,Baym:2017whm,Li:2018ayl}. \\

\subsection* {Crossover transitions}
\label{sec:crs} 

As it is not clear that a first-order phase hadron-to-quark transition at finite baryon density is demanded by fundamental considerations, crossover or second-order transitions have also been explored recently; see e.g. Refs.~\cite{Baym:2017whm,Dexheimer:2014pea,McLerran:2018hbz}. As details of results ensuing from the model of Ref.~\cite{Dexheimer:2014pea} have been recorded earlier in Refs.~\cite{Dexheimer:2009hi,Negreiros:2010hk}, we will only examine the cases of interpolated and quarkyonic models in what follows. 

\subsubsection*{Interpolated EoS}

We follow the simple recipe in Ref.~\cite{Masuda:2012ed} where the interpolated EoS in the hadron-quark crossover region is characterized by its central value $\bar{n}$ and  width $2\,\Ga$. Pure hadronic matter exists for $n\lesssim\bar{n}-\Ga$, whereas a phase of pure quark matter is found for $n\gtrsim\bar{n}+\Ga$. In the crossover region, $\bar{n}-\Ga\lesssim n \lesssim \bar{n}+\Ga$, strongly interacting hadrons and quarks coexist in prescribed proportions. The interpolation is performed for pressure vs baryon number density according to
\ba
\label{eqn:p_crs} 
P(n)&=& P_{H} (n) f_{-}(n) + P_{Q}(n) f_{+}(n) \\ 
f_{\pm} (n) &=&\frac{1}{2} \left[ 1\pm \tanh \left( \frac{n-\bar{n}}{\Ga}\right)\right ] \,,
\ea
where $P_{H}$ and $P_{Q}$ are the pressure in pure hadronic and pure quark matter, respectively. The interpolated EoS for the crossover, \Eqn{eqn:p_crs}, is different from that of the Gibbs construction within the conventional picture of a first-order phase transition in that the pressure equality between the two phases has been abandoned. Also, $f_{-}$ and $f_{+}$ are not solved for, but chosen externally. (Alternative forms of interpolation have also been suggested in Refs.~\cite{Fukushima:2015bda,Kojo:2014rca}, but do not qualitatively change the outcome.) The energy density $\ep$ vs $n$ is obtained by integrating $P=n^2 \partial (\ep/n) / \partial n$:
\ba
\label{eqn:eps_crs} 
\ep(n)&=& \ep_{H} (n) f_{-}(n) + P_{Q} (n) f_{+}(n) + \De\ep  \nonumber \\ 
\De\ep&=&n \int_{\bar{n}}^{n} \mathrm{d} n^{'} \left[ \ep_{H} (n^{'})-\ep_{Q} (n^{'})\right ] \frac{g(n^{'})}{n^{'}} \,, \nonumber \\ 
g(n^\prime) &=&  \frac {{\rm sech}^2 X}{2\,\Ga}\,, \quad X=\frac {n^\prime- \bar {n}}{\Ga} \,.
\ea

\subsubsection* {Quarkyonic matter}

The transition to matter containing quarks in the model termed quarkyonic matter \cite{McLerran:2007qj,McLerran:2018hbz} is of second or higher order, depending on the  behavior of the squared speed of sound $c_s^2=dP/d\ep = d \ln \mu/d \ln \nb = (n/\mu)  (d^2P/d \mu^2)^{-1}$ with $\nb$. The order of the phase transition is not determined by the quarkyonic matter scenario a priori, but depends on its specific implementation. In the model proposed in Ref.~\cite{McLerran:2018hbz}, $c_s^2$ exhibits a kink at the onset of the transition, hence its derivative with respect to $\nb$ is discontinuous. It is in this sense that the transition is of second order for Ref.~\cite{McLerran:2018hbz} which may not be the case in other implementations of the quarkyonic matter scenario. This model is a departure from the first-order phase transition models insofar as once quarks appear, both nucleons and quarks are present until asymptotically large densities when the nucleon concentrations vanish. Keeping the structure of the quarkyonic matter model as in Refs.~\cite{McLerran:2007qj,McLerran:2018hbz} in which isospin symmetric nuclear matter (SNM) and pure neutron matter (PNM) were considered, we present below its generalization to charge-neutral and beta-equilibrated neutron star matter. 
In quarkyonic matter, the appearance of quarks is subject to the threshold condition \cite{McLerran:2018hbz}
\ba
k_{Fq} = \frac {(k_{FB} - \De)}{N_c}~\Theta (k_{FB} - \De) \,, 
\label{Qthresh}
\ea
where $k_{FB}$ is the baryon momentum, $N_c=3$ is the number of colors, and the momentum threshold $\De$ is chosen to be
\ba
\De = \frac {\La_Q^3}{k_{FB}^2} + \kappa \frac {\La_Q}{N_c^2} \,.
\ea
Above, $\La_Q \sim \La_{\rm QCD} \simeq 300-500$ MeV, and $\kappa \simeq 0.1-0.3$ is suitably chosen to preserve causality. In PNM, the transition density, $\ntrans$, for the appearance of quarks is $0.77\,(3.55)\, n_0$ for $\Delta = 300\,(500)$ MeV and $\kappa=0.3$, where $n_0$ is the SNM equilibrium density. The corresponding values for $\kappa=0.1$ are $0.75\, n_0$ and $3.47\, n_0$, respectively, and show weak dependence of $\ntrans$ on $\kappa$. Unlike in the other approaches, the transition density at which quarks begin to appear in this model is independent of the EoSs used in the hadronic and quark sectors, being dependent entirely on $\La_{\rm QCD}$ and large $N_c$ physics.

The total baryon density of quarkyonic matter is
 \ba
 \nb = \sum_{i=n,p} 2 \int_{N_c k_{Fq}}^{k_{Fi}} \frac {d^3k}{(2\pi)^3} + 
 \sum_{q=u,d,s} \frac {2 N_c}{3}   \int_{0}^{k_{Fq}} \frac {d^3k}{(2\pi)^3} \,. \nonumber \\
  \ea
Notice that once quarks appear, the shell width $\De$ in which nucleons reside decreases with density as $\nb^{-2/3}$, yielding the preponderance of quarks with increasing $\nb$. Including leptonic (electron and muon) contributions $\ep_\ell$, the total energy density is
\ba
\ep &=& \sum_{i=n,p }2 \int_{N_c k_{Fq}}^{k_{Fi}} \frac {d^3k}{(2\pi)^3}~e_k  + \ep_{\rm int}(n_n,n_p) \nonumber \\
&+& \sum_{q=u,d,s} 2 N_c \int_0^{k_{Fq}} \frac {d^3k}{(2\pi)^3}~{\sqrt {k^2 + M_q^2}} + \ep_{\rm int} (q_k,q_\ell) \nonumber \\
&+& \sum_{\ell=e^-,\mu^- }\ep_\ell \,,
\label{epsQua}
\ea
where $e_k$ is the single particle kinetic energy inclusive of the rest mass energy. The nucleonic part of the energy density for $n \gtrsim 0.5\,n_0$ can be taken from a suitable potential or field-theoretical model that is constrained by nuclear systematics near nuclear densities, and preserves causality at high densities. Below $0.5\, n_0$, the energy density is that of crustal matter as in e.g. Refs.~\cite{Negele73ns,Baym71tg}. It is important to realize that the term $\ep_{\rm int}(n_n,n_p)$ contributes in regions where $k_{FB} < \De$ as well as where $k_{FB} > \De$.

The chemical potentials and pressure are obtained from 
\ba
\mu_k = \left. \frac {\partial \ep}{\partial n_k} \right|_{n_j}\,,  ~ 
P &=& \nb^2 \frac {\partial (\ep/\nb)}{\partial \nb} +\sum_{\ell=e^-,\mu^- } P_\ell \nonumber \\
&=& \sum_k n_k \mu_k - \ep \,, 
\ea
where the sum above runs over all fermions.

As with nucleons, an appropriate  choice of the quark EoS is also indicated. Reference~\cite{McLerran:2018hbz} set $\ep_{\rm int} (q_k,q_\ell) = 0$, and the quark masses $M_q$ were taken as $M_n/3$. The use of the nucleon constituent quark masses takes account of quark-gluon interactions to a certain degree as has been noted in the case of finite temperature QCD as well. This procedure however, omits density-dependent contributions from interactions between quarks. In our work, we will employ quark models (such as vMIT, vNJL) in which contributions from interacting quarks are included. Subtleties involved in the calculation of  the kinetic part of the nucleon chemical potentials and in satisfying the thermodynamic identity are detailed in Appendix~\ref{sec:TI}.   

This model has a distinct behavior for $c_s^2=dP/d\ep$ vs $\nb$ in that $c_s^2$ exhibits a maximum (its location controlled by $\La_{Q}$, and the magnitude depending on both $\La_{Q}$ and $\kappa$) before approaching the value of 1/3 characteristic of quarks at asymptotically large densities~\cite{McLerran:2018hbz}.

\begin{figure*}[htb]
\parbox{0.35\hsize}{
\includegraphics[width=\hsize]{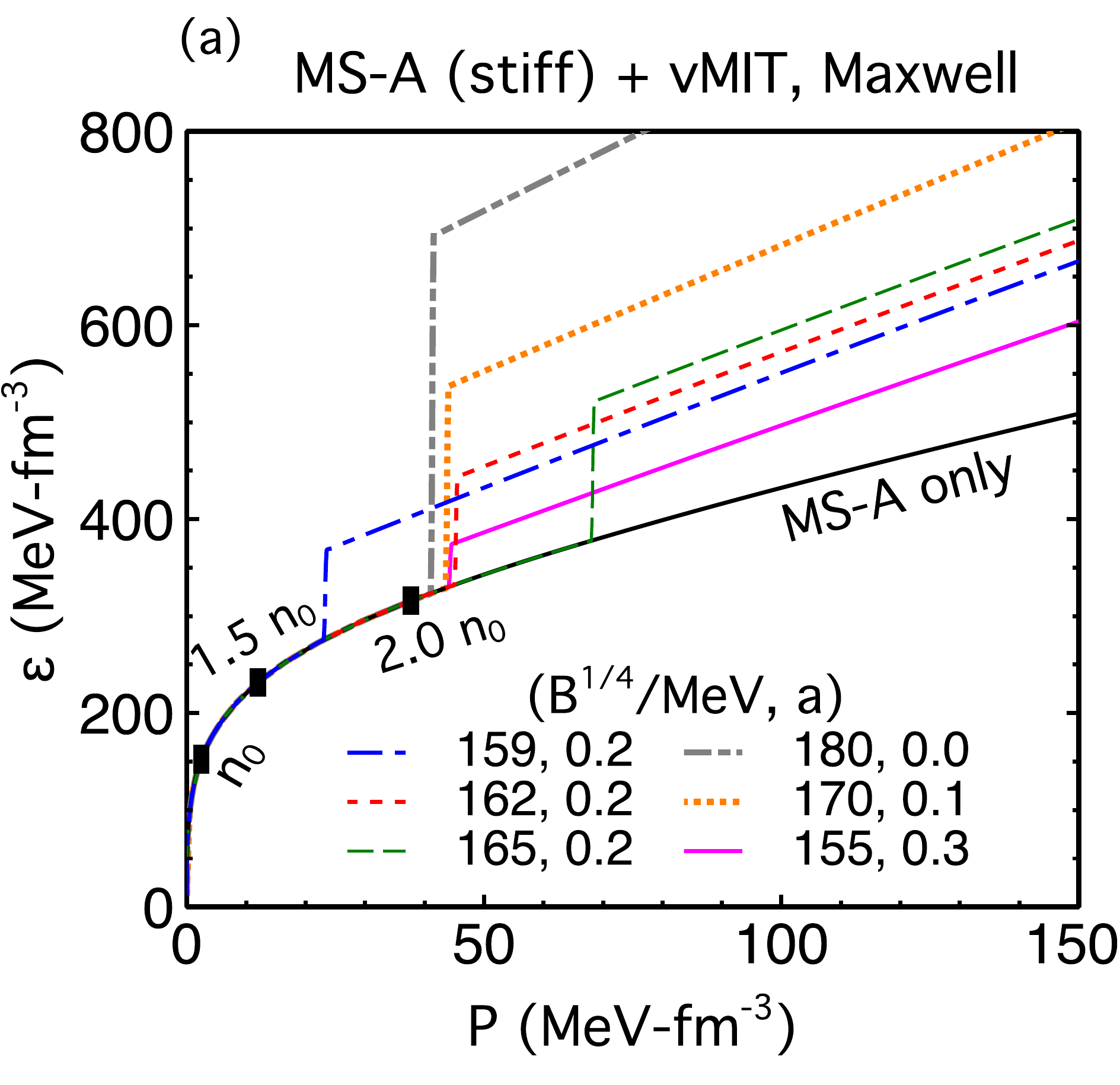}\\[-2ex]
}\parbox{0.35\hsize}{
\includegraphics[width=\hsize]{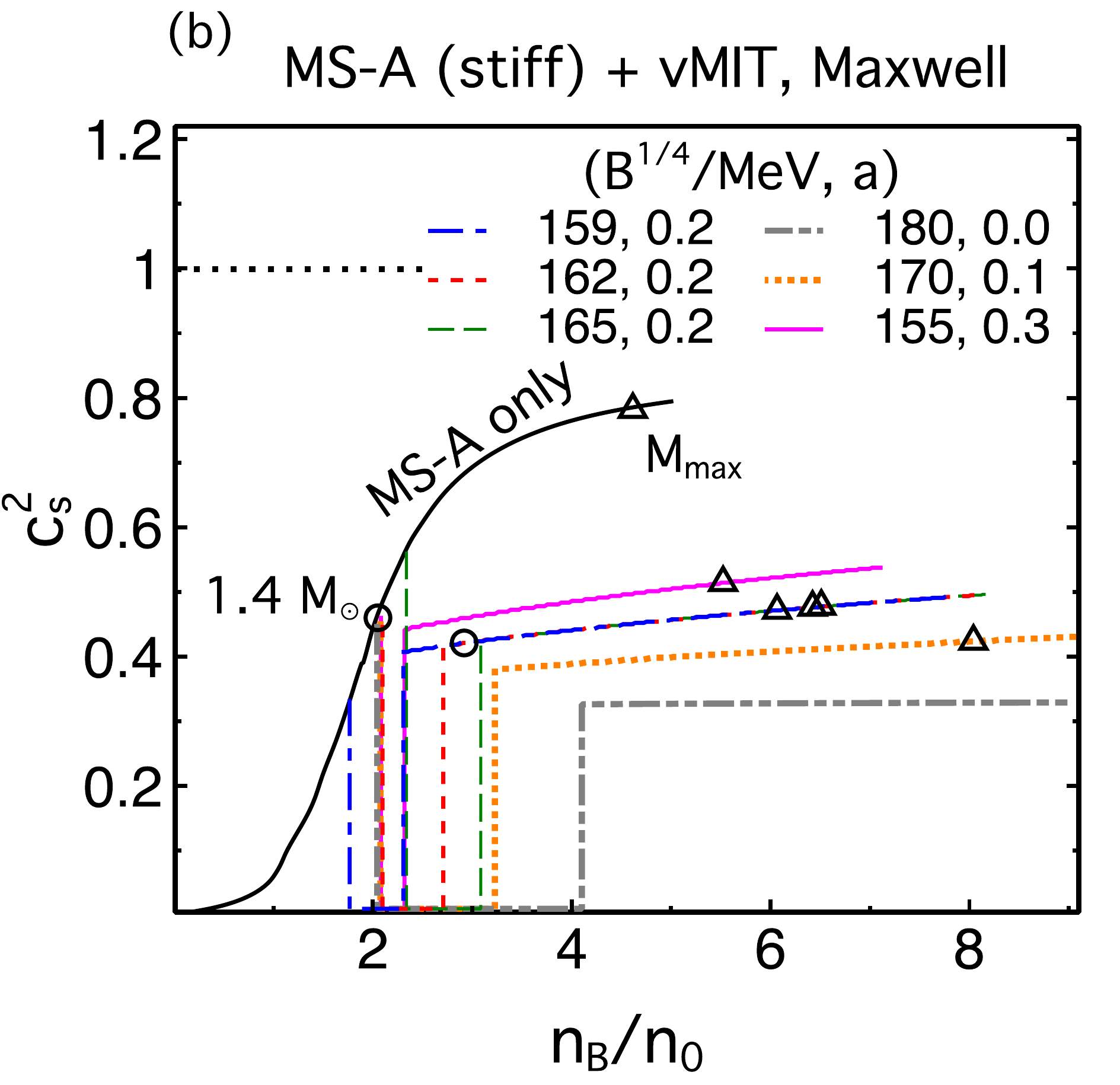}\\[-2ex]
}\parbox{0.35\hsize}{
\includegraphics[width=\hsize]{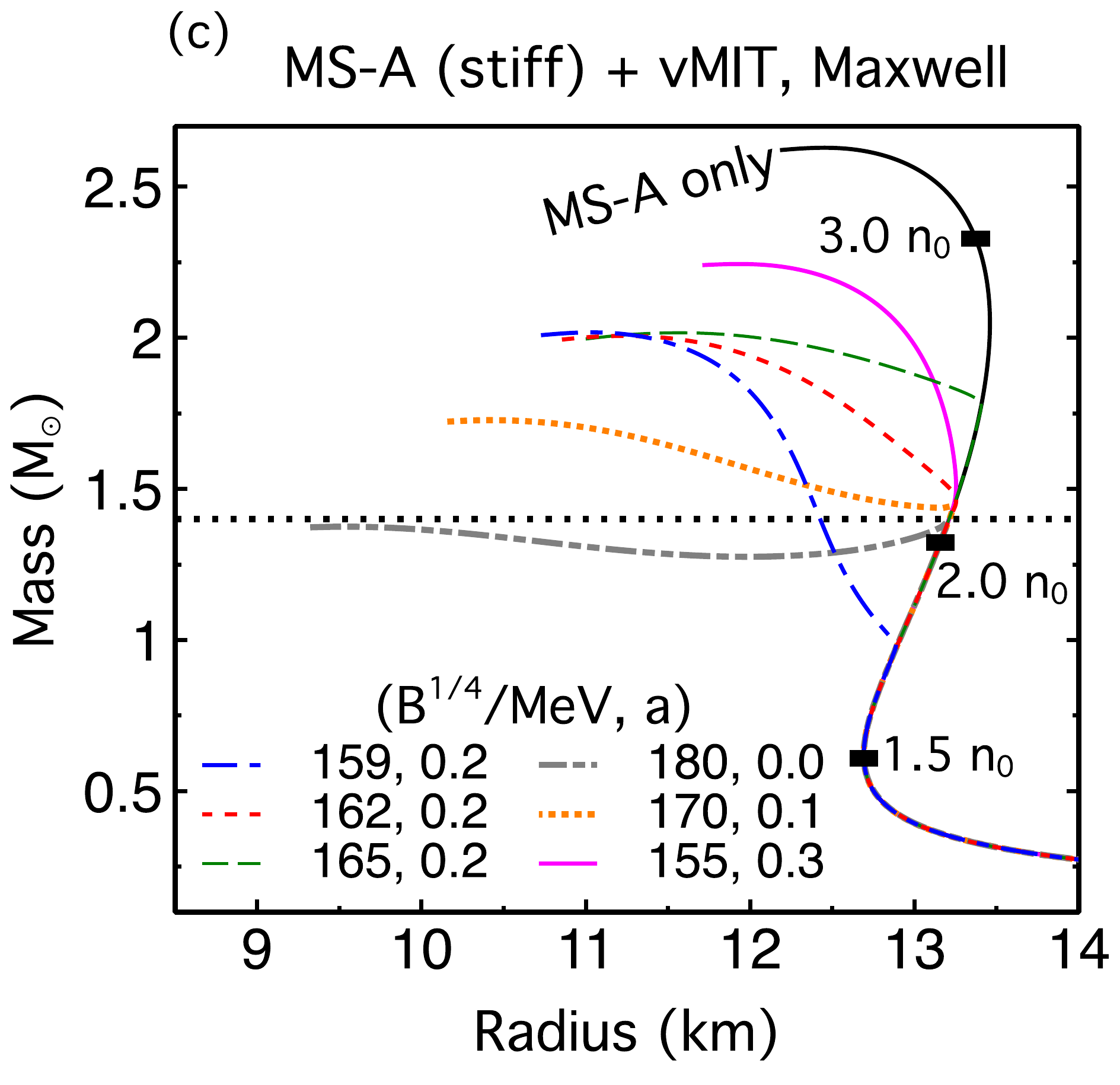}\\[-2ex]
}\\[3ex]
\parbox{0.35\hsize}{
\includegraphics[width=\hsize]{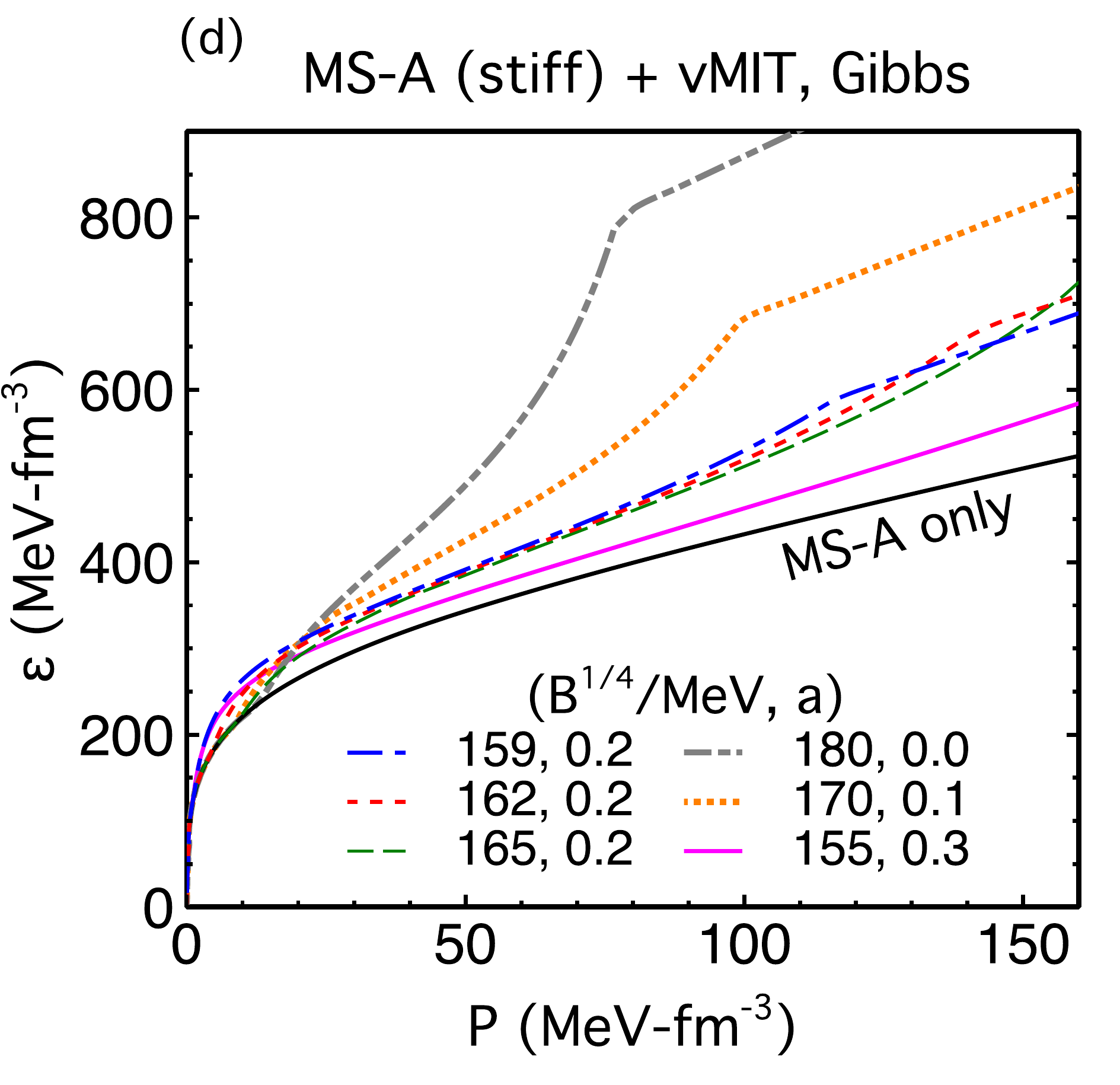}\\[-2ex]
}\parbox{0.35\hsize}{
\includegraphics[width=\hsize]{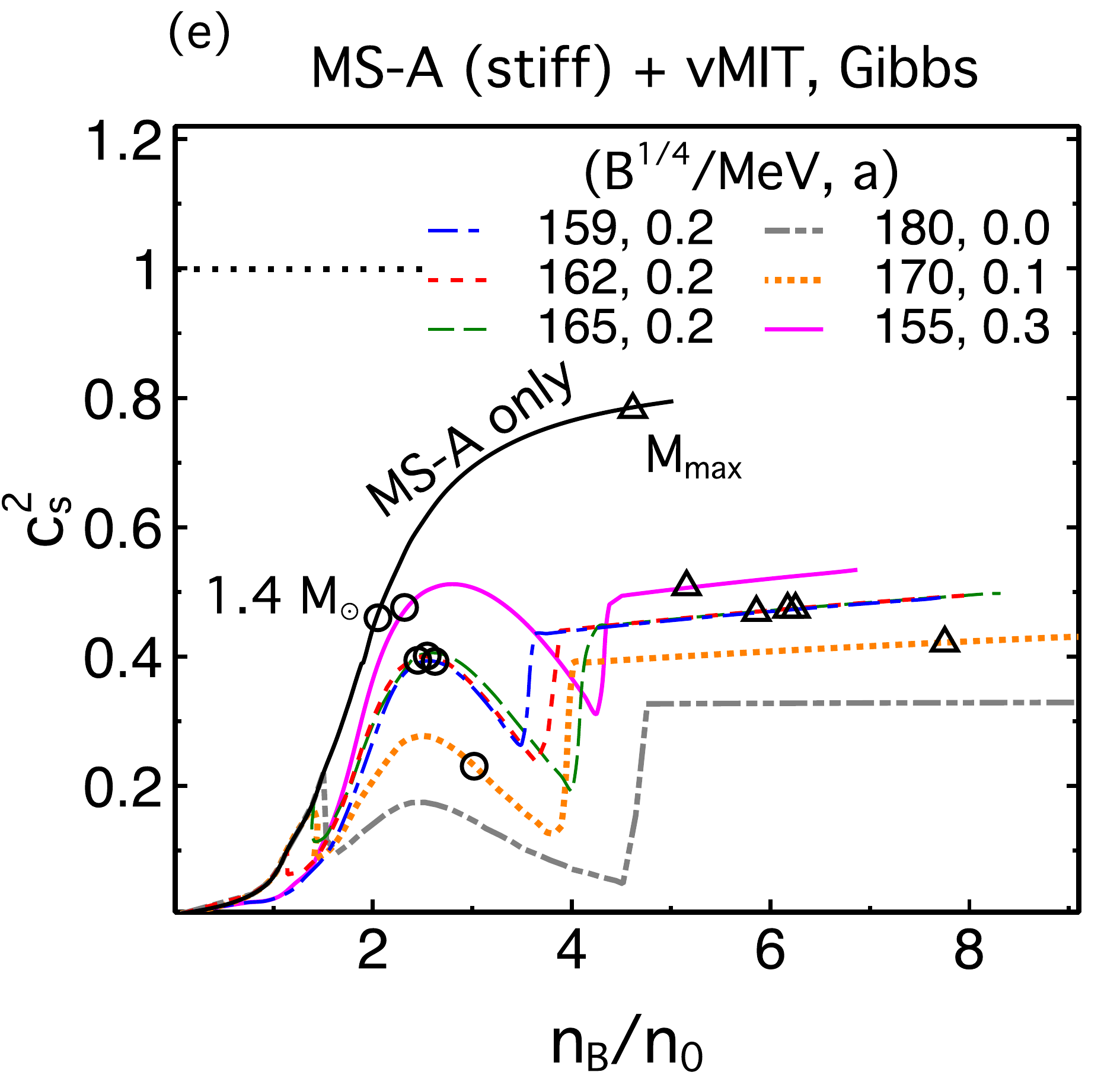}\\[-2ex]
}\parbox{0.35\hsize}{
\includegraphics[width=\hsize]{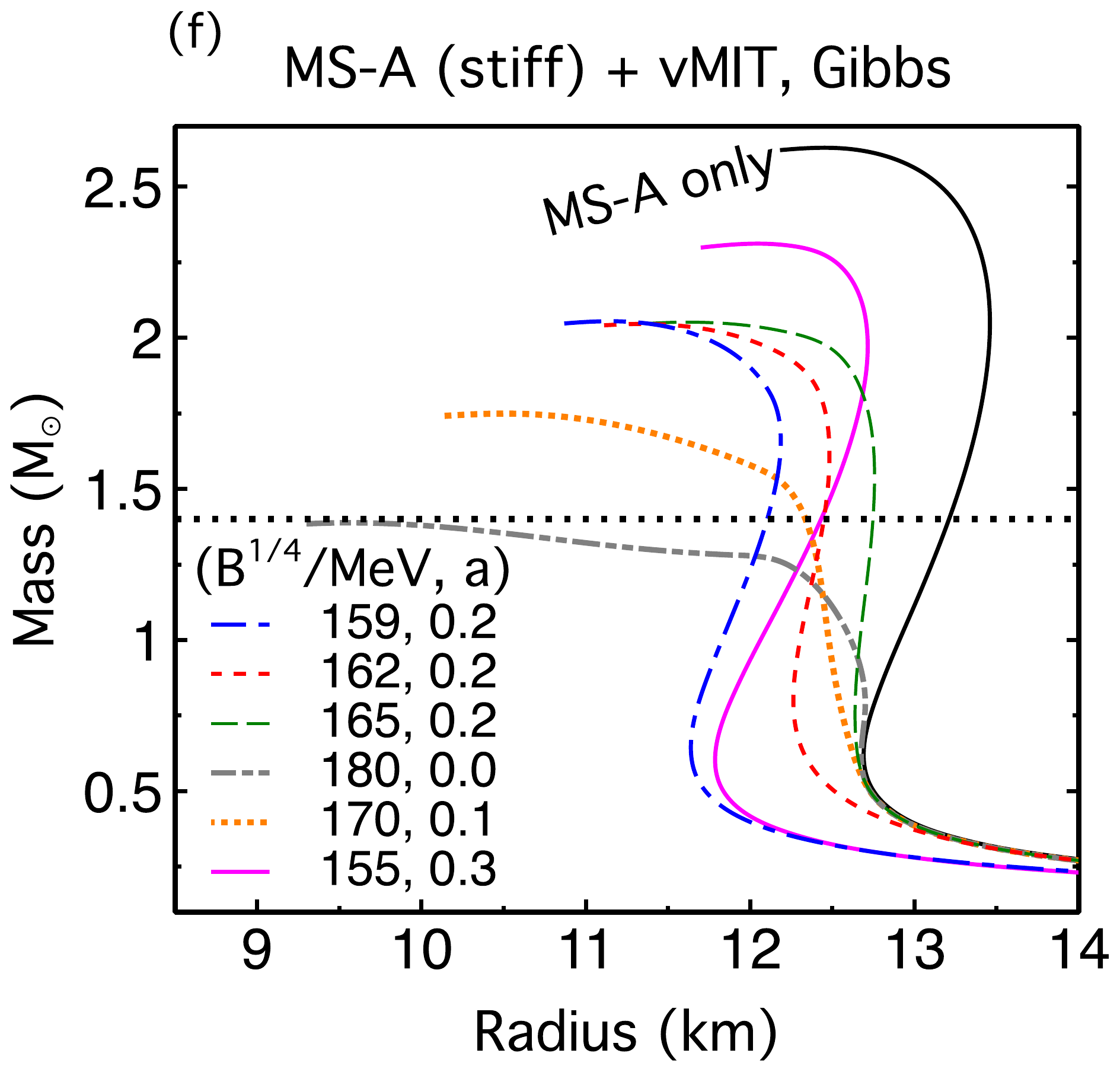}\\[-2ex]
}
\caption{(Color online) Energy density vs pressure, squared speed of sound vs ratio of baryon density to the nuclear equilibrium density, and mass vs radius curves for the models indicated. Panels (a)-(c) are for Maxwell construction, whereas (d)-(f) are for Gibbs construction; the quark model parameters used are in the inset and results are for beta-equilibrated matter. The $1.4\,\Msolar$ and maximum-mass stars are marked with open circles and triangles respectively in $c_s^2 (\nb)$ plots. 
}
\label{fig:MR-eos_vMIT}
\end{figure*}

\section{Results with Phase Transitions}
\label{sec:Results}

The hadronic EoSs chosen in this study satisfy the available nuclear systematics near the nuclear equilibrium density (see Tables \ref{tab:Couplings1}-\ref{tab:Struc1}). Their supra-nuclear density behavior can however, be varied to yield a soft or stiff EoS by varying the parameters in the chosen model. Depending on the quark EoS examined such as vMIT, vNJL or of quarkyonic matter, the examination of a broad range of transitions into quark matter - soft-to-soft, soft-to-stiff, stiff-to soft and stiff-to-stiff - become possible. 
For both first-order and crossover transitions, we calculate the mass-radius curves and tidal deformabilities, and then discuss the results in view of the existing observational constraints.  Of particular relevance to the zero-temperature EoS is the limit set by the data on the binary tidal deformability \cite{Flanagan:2007ix,Favata:2013rwa}
\be
\tilde{\La}=\frac{16}{13}\frac{(m_1+12m_2)m_1^4\,\La_1+(m_2+12m_1)m_2^4\,\La_2}{(m_1+m_2)^5}.
\label{CLam1}
\ee
For each star, the dimensionless tidal deformability (or induced quadrupole polarizability) is given by \cite{L1909}
\be
\Lambda_{1,2} = \frac {2}{3} k_2^{(1,2)} \left( \frac{R_{1,2}\,c^2}{G\,m_{1,2}} \right)^5 \,,
\label{CLam2}
\ee
where the second tidal Love number $k_2^{(1,2)}$ depends on the structure of the star, and therefore on the mass and the EoS. Here $G$ is the gravitational constant and $R_{1,2}$ are the radii. The computation of $k_2^{(1,2)}$ with input EoSs is described in Refs.~\cite{TC67,Hinderer:2007mb,Damour:2009vw}. For a wide class of neutron star EoSs, $k_2\simeq 0.05-0.15$ \cite{Hinderer:2007mb,Hinderer:2009ca,Postnikov:2010yn}.

Combining the electromagnetic (EM)~\cite{GBM:2017lvd} and gravitational wave (GW) information from the binary neutron star (BNS) merger GW170817, Ref.~\cite{Margalit:2017dij} provides constraints on the radius $R_{\rm ns}$ and maximum gravitational mass $M^g_{\rm max}$ of a neutron star: 
 \ba
 M^g_{\rm max} &\lesssim & 2.17\, \Msolar \,, \nonumber \\
 R_{1.3} &\gtrsim & 3.1\,GM^g_{\rm max} \simeq 9.92~{\rm km} \,,
 \ea
where $R_{1.3}$ is the radius of a $1.3\,\Msolar$ neutron star and its numerical value above corresponds to $M^g_{\rm max}=2.17\,\Msolar$. These estimates have been revisited in a recent analysis of Ref.~\cite{Shibata:2019ctb} where a weaker constraint on the upper limit of the maximum mass  $M^g_{\rm max} \lesssim  2.3 \,\Msolar$ has been reported. Combining the total mass measurement of $2.74^{+0.04}_{-0.01}\,\Msolar$ from GW170817 with (empirical) universal relations between the baryonic and the maximum rotating and non-rotating masses of neutron stars, Ref.~\cite{Rezzolla:2017aly} constrains the maximum non-rotating neutrons star mass in the range $2.01^{+0.04}_{-0.01}\,\Msolar\leq M_{\rm max}^{\rm nonrot} \leq 2.16^{+0.17}_{-0.15}\,\Msolar$.

\subsection* {First-order transitions: Maxwell vs. Gibbs}
\label{sec:1st}

\begin{figure*}[htb]
\parbox{0.35\hsize}{
\includegraphics[width=\hsize]{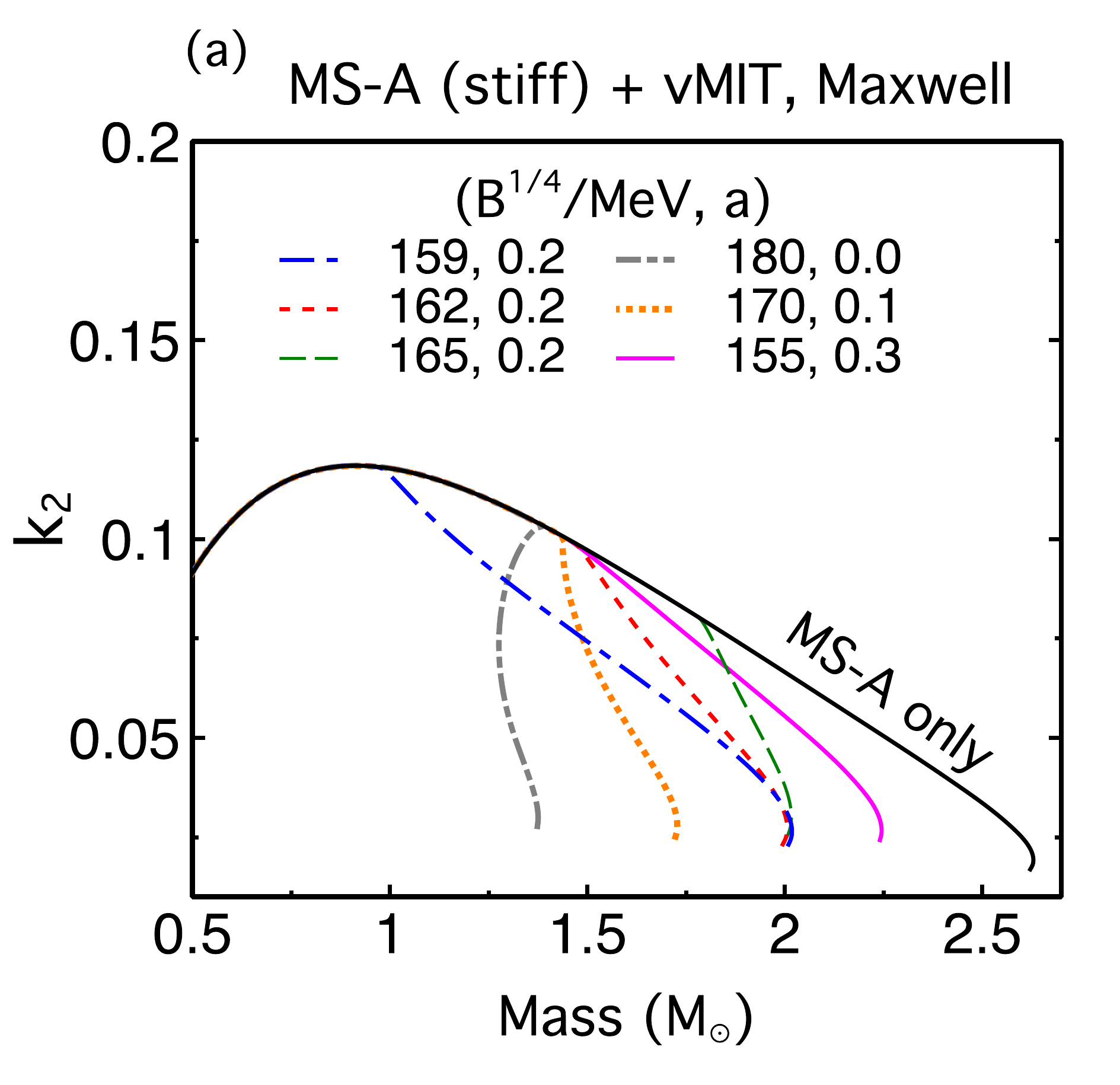}\\[-2ex]
}\parbox{0.35\hsize}{
\includegraphics[width=\hsize]{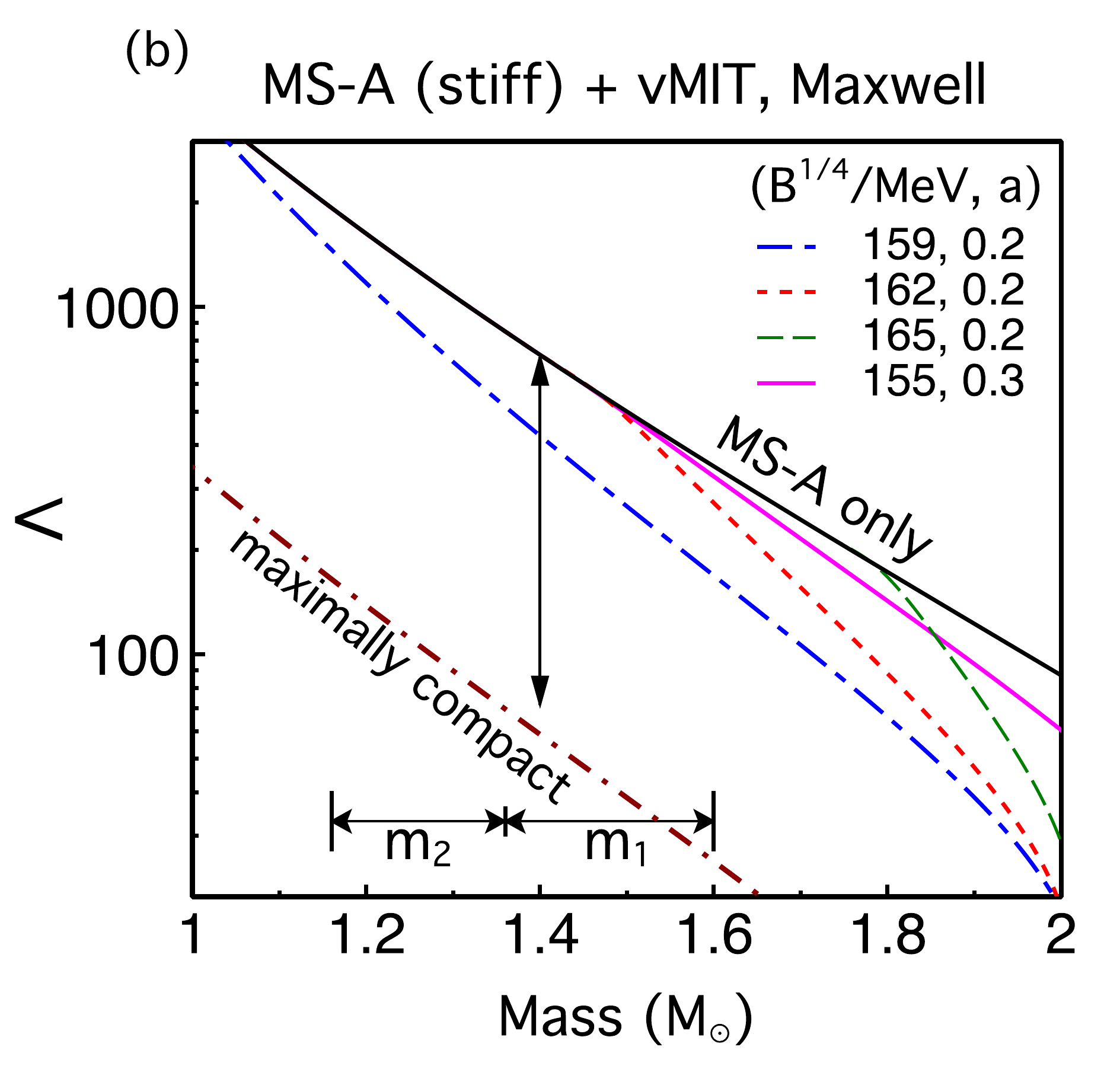}\\[-2ex]
}\parbox{0.35\hsize}{
\includegraphics[width=\hsize]{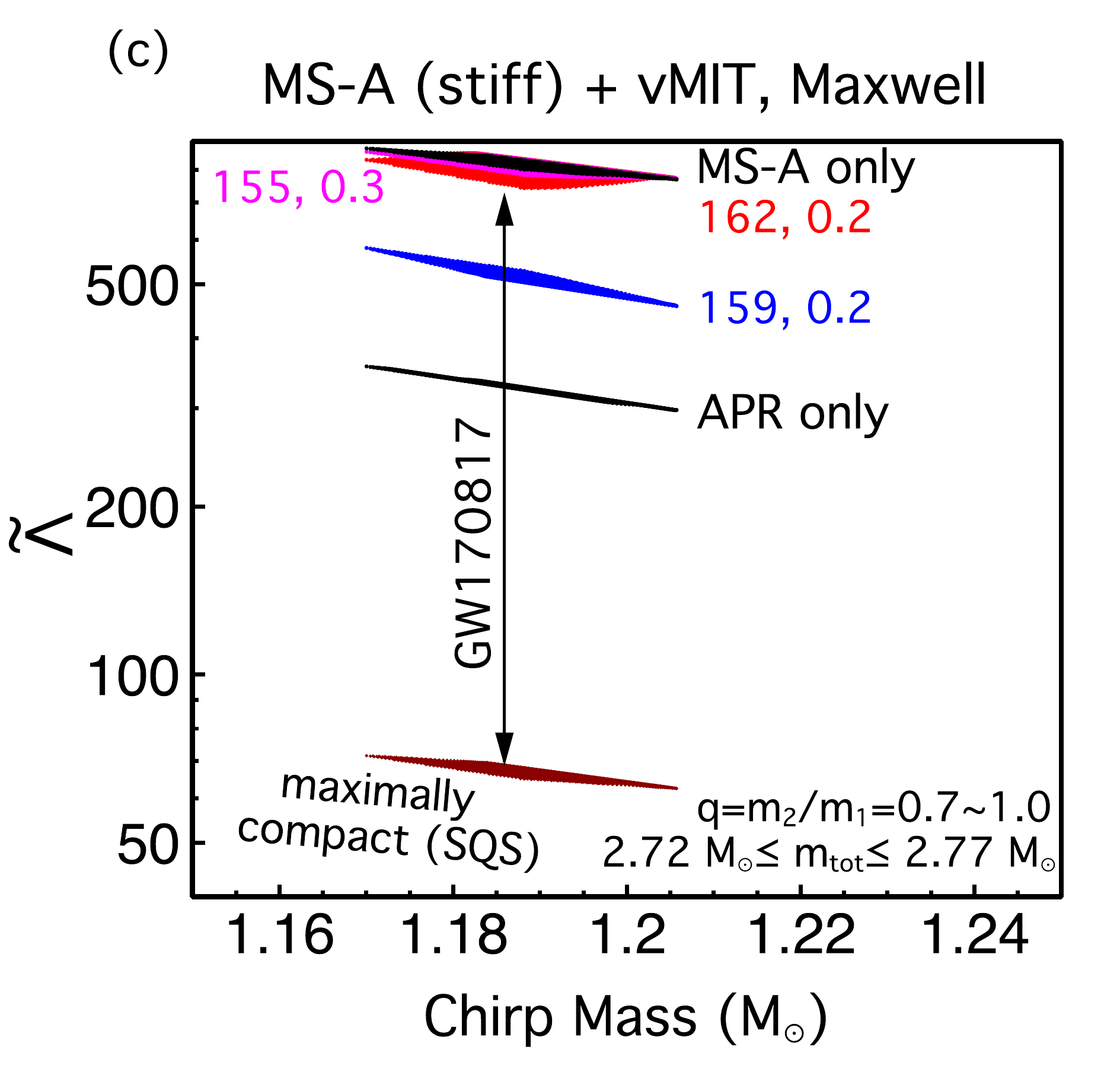}\\[-2ex]
}\\[3ex]
\parbox{0.35\hsize}{
\includegraphics[width=\hsize]{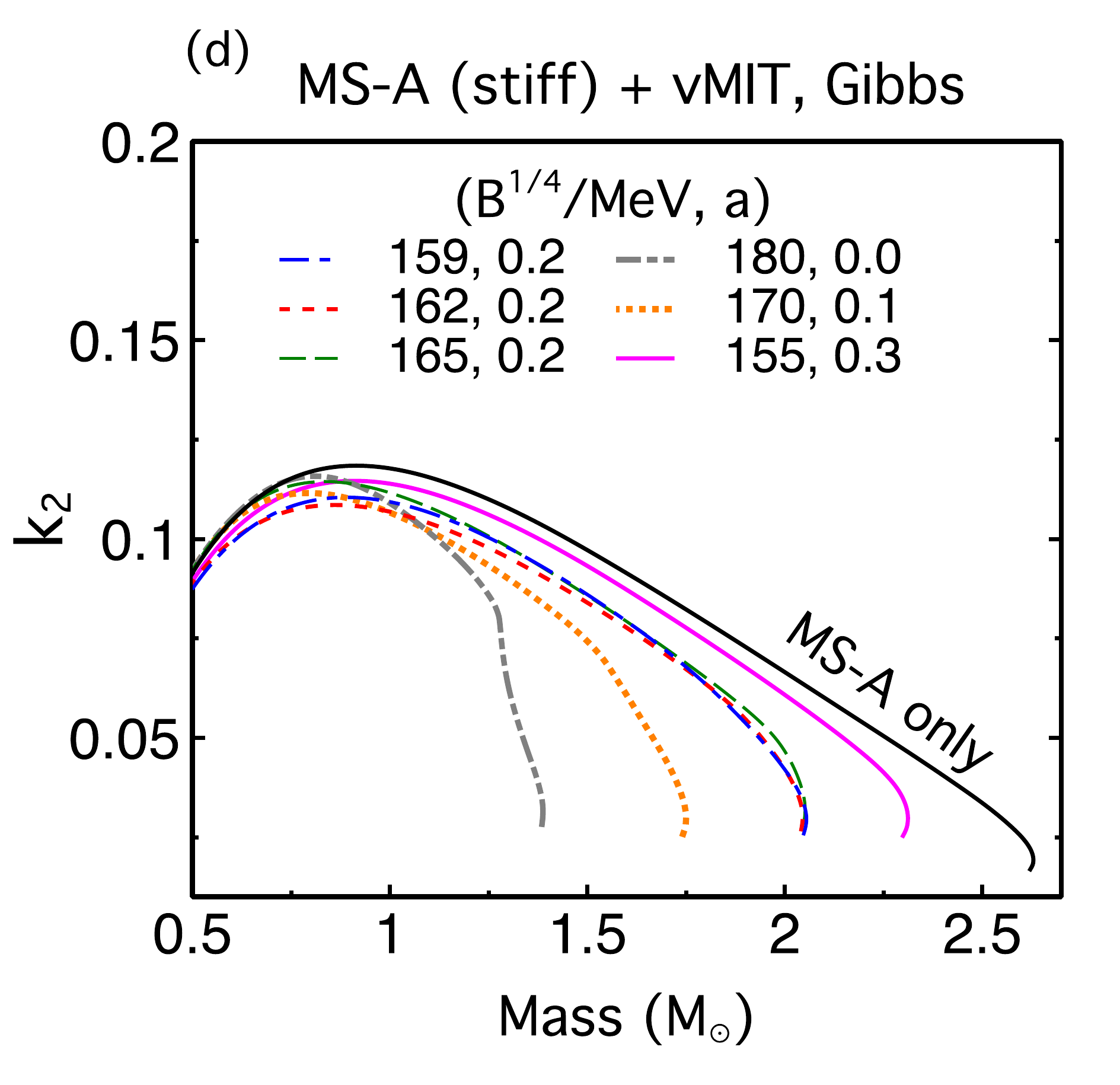}\\[-2ex]
}\parbox{0.35\hsize}{
\includegraphics[width=\hsize]{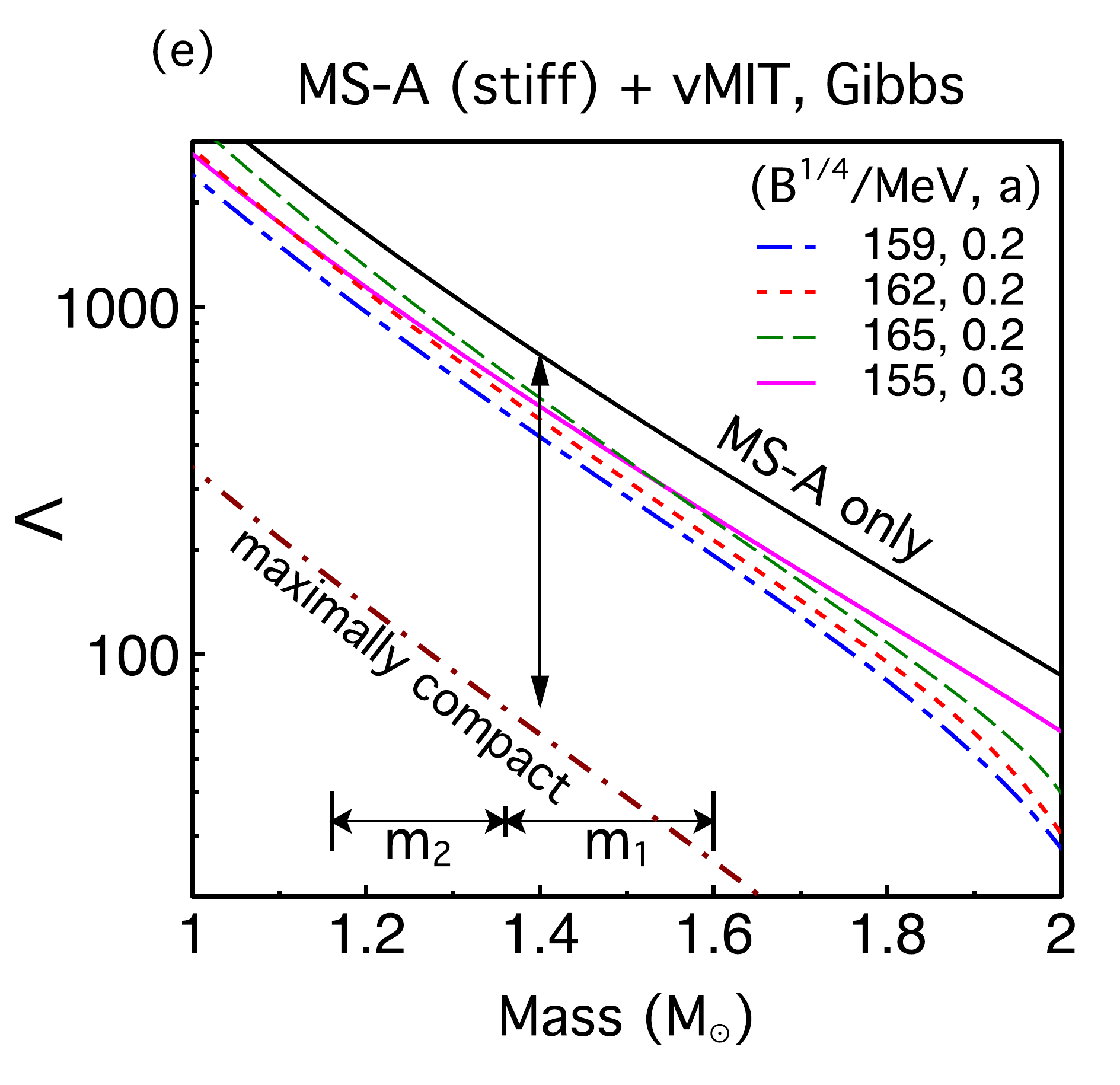}\\[-2ex]
}\parbox{0.35\hsize}{
\includegraphics[width=\hsize]{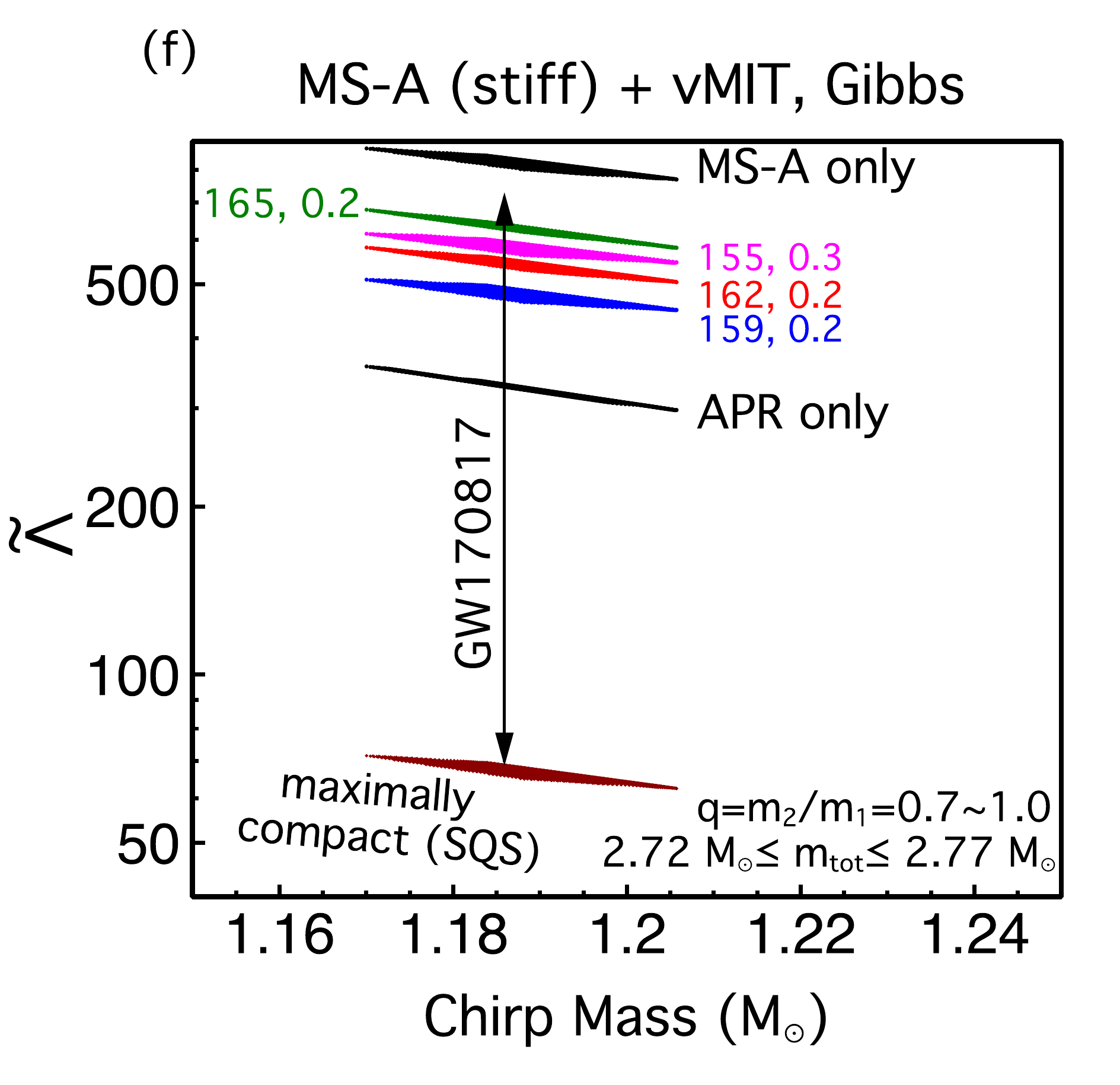}\\[-2ex]
}
\caption{(Color online) Tidal deformability parameters $k_2$ (Love number), $\La$ and $\tilde \La$ in Eqs. (\ref{CLam1}) and (\ref{CLam2}) as functions of the indicated masses. For comparison, results of $\tilde \Lambda$ for the EoS of the APR model are also shown; the bottom dash-dotted lines in (b) \& (e), as well as the dark red bands at the bottom in (c) \& (f), refer to the ``maximally compact EoS'' for self-bound strange quark stars (SQSs)~\cite{Lattimer:2015nhk}.
}
\label{fig:k2-Lam_vMIT}
\end{figure*}

\begin{figure*}[htb]
\parbox{0.34\hsize}{
\includegraphics[width=\hsize]{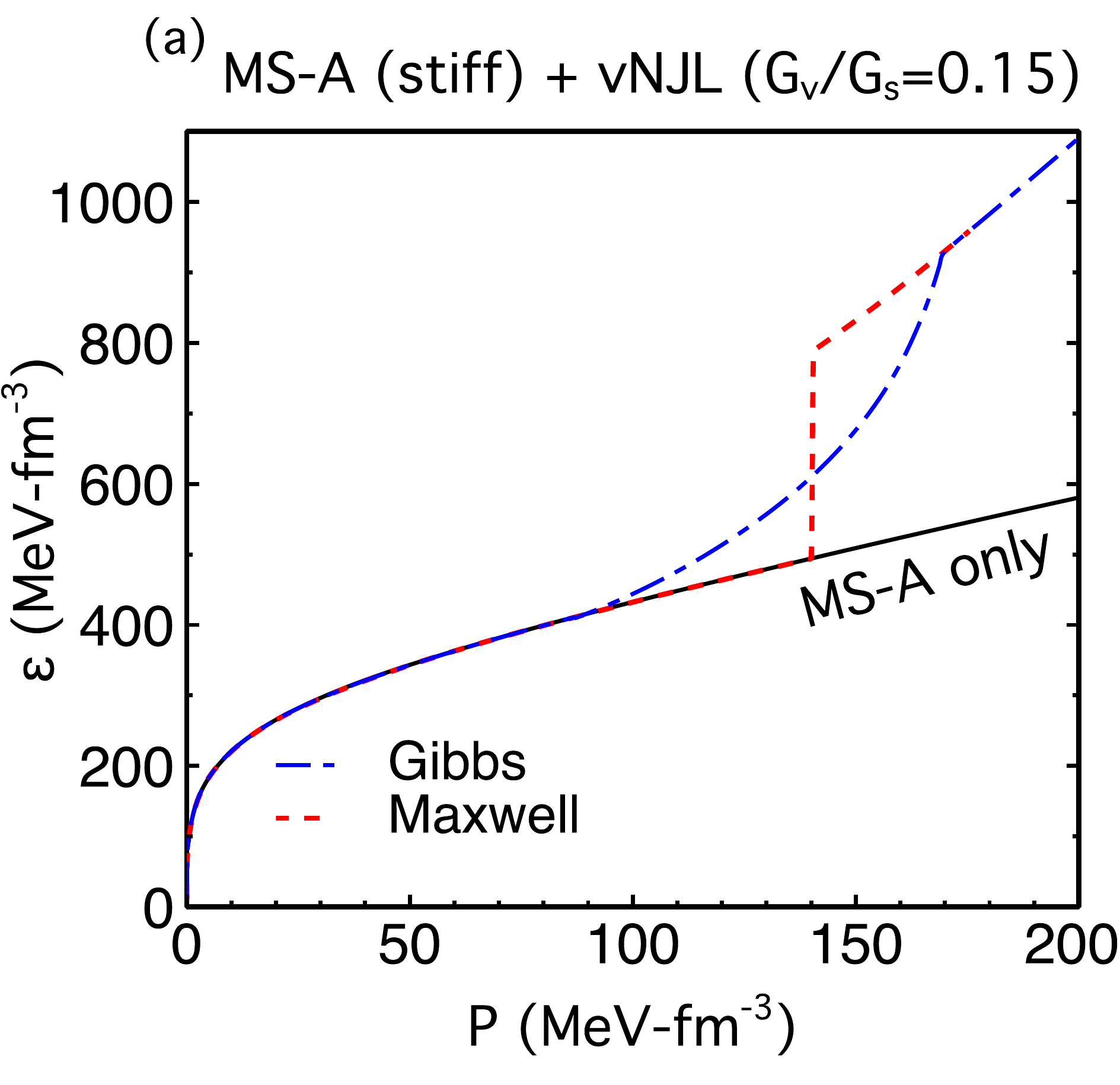}\\[-2ex]
}\parbox{0.34\hsize}{
\includegraphics[width=\hsize]{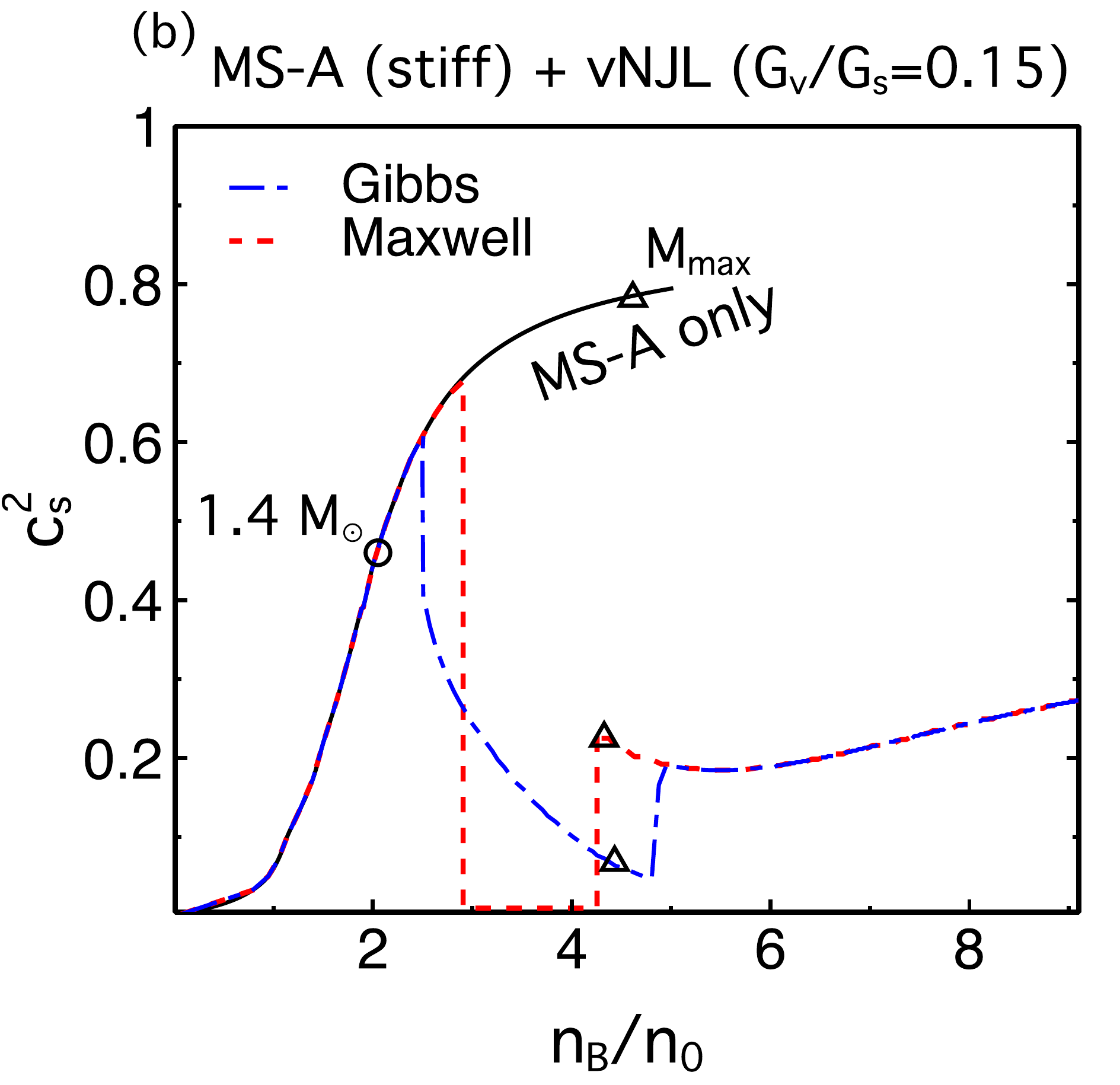}\\[-2ex]
}\parbox{0.34\hsize}{
\includegraphics[width=\hsize]{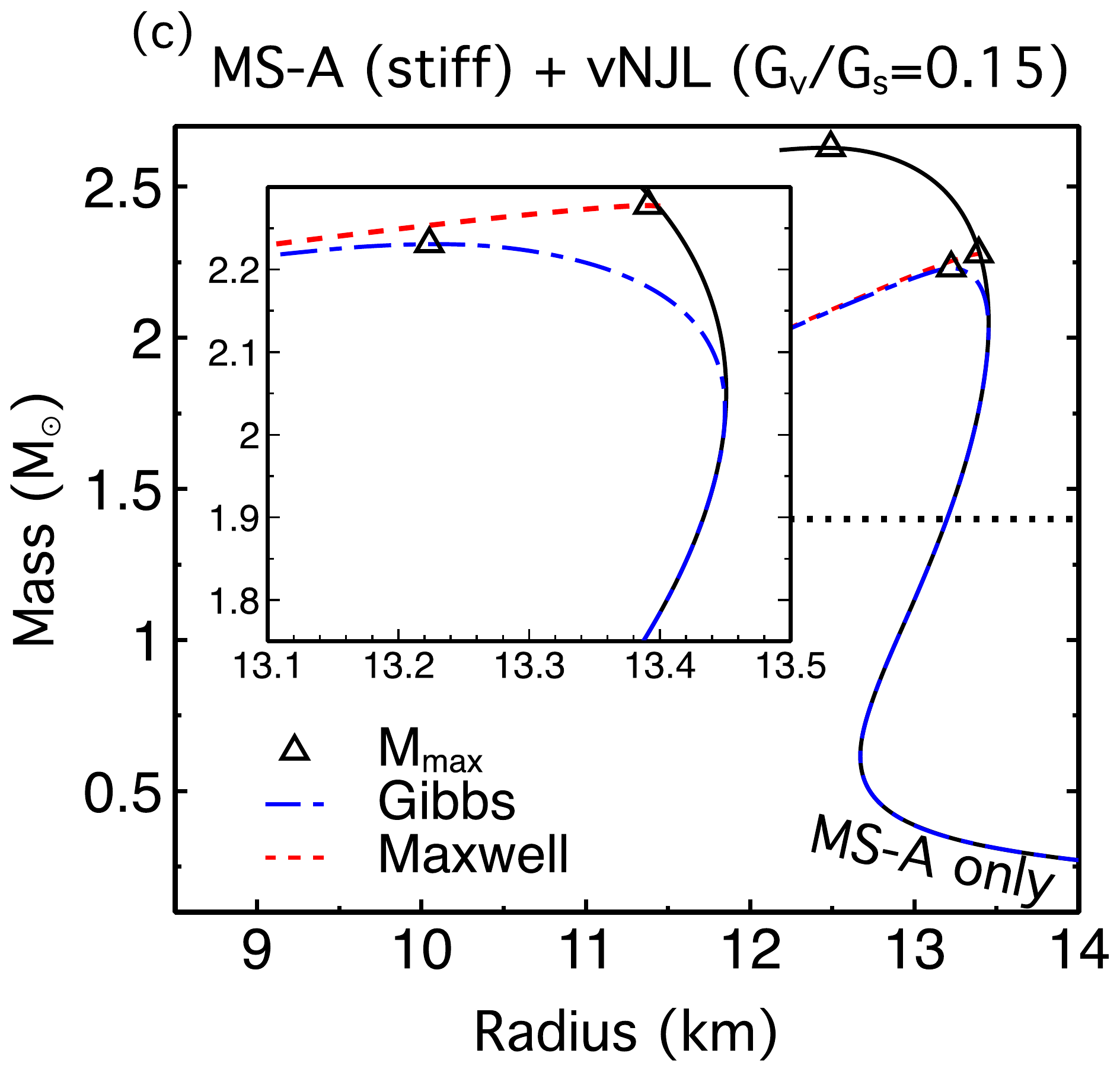}\\[-2ex]
}
\caption{(Color online) As in Fig.~\ref{fig:MR-eos_vMIT}, but for Maxwell and Gibbs constructions in MS-A (stiff) + vNJL (soft) models; the $1.4\,\Msolar$ and maximum-mass stars are marked with open circle and triangles respectively in the $c_s^2 (\nb)$ plot. Here, BNS merger observables with $m_1, m_2 = 1.0-1.6 \,\Msolar$ only constrain the hadronic matter EoS as the onset density for quarks is too high ($\Mtrans\gtrsim 1.7\,\Msolar$). Since the stiff MS-A is ruled out by $\tilde\La(\Mchirp)$ of GW170817 (see Fig.~\ref{fig:k2-Lam_vMIT}), this indicates that vNJL model (or NJL-type models) is ruled out in the first-order transition scenario. Resorting to the crossover scenario is inevitable for it to survive.
}
\label{fig:MR-eos_vNJL}
\end{figure*}

We first survey the allowed parameter space for valid first-order phase transitions, namely, a critical pressure exists above which quark matter is energetically favored. We then proceed with both Maxwell and Gibbs constructions, calculating quantities to be compared with observational constraints. Our results are summarized in Figs.~\ref{fig:MR-eos_vMIT}-\ref{fig:MR-eos_vNJL}. Where possible, we also characterize the behavior of the hadron-to-quark transition with quantities introduced in the ``Constant-Sound-Speed (CSS)'' approach in Ref.~\cite{Alford:2013aca}. This approach can be viewed as the lowest-order Taylor expansion of the high-density EoS about the transition pressure $\ptrans$, by specifying the discontinuity in energy density $\De\ep$ at the transition, and the density-independent squared sound speed $\cQMsq$ in quark matter. This generic parametrization has been widely used in recent studies on the manifestation of a first-order phase transition with Maxwell construction in neutron star phenomenology, see e.g. \cite{Paschalidis:2017qmb,Burgio:2018yix,Christian:2018jyd,Montana:2018bkb}. Despite different choices of the baseline hadronic EoS, comparison between separate works is afforded by mapping onto the CSS parameter space. To facilitate such a comparison, we list in Table~\ref{tab:mapping-CSS} the corresponding CSS parameter values for calculations from our physically based models. \\

\noindent {\bf MS-A + vMIT (stiff $\to$ soft/stiff)} \\

Fixing the hadronic EoS to be the stiff model MS-A, we choose in the vMIT model six parameter sets of ($B^{1/4}, a$), where $B$ is the bag constant and $a=(G_v/m_v)^2$ measures the strength of vector interactions between quarks. The bag constant is adjusted so that the transition to quark matter occurs at $\ntrans = 1.5-2.4 \,n_0$, and the finite vector coupling $a$ stiffens the quark matter EoS.
Soft hadronic EoSs are not applied, as they either (with softer quark EoSs) violate the $\Mmax\geq 2\,\Msolar$ constraint or (with stiffer quark EoSs) cannot establish a valid first-order phase transition, i.e., there is no intersection between the two phases in the $P$-$\mu$ plane. We note that this limitation (hadronic matter being stiff) does not necessarily hold if a generic parameterization such as CSS is utilized instead of specific quark models to perform first-order transitions. \\

In the vMIT model, the sound speed varies little even with the inclusion of vector repulsive interactions within the star (see Fig.~\ref{fig:MR-eos_vMIT} (b) and (e)) and can be approximated as being density independent. The mass-radius topology with the Maxwell construction is determined by the three parameters ($\ptrans/\etrans$, $\De\ep/\etrans$, $\cQMsq$) in CSS, giving rise to either connected, disconnected (i.e. twin stars or third-family stars) or both branches of stable hybrid stars; $\ptrans$ and $\etrans$ are the pressure and energy density in hadronic matter at the transition, respectively, $\De\ep$ is the discontinuity in energy density at $\ptrans$, and $\cQMsq$ is the squared speed of sound in quark matter just above $\ptrans$. The threshold value $\decrit$ below which there is always a stable hybrid branch connected to the purely-hadronic branch is given by $\decrit/\etrans = \frac{1}{2} + \frac{3}{2} \ptrans/\etrans$~\cite{Seidov:1971,Schaeffer:1983,Lindblom:1998dp}. The relevant quantities for the mapping between the stiff MS-A+vMIT model (Maxwell) and the CSS parametrization are listed in Table \ref{tab:mapping-CSS}. \\

After extensively varying all parameters and calculating the  corresponding mass-radius relations, we find that $a=0.18$ is most likely the smallest value (corresponding to $\cQMsq\approx 0.4$) that barely ensures $\Mmax\simeq 2\,\Msolar$. When $a$ is increased from zero, the energy density discontinuity becomes progressively smaller ($\De\ep/\etrans\lesssim0.5$) and eventually the twin-star solutions disappear, roughly at $a\geq 0.15$. Within the range $a=0.18-0.3$, the $M(R)$ curves of stable hybrid stars obtained are continuous, and quarks can appear at $1.0\leq\Mtrans\leq1.8 \,\Msolar$, pertinent to the range of component masses in BNS mergers. For too large vector interaction couplings e.g. $a=0.5$, the onset for quarks is beyond the central density of the maximum-mass hadronic star, and thus no stable quark cores would be present even though QM is sufficiently stiff.  \\

Fig.~\ref{fig:MR-eos_vMIT} (c) shows that requiring $\Mmax\geq 2\,\Msolar$ excludes certain twin-star solutions obtained from EoSs with zero (gray dash-dot-dotted) and small (orange dotted) repulsive vector interactions between quarks, mainly due to the insufficient stiffness of the quark matter EoS. By invoking very stiff EoSs with $c_s^2 \rightarrow 1$ in the quark sector and using the CSS parametrization coupled with hadronic EoSs at low density, recent works have reported twin stars compatible with the constraint $M_{\rm max} \geq 2\,\Msolar$ and bounds on $\tilde\La$ from GW170817 (see e.g. Refs.~\cite{Christian:2018jyd,Sieniawska:2018zzj,Burgio:2018yix,Han:2018mtj,Montana:2018bkb} and references therein). Moreover, the typical neutron star radius $\Rtyp$ can be observationally constrained by radius estimates from x-ray emission and/or tidal deformability ($\La$) measurements in pre-merger gravitational-wave detections. For hybrid EoSs with a sharp phase transition, the value of $\Rtyp$ or $\La_{1.4}$ is sensitive to the onset density $\ntrans$, above which $M(R)$ and $\La(M)$ deviate from normal hadronic EoSs without  a sharp transition. We demonstrate this effect in Fig.~\ref{fig:k2-Lam_vMIT} by confronting calculated tidal deformations with inferred bounds from the first BNS event GW170817 \cite{LIGO:2017qsa,LIGO:2018exr,De:2018uhw}. \\

\begin{table}[htb]
\begin{center}
\begin{tabular}{c|c|c|c|c|c|c}
\hline 
&&&&&&\\[-2ex]
$\left( \frac{B^{1/4}}{{\rm MeV}}, a \right)$ & $\frac{\ntrans}{n_0}$  & $\frac{\ptrans}{\etrans}$ & $\frac{\De\ep}{\etrans}$ & $\cQMsq$ & $\frac{\De\ep_{\rm crit}}{\etrans}$ & $M(R)$ \\[0.5ex]
\hline  
&&&&&&\\[-2ex]
(159, 0.2) & 1.77 & 0.084 & 0.33 & 0.407 & 0.626 & Connected \\[0.5ex]
(162, 0.2) & 2.10 & 0.136 & 0.33 & 0.416  & 0.704 & Connected \\[0.5ex]
(165, 0.2) & 2.34 & 0.180 & 0.38 & 0.424   & 0.77 & Connected \\[0.5ex]
(180, 0.0) & 2.04 & 0.127 & 1.13 & 0.326  & 0.691 & Disconnected \\[0.5ex]
(170, 0.1) & 2.08 & 0.133 & 0.63 & 0.380  & 0.70 & Both \\[0.5ex]
(155, 0.3) & 2.08 & 0.134 & 0.13 & 0.442  & 0.701 & Connected \\[0.5ex]
\hline  
&&&&&&\\[-2ex]
$\left( \frac{B_{\rm eff}^{1/4}}{{\rm MeV}}, \frac{G_v}{G_s}\right)$ & $\frac{\ntrans}{n_0}$  & $\frac{\ptrans}{\etrans}$ & $\frac{\De\ep}{\etrans}$ & $\cQMsq$ & $\frac{\De\ep_{\rm crit}}{\etrans}$ & $M(R)$ \\[0.5ex]
\hline  
&&&&&&\\[-2ex]
(218.3, 0.15) & 2.91 & 0.284 & 0.594 & 0.236 & 0.926 & Connected~\footnote{This connected branch is tiny $(\Mmax-\Mtrans) \lesssim 10^{-3} \,\Msolar$ (invisible on the magnified $M(R)$ plot; see Fig.~\ref{fig:MR-eos_vNJL} (c)) and thus hybrid stars are undetectable.} 
\\[0.5ex]
\hline
\end{tabular}
\end{center}
\caption{Mapping onto the CSS phase transition parameters \cite{Alford:2013aca} for the stiff MS-A+vMIT/vNJL hybrid EoSs with Maxwell construction; see also Fig.~\ref{fig:MR-eos_vMIT}, panels (a)-(c). Meanings of the various entries are explained in the associated text.}
\label{tab:mapping-CSS}
\end{table}  

With high accuracy, the chirp mass $\Mchirp=(m_{1}m_{2})^{3/5}/(m_1+m_2)^{1/5}$, where $m_{1,2}$ are the masses of the merging neutron stars, was determined to be $\Mchirp=1.186_{-0.001}^{+0.001}\Msolar$ in GW170817 \cite{LIGO:2018wiz}. This event also  revealed information on the binary tidal deformability, $\tilde{\La}(\Mchirp=1.186_{-0.001}^{+0.001}\Msolar)=300_{-230}^{+420}$ for low-spin priors (using a 90\% highest posterior density interval). Furthermore, by assuming a linear expansion of $\La(M)$, which holds fairly well for normal hadronic stars without sharp transitions, limits on the dimensionless tidal deformability of a $1.4 \,\Msolar$ NS were derived~\cite{LIGO:2018exr}: $70\leq\La_{1.4}\leq 580$ for low spin priors (at $90\%$ confidence level). This single detection of GW170817 rules out purely-hadronic EoSs that are too stiff and correlated with large tidal deformabilities, as shown in Fig.~\ref{fig:k2-Lam_vMIT} (b) and (c). The stiff MS-A model by itself is incompatible with the estimated ranges of $\La_{1.4}$ and $\tilde{\La}$. The only solution to rescue such a stiff hadronic EoS is to introduce a phase transition at not-too-high densities, e.g., a possible smaller $\La$ can be achieved in a hybrid star that already exists in the pre-merger stage. For Maxwell constructions, one of the six parameter sets, $(B^{1/4}, a)=(159, 0.2)$ (blue dash-dotted) with $\ntrans/n_0=1.77$ (see Table \ref{tab:mapping-CSS}) is successful to survive the LIGO constraint. Together with the maximum-mass constraint, the parameter space for sharp phase transitions is severely limited.

\begin{figure*}[htb]
\parbox{0.35\hsize}{
\includegraphics[width=\hsize]{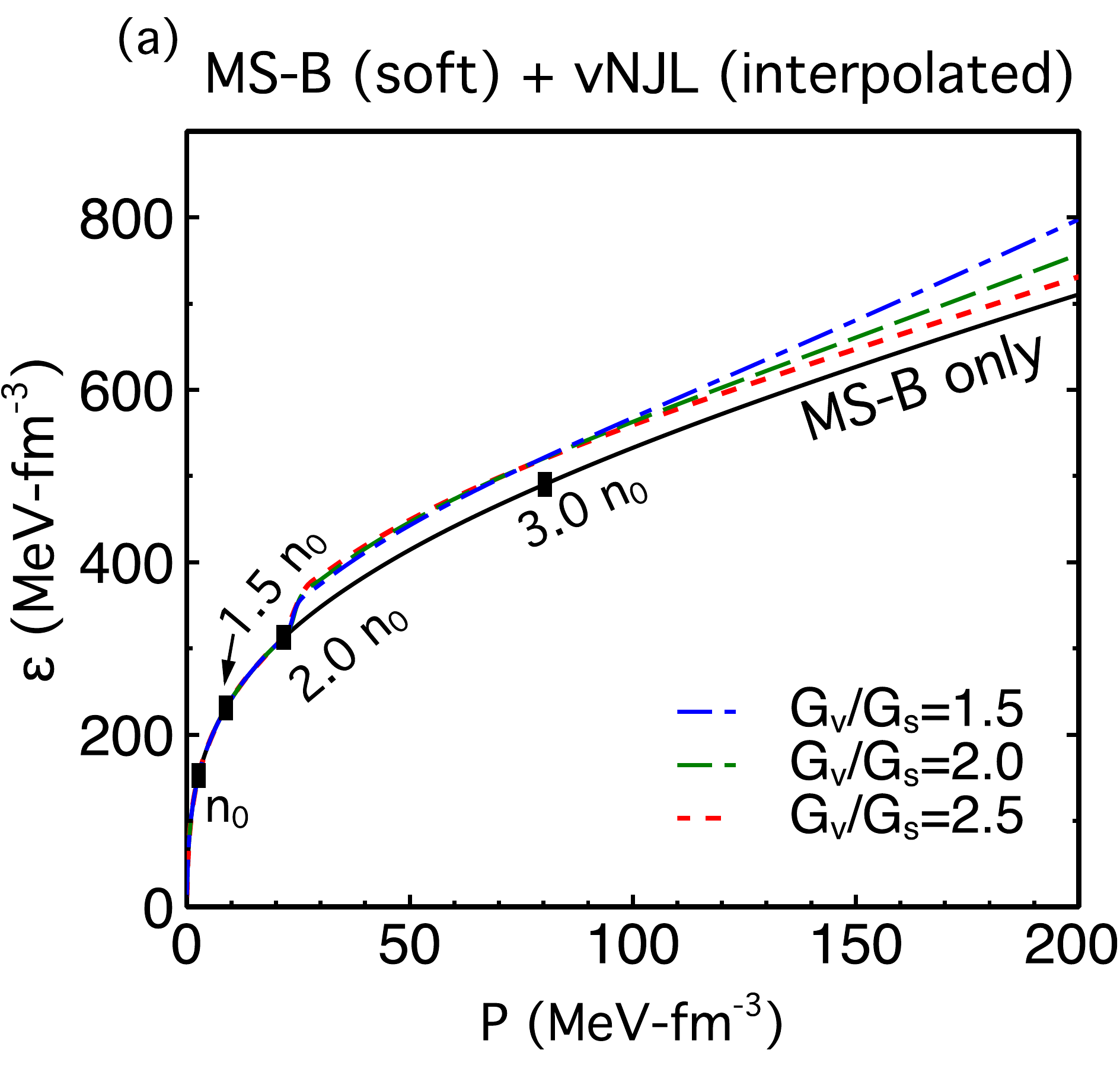}\\[-2ex]
}\parbox{0.35\hsize}{
\includegraphics[width=\hsize]{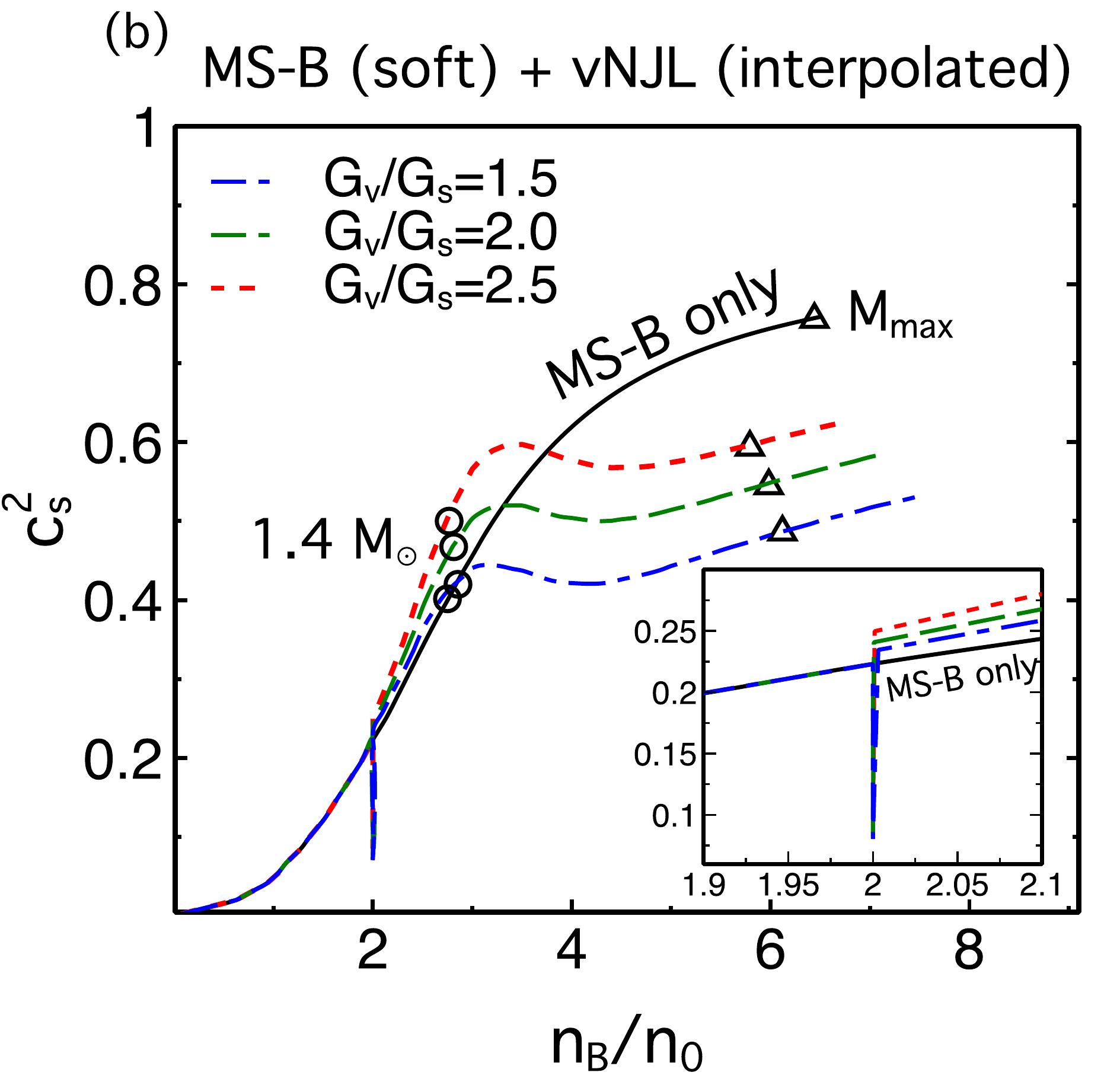}\\[-2ex]
}\parbox{0.35\hsize}{
\includegraphics[width=\hsize]{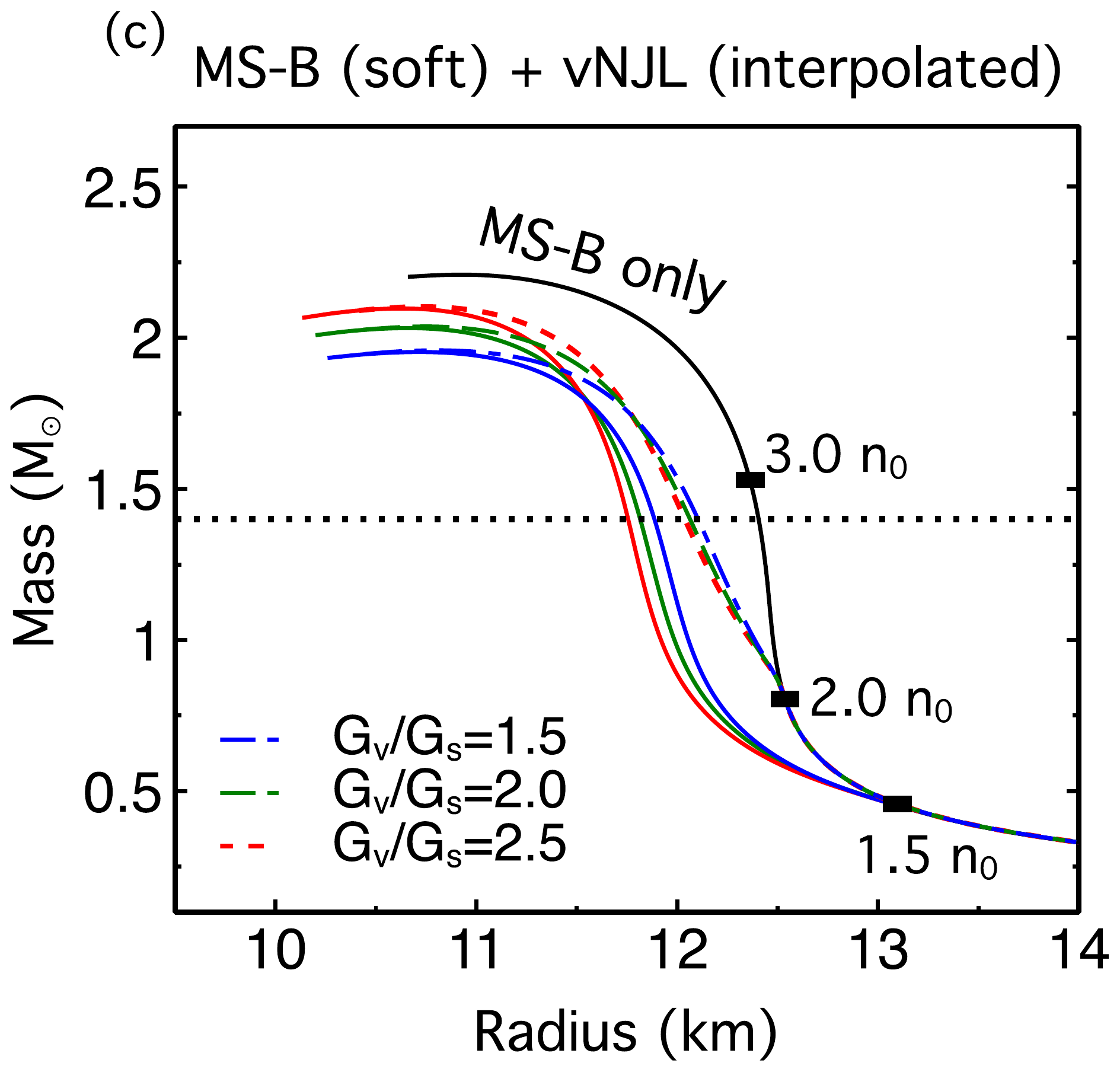}\\[-2ex]
}\\[3ex]
\parbox{0.35\hsize}{
\includegraphics[width=\hsize]{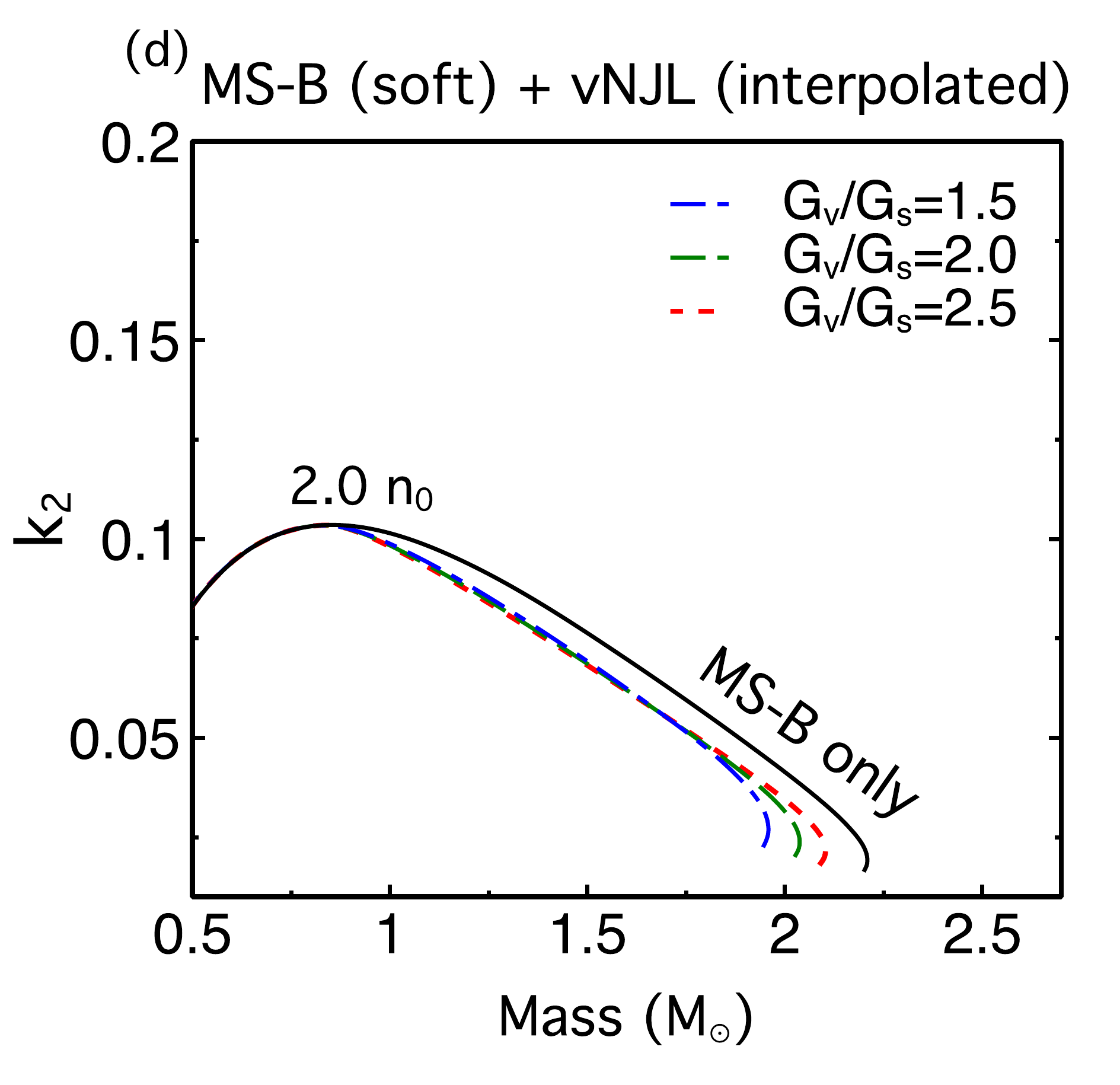}\\[-2ex]
}\parbox{0.35\hsize}{
\includegraphics[width=\hsize]{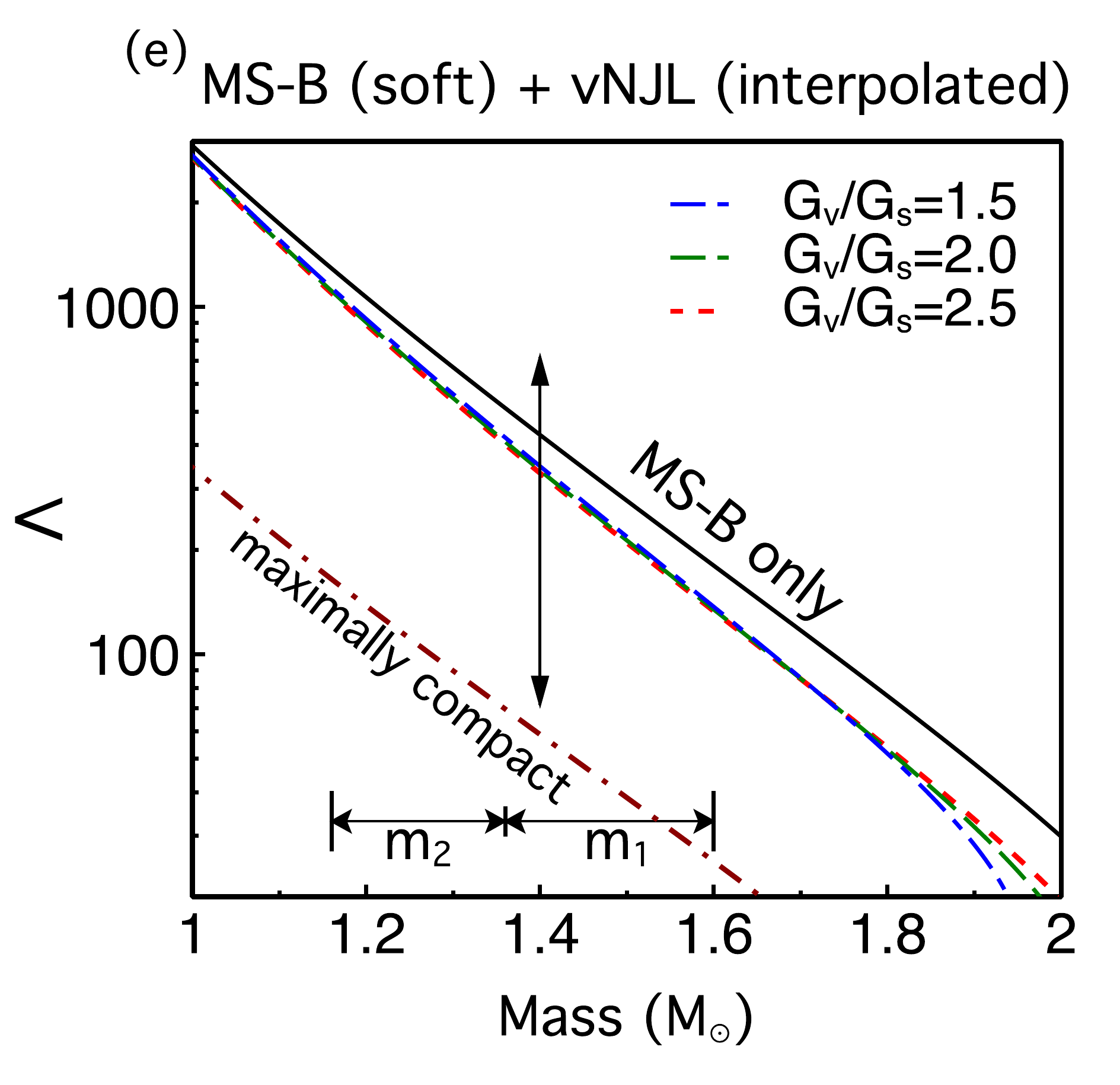}\\[-2ex]
}\parbox{0.35\hsize}{
\includegraphics[width=\hsize]{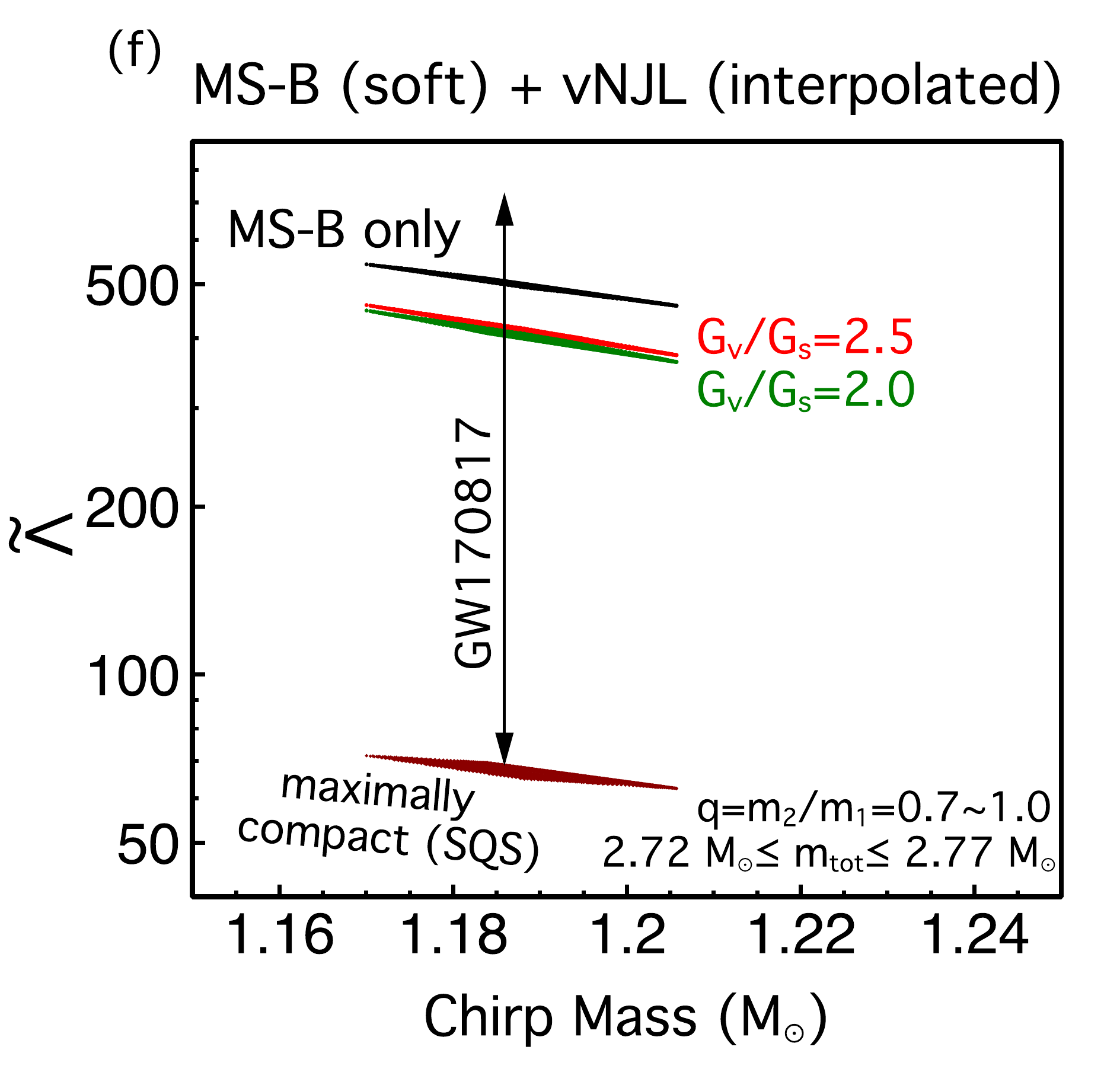}\\[-2ex]
}
\caption{(Color online) The EoS MS-B is a soft version within MS models, but can be stiffer than normal hadronic EoSs based on other models such as Skyrme or APR; MS-A is not applied here because its stiff hadronic part leads to violation of the tidal deformability constraint $\tilde\La (\Mchirp=1.186\,\Msolar)\leq720$ \cite{LIGO:2018wiz}. The $1.4\,\Msolar$ and maximum-mass stars are marked with open circles and triangles respectively in the $c_s^2 (\nb)$ plot. In the $\tilde\La(\Mchirp)$ plot, only EoSs that satisfy $\Mmax\geq 2\,\Msolar$ are shown. In the interpolation picture, although the maximum mass is mostly determined by the high-density quark part and increases with its stiffness, changes in radii are flexible depending on e.g., the choice of window parameters and the low-density hadronic part (for an extensive exploration, see Ref.~\cite{Masuda:2012ed}). Panel (c) also shows $M(R)$ for a lower cutoff density $\ntrans=1.5\,n_0$ (solid colored curves).
}
\label{fig:int-MS_crossover}
\end{figure*}

The panels (e)-(f) of Figs.~\ref{fig:MR-eos_vMIT}-\ref{fig:k2-Lam_vMIT} represent results for the stiff MS-A+vMIT model with Gibbs constructions, for which the model parameters  remain the same as in their Maxwell counterparts (panels (a)-(c)). The smooth feature of the Gibbs construction  advances the appearance of quarks in the mixed phase to lower densities, while it defers the region of the purely quark phase to higher densities. These features are also manifested in the corresponding $\ep(P)$ relation and its finite speed-of-sound behavior (Fig.~\ref{fig:MR-eos_vMIT} (e) and (f)). Effectively, the softening due to (Gibbs) phase transition occurs earlier, smoothly decreasing the NS radii and tidal deformabilities for a broader range of masses, which gives rise to increased compatibility with observational constraints. Three more parameter sets of the stiff MS-A+vMIT model that satisfy $\Mmax\geq 2\,\Msolar$ are now  consistent with the tidal deformability constraint (Fig. \ref{fig:k2-Lam_vMIT} (e) and (f)), in contrast to the only candidate that qualifies in Maxwell constructions. In this respect, applying Gibbs construction is advantageous to enlarging the quark model parameter space that suitably satisfies the current constraints  from observation (and also revives previously-excluded stiff hadronic models). However, the clear-cut distinction between hybrid and purely-hadronic branches in terms of $M(R)$ and $\La (M)$ diminishes: the drastic effect from a sharp hadron/quark transition is toned down, and thus distinguishability of quarks with regard to global observables becomes less feasible if they take the form of a mixture with hadrons. This feature accentuates the significance of dynamical properties such as NS cooling and spin-down, and the evolution of merger products.  \\

\noindent {\bf MS-A + vNJL (stiff $\to$ soft)} \\

In the vNJL model, pressures at $\lesssim 2\,n_0$ exhibit an unphysical behavior (of being negative and/or decreasing with density) which forbids attempts to shift $\ntrans$ to low densities. If a finite vector coupling $G_v$ is introduced, the onset of quarks is typically reached at $\ntrans\gtrsim 2.3 \,n_0$ ($\Mtrans\gtrsim 1.7 \,\Msolar$), leading to a short stable hybrid branch that obeys $\Mmax\geq 2\,\Msolar$ because of the stiff hadronic EoS. We display one such example in Fig.~\ref{fig:MR-eos_vNJL} for both Maxwell and Gibbs constructions. Note that the speed of sound in the quark phase remains small, restricted by the fact that a too large $G_v$ (correlated with stiffer QM) significantly delays the onset for quarks which yields no stable hybrid stars. Some relevant points to note are: \\

\noindent (i) $\Mtrans\gtrsim 1.7\,\Msolar$ indicates that most likely there will be no quarks in e.g., the component neutron stars of a binary before they coalesce. Thus,  tidal properties are not shown in Fig.~\ref{fig:MR-eos_vNJL} due to the high onset density for quarks: i.e., in this case  BNS observables are irrelevant; \\
\noindent (ii) A small $G_v$ has little effect on stiffening quark matter ($\cQMsq\lesssim 1/3$), which is not desirable in terms of supporting $2\,\Msolar$ mostly by quarks; and \\
\noindent (iii) Gibbs construction helps maintaining slightly more quark content than Maxwell in the most massive stars, but quarks are effectively ``invisible'' even if they exist. \\

Note that the tidal deformability constraint rules out a very stiff hadronic EoS, e.g., MS-A. This stiffness in the hadronic EoS is nevertheless a prerequisite for vNJL to construct a valid first-order transition; stable hybrid stars that are consistent with observation do not exist in this scenario. There is no solution other than an alternative treatment, such as a  crossover transition to which we turn below. It is noteworthy that the only successful scenario we find for first-order phase transitions to be compatible with observations is a stiff HM $\to$ stiff QM transition (see summary in Table~\ref{tab:pt-summary}). This conclusion agrees fairly well with those from other previous studies in which specific models of quark matter were used, see e.g. \cite{Nandi:2017rhy,Paschalidis:2017qmb,Alvarez-Castillo:2018pve,Gomes:2018eiv}. \\

\subsection* {Crossover transitions: Interpolatory procedures and quarkyonic matter}
\label{result:crs}
In obtaining the results shown below in Figs. \ref{fig:int-MS_crossover} and \ref{fig:qyc-MS_crossover}, we have followed the  methods detailed in Sec.~\ref{sec:treat} for constructing  crossover hadron-to-quark transitions. Although the generalization of the quarkyonic matter model to beta-equilibrated stars is presented in that section, results shown here for this case are for pure neutron matter only, in order to provide a direct comparison with the results of Ref.~\cite{McLerran:2018hbz}.  \\

\noindent {\bf Interpolated EoSs} \\

The results shown for this case correspond to a smooth interpolation in the window $(\bar{n}, \Ga) = (3\,n_0, n_0)$ between the soft hadronic EoS MS-B and stiff quark EoSs in the vNJL model with $G_v/G_s = 1.5, 2.0, 2.5$. Outside this window in density, pure hadronic and pure quark phases are expected to exist. Due to the abrupt cutoff imposed in the boundary condition, there is a finite jump in $c_s^2$ at the lower-end of the crossover window $\bar{n}-\Ga=2\,n_0\equiv \ntrans$ below which only pure hadronic phase is present. At the higher-end and above, we continue to use the interpolated form. This is different from Ref.~\cite{Masuda:2012ed}, where the interpolated form extended to all densities. As we will see below, the cutoffs are important to typical radii and thus could be significant. \\

Effects of introducing quarks above $\ntrans=2\,n_0$ through smooth interpolations in the EoS are shown in Fig.~\ref{fig:int-MS_crossover} (a)-(c). The maximum mass is primarily determined by the stiffness above $4\,n_0$, hence the use of large vector-coupling strengths in vNJL. Consequently, one can derive a constraint on $G_v/G_s$ from $\Mmax\geq2\,\Msolar$ if other parameters are fixed, e.g., $G_v/G_s=1.5$ is probably ruled out. 
On the other hand, typical radii for $1.0-1.6 \,\Msolar$ stars are sensitive to the stiffness in the hadronic phase at $\nb \lesssim 2\,n_0$, as well as to the choice of the threshold density. For instance, we have found that for $\ntrans=1.5\,n_0$ instead of $2\,n_0$, $\Rtyp$ decreases by about $0.3$ km. Note that the hyperbolic construction results in admixtures of the hadron and quark EoSs in the interpolated region. This feature causes a finite discontinuity in $c_s^2 (\nb)$ at low density, which is an artifact of the scheme. Alternative forms of interpolation suggested in e.g. Refs.~\cite{Fukushima:2015bda,Kojo:2014rca} do not allow for spillovers into the region of interpolation. Use of such forms, however, does not qualitatively change the outcome: while $2\,\Msolar$ NS can still be produced, the constraints on $\Rtyp$ and $\tilde\La$ cannot be easily transformed into constraints on the parameters of interpolated EoSs. If, however, a stiff hadronic matter EoS such as MS-A  in Sec.~\ref{result:1st} is applied, the resulting radius and tidal deformability are apparently too large and violate the condition $\tilde\La (\Mchirp=1.186\,\Msolar)\leq720$ \cite{LIGO:2018wiz}. \\

As can be seen from Fig.~\ref{fig:int-MS_crossover} (d)-(f), the softer MS-B EoS is by itself compatible with the current constraint on the binary tidal deformability. Implementing the crossover region through interpolation further enhances the compatibility. Better measurements of $\tilde\La(\Mchirp)$ from multiple merger detections in the future might help in limiting the relevant interpolation parameters. Recall that such ``soft HM $\to$ stiff QM'' combination is usually forbidden in a first-order transition, given the absence of an intersection in the $P$-$\mu$ plane between pure hadronic and pure quark phases (see summary in Table~\ref{tab:pt-summary}). \\

\noindent {\bf Quarkyonic matter} \\

\begin{figure*}[htb]
\parbox{0.35\hsize}{
\includegraphics[width=\hsize]{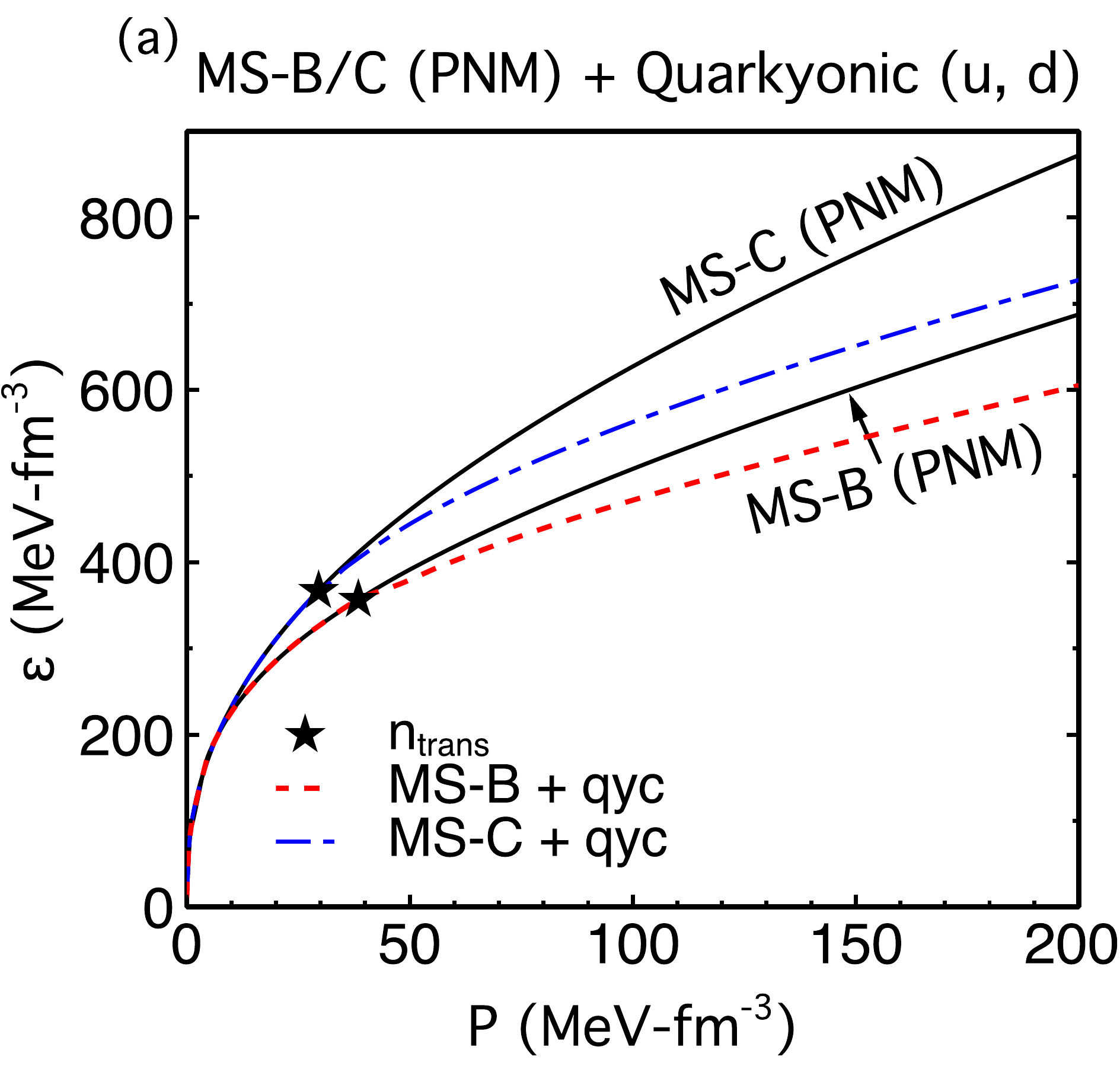}\\[-2ex]
}\parbox{0.35\hsize}{
\includegraphics[width=\hsize]{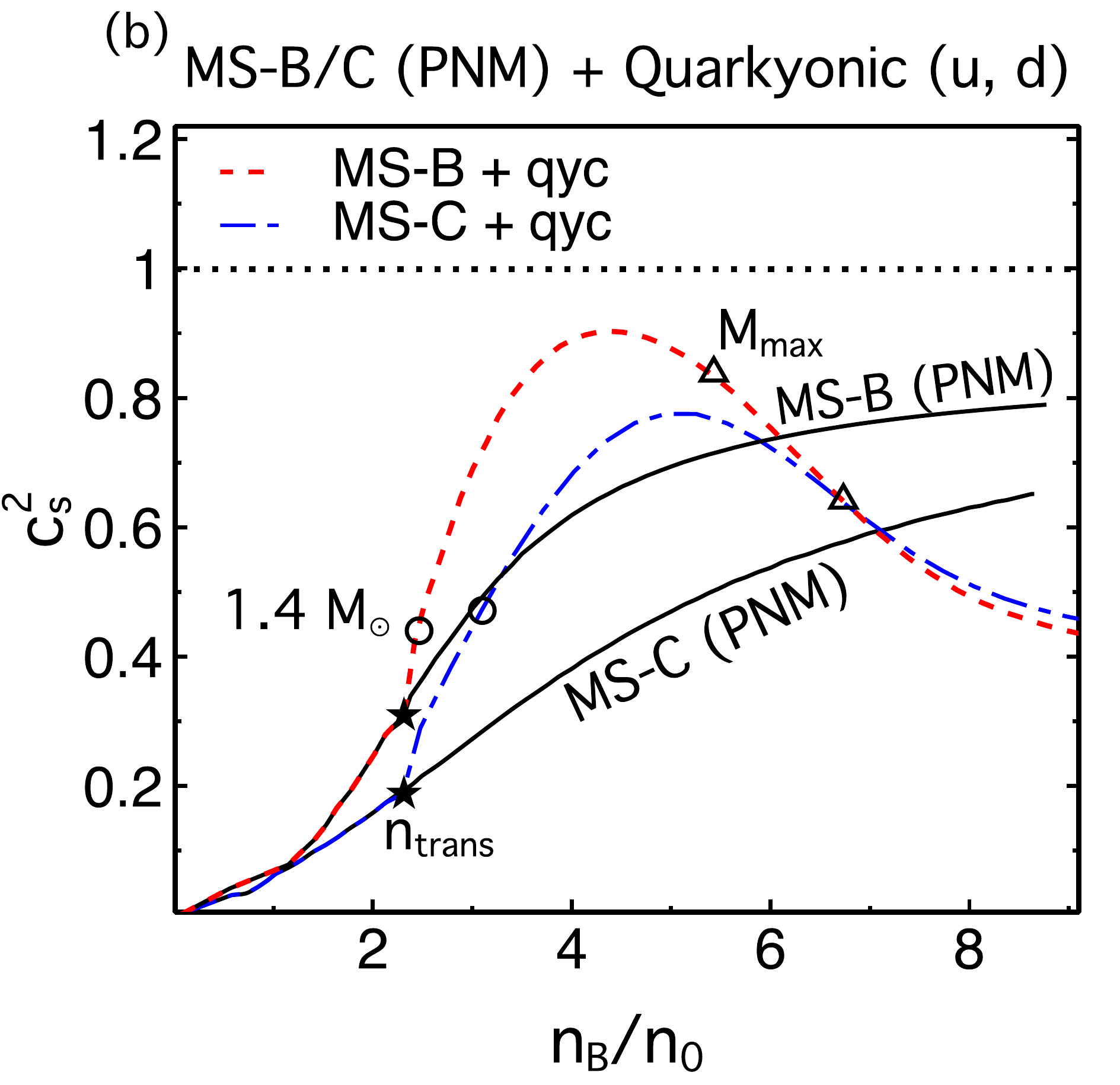}\\[-2ex]
}\parbox{0.35\hsize}{
\includegraphics[width=\hsize]{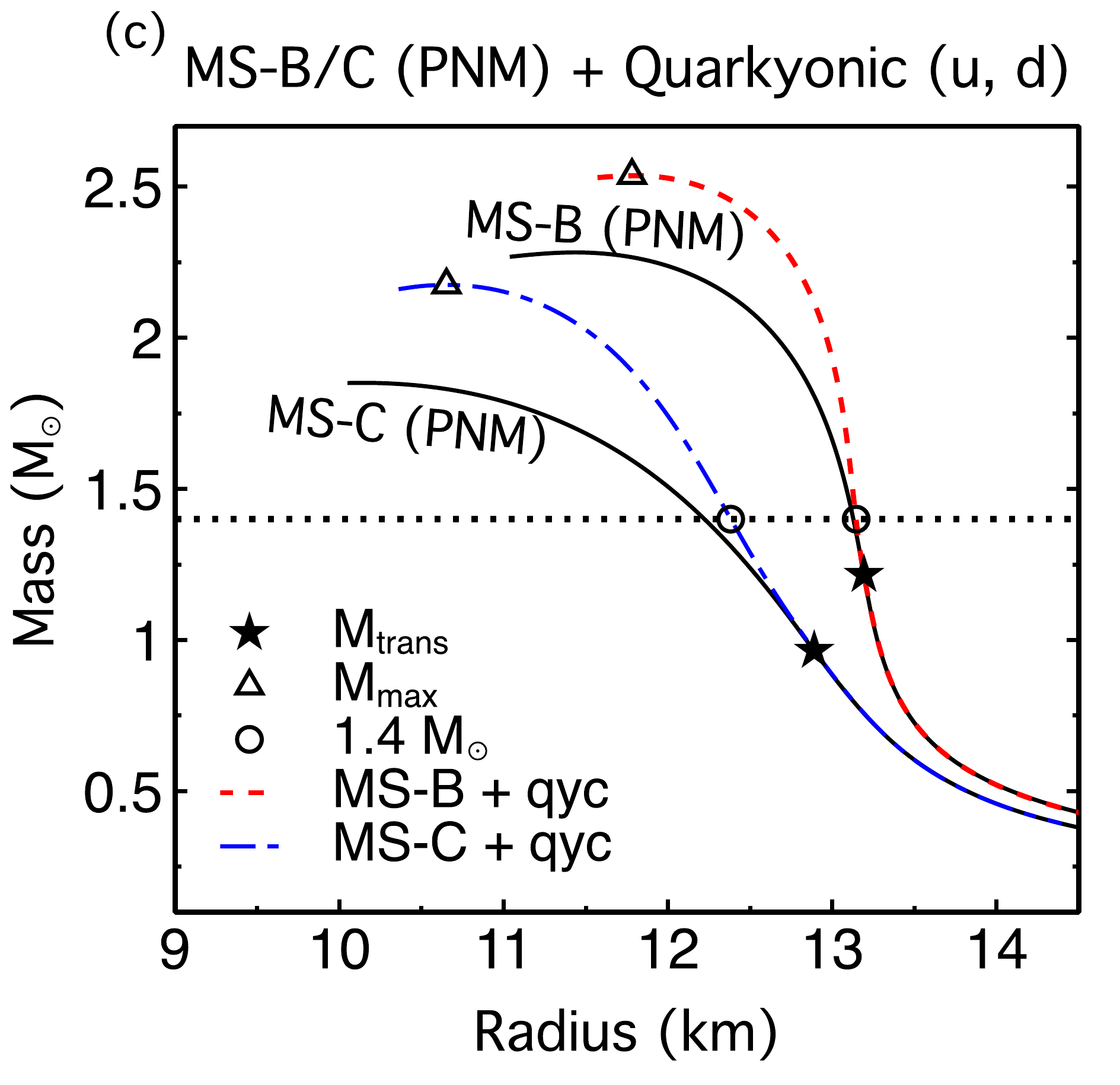}\\[-2ex]
}\\[3ex]
\parbox{0.35\hsize}{
\includegraphics[width=\hsize]{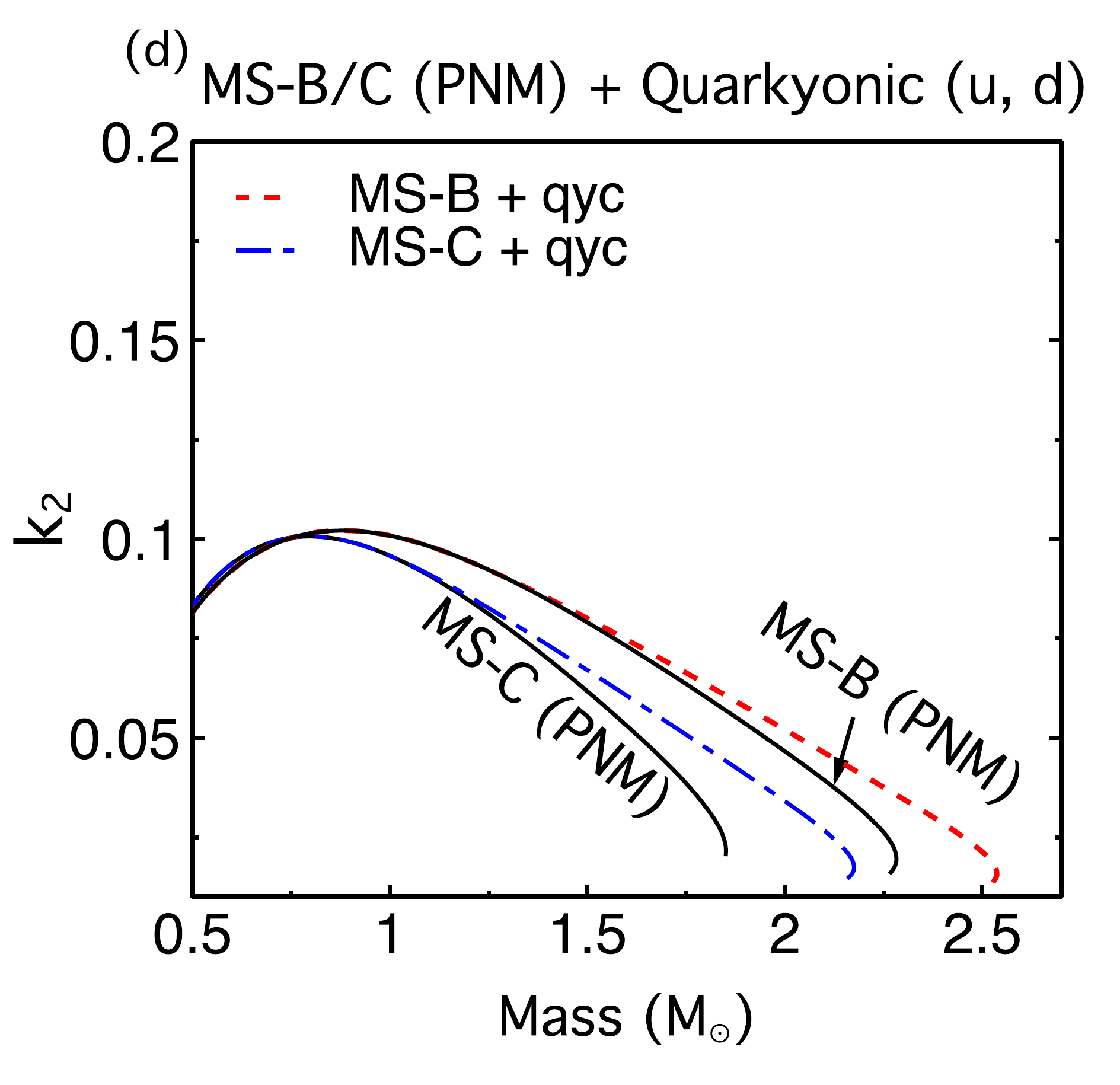}\\[-2ex]
}\parbox{0.35\hsize}{
\includegraphics[width=\hsize]{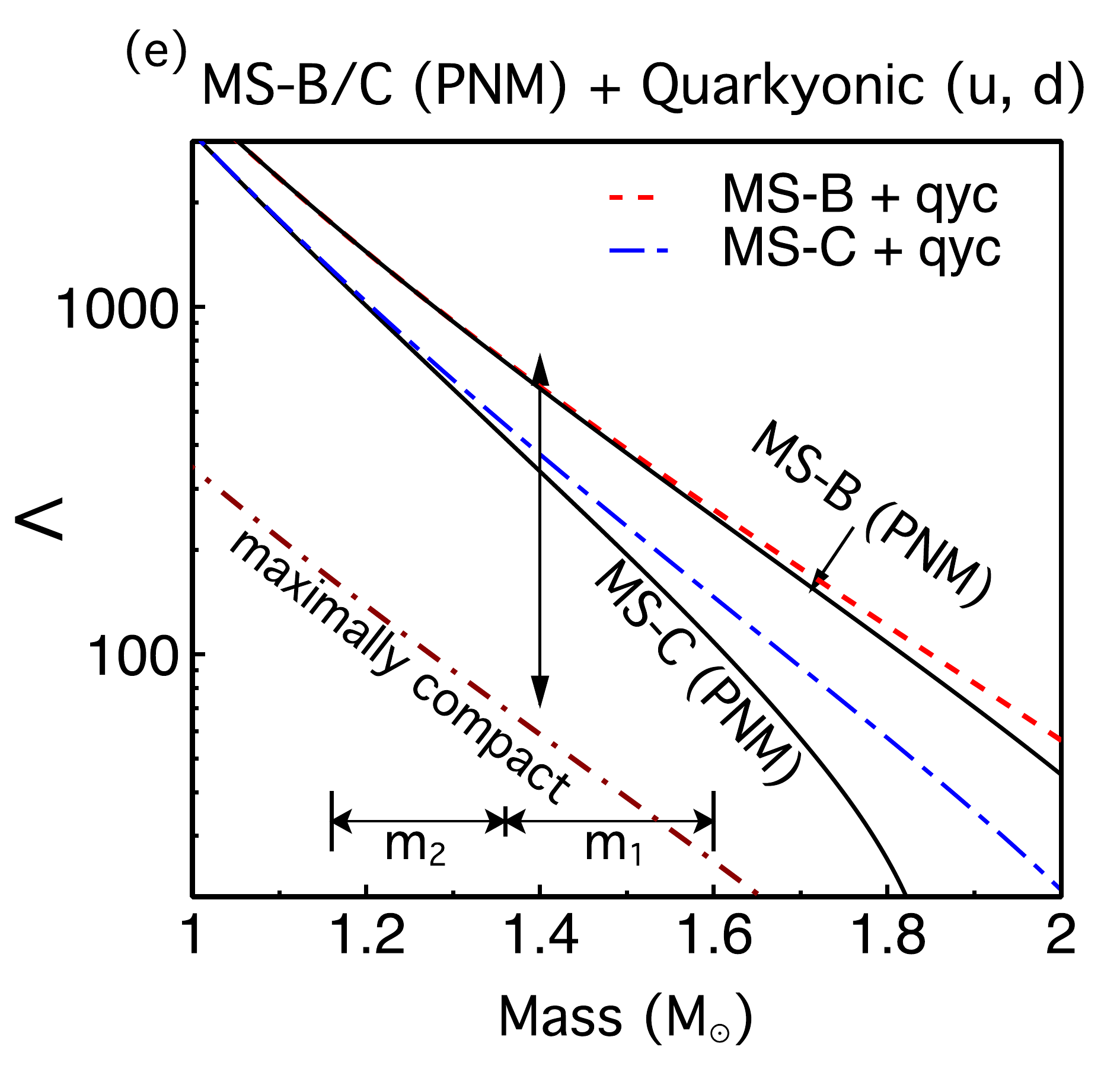}\\[-2ex]
}\parbox{0.35\hsize}{
\includegraphics[width=\hsize]{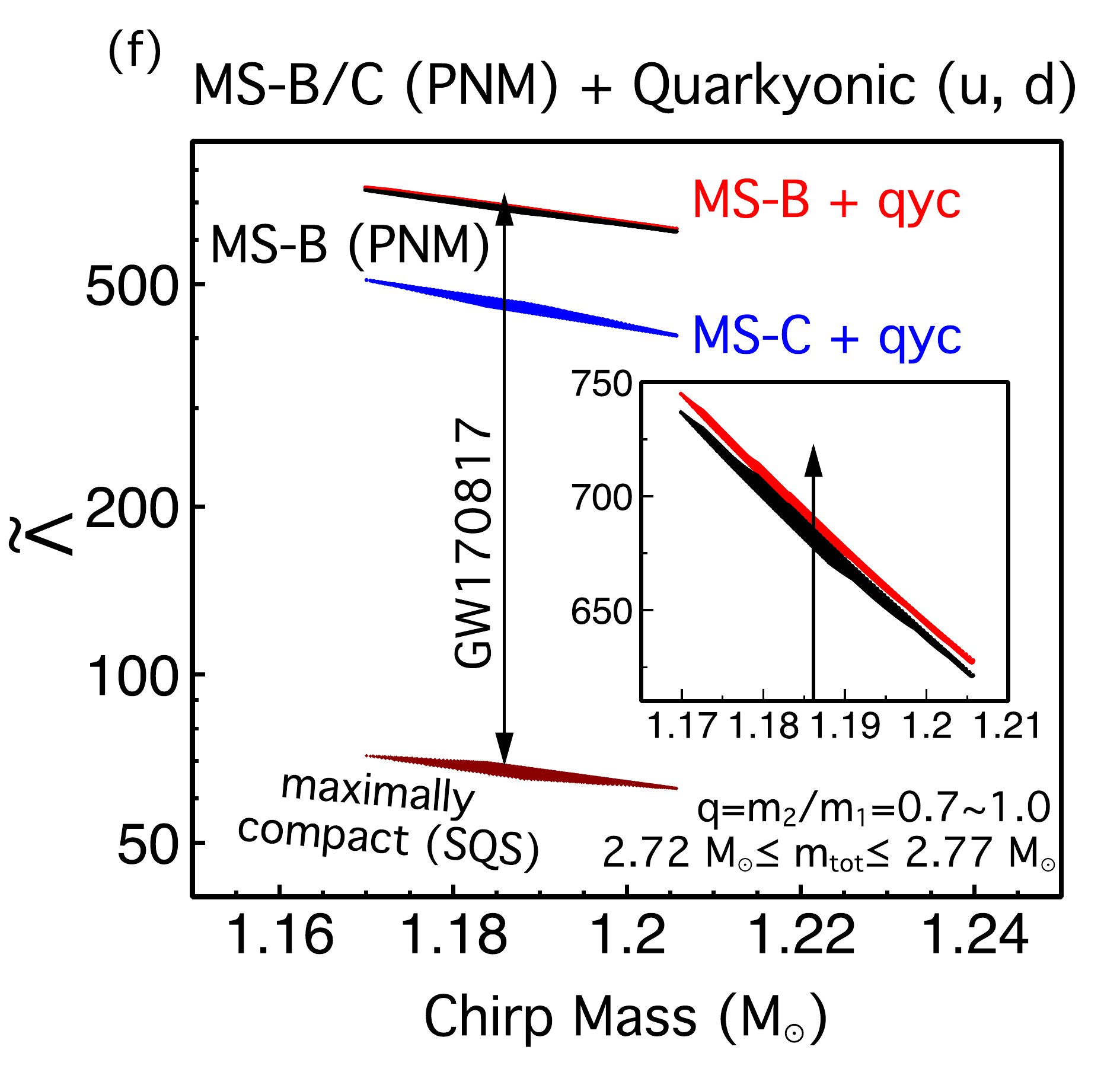}\\[-2ex]
}
\caption{(Color online) The EoS MS-C is an example of even softer EoSs within the same model that cannot support $2\,\Msolar$ stars by themselves; MS-A is not applied here because its stiff hadronic part leads to violation of the tidal deformability constraint $\tilde\La (\Mchirp=1.186\,\Msolar)\leq720$ \cite{LIGO:2018wiz}. The $1.4\,\Msolar$ and maximum-mass stars are marked with open circles and triangles respectively in the $c_s^2 (\nb)$ plot. In the two-flavor quarkyonic picture, switching to the smooth crossover region increases both the maximum masses and typical radii (hence the tidal deformabilities). Note that here quark masses are dynamically generated here with vNJL ($G_v/G_{s}=0.5$) instead of the original  assumption that $M_q=M_n/3$ as in the Fermi gas model of Ref.~\cite{McLerran:2018hbz}. In the $\tilde\La(\Mchirp)$ plot, only EoSs that satisfy $\Mmax\geq 2\,\Msolar$ are shown.
}
\label{fig:qyc-MS_crossover}
\end{figure*}

In this case, we present results obtained by using the hadronic EoSs MS-B/C for pure neutron matter, and two-flavor quark EoSs with and without interactions between quarks when they appear. The main reason for the rapid increase in pressure at supra-nuclear densities and the attendant behavior of $c_s^2$ vs $\nb$ is also elucidated in more detail than was done in Ref.~\cite{McLerran:2018hbz}.

In the quarkyonic picture, both the maximum mass and typical radii are larger than those obtained by EoSs with neutrons only. In fact, some EoSs that are too soft to survive the $\Mmax\geq 2\,\Msolar$ constraint can be rescued by the boost in stiffness once quarkyonic matter appears; see e.g. MS-C (PNM) in Fig.~\ref{fig:qyc-MS_crossover} (a)-(c). However, for a stiff neutrons-only EoS, if a transition into quarkyonic matter takes place, compatibility with binary tidal deformability constraint from GW170817 becomes reduced, because of the tendency to also increase $R$ and therefore $\La$. These increases put the model at more risk of breaking the upper limit on $\La$. This is evident in  Fig.~\ref{fig:qyc-MS_crossover} (d)-(f), where the MS-B (PNM) EoS is at the edge of exclusion, and with quarkyonic matter the situation is slightly worse.

\begin{figure*}[htb]
\parbox{0.35\hsize}{
\includegraphics[width=\hsize]{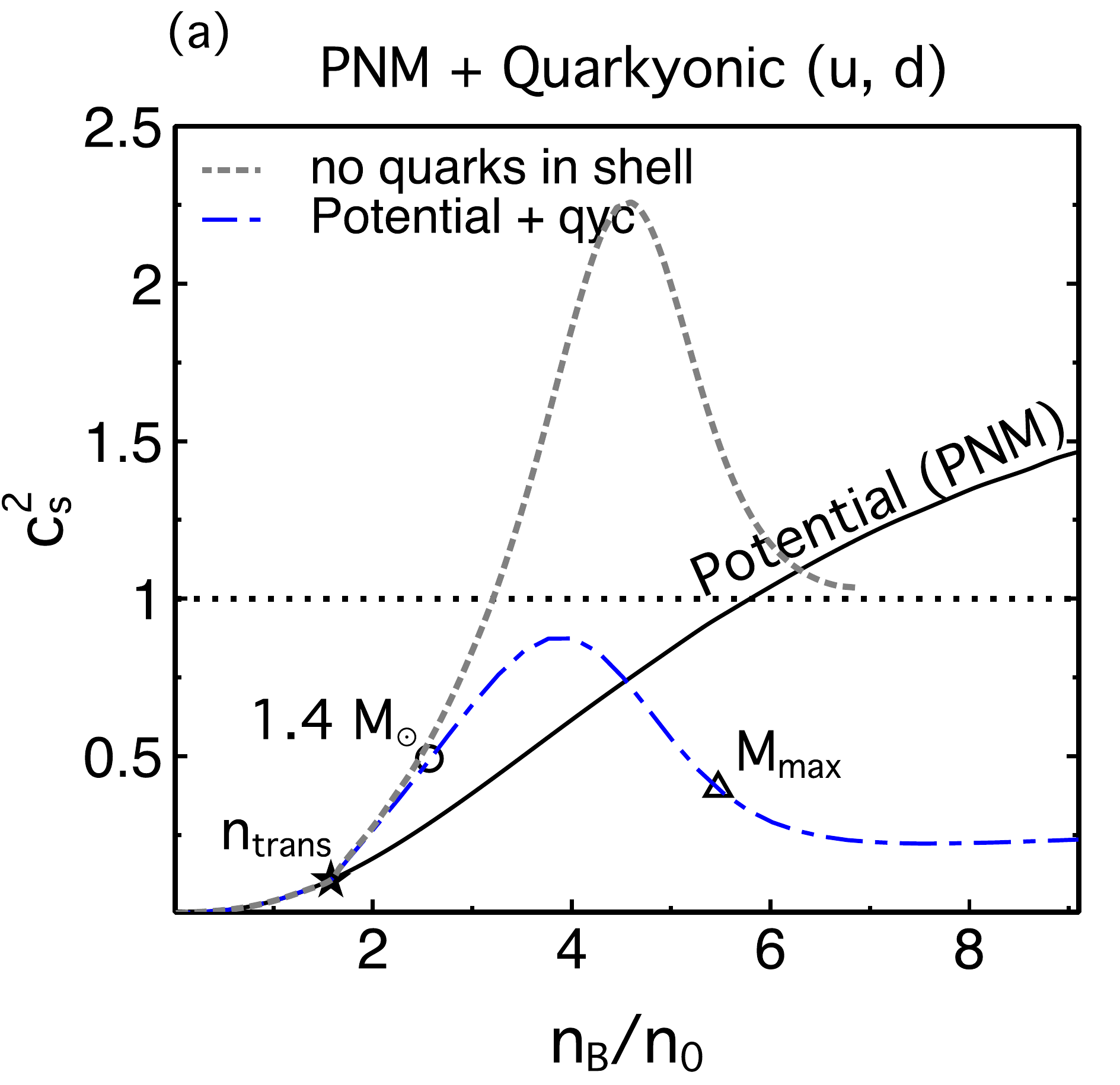}\\[-2ex]
}
\hspace{0.8cm}
\parbox{0.35\hsize}{
\includegraphics[width=\hsize]{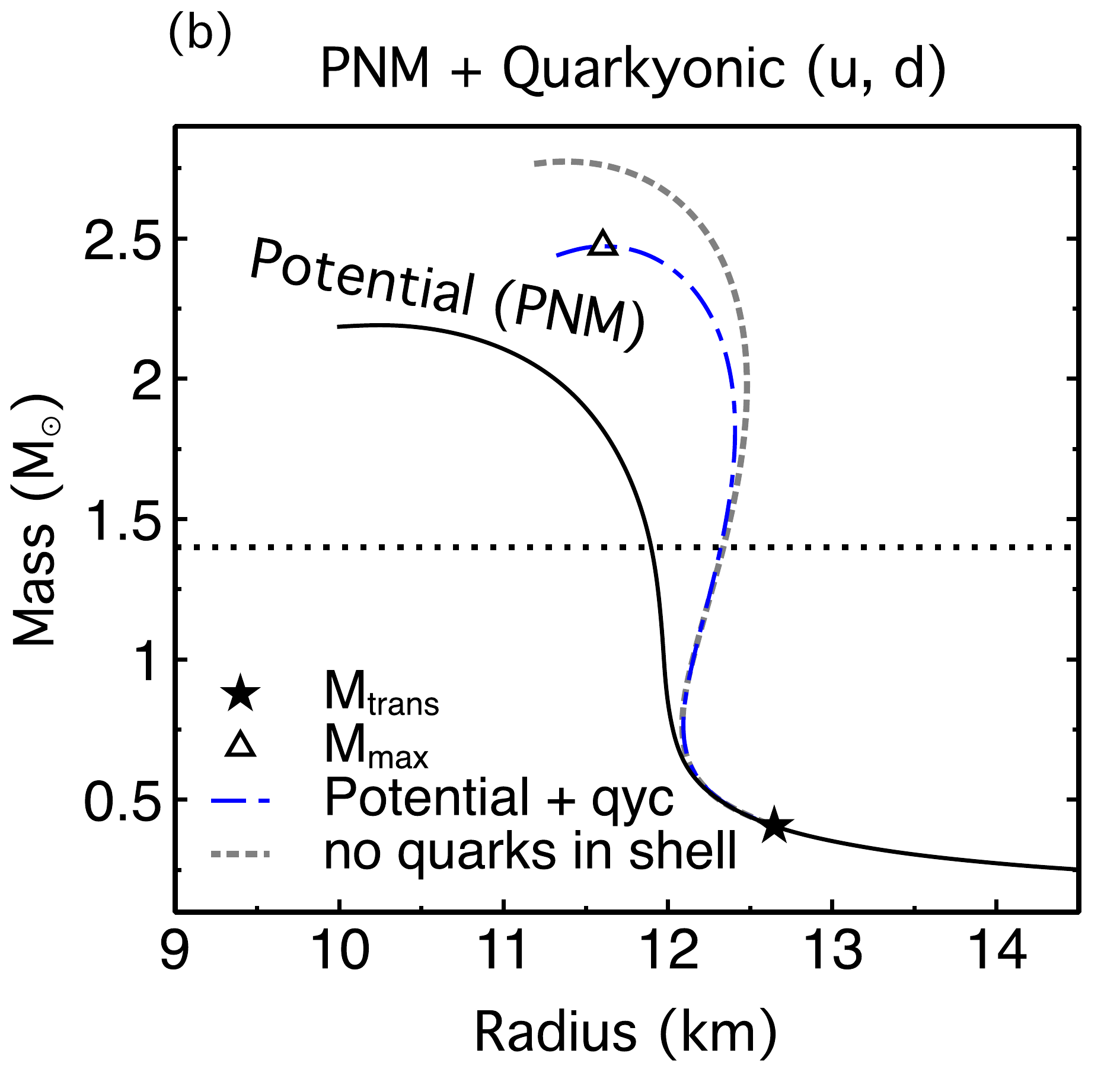}\\[-2ex]
}\\[3ex]
\parbox{0.35\hsize}{
\includegraphics[width=\hsize]{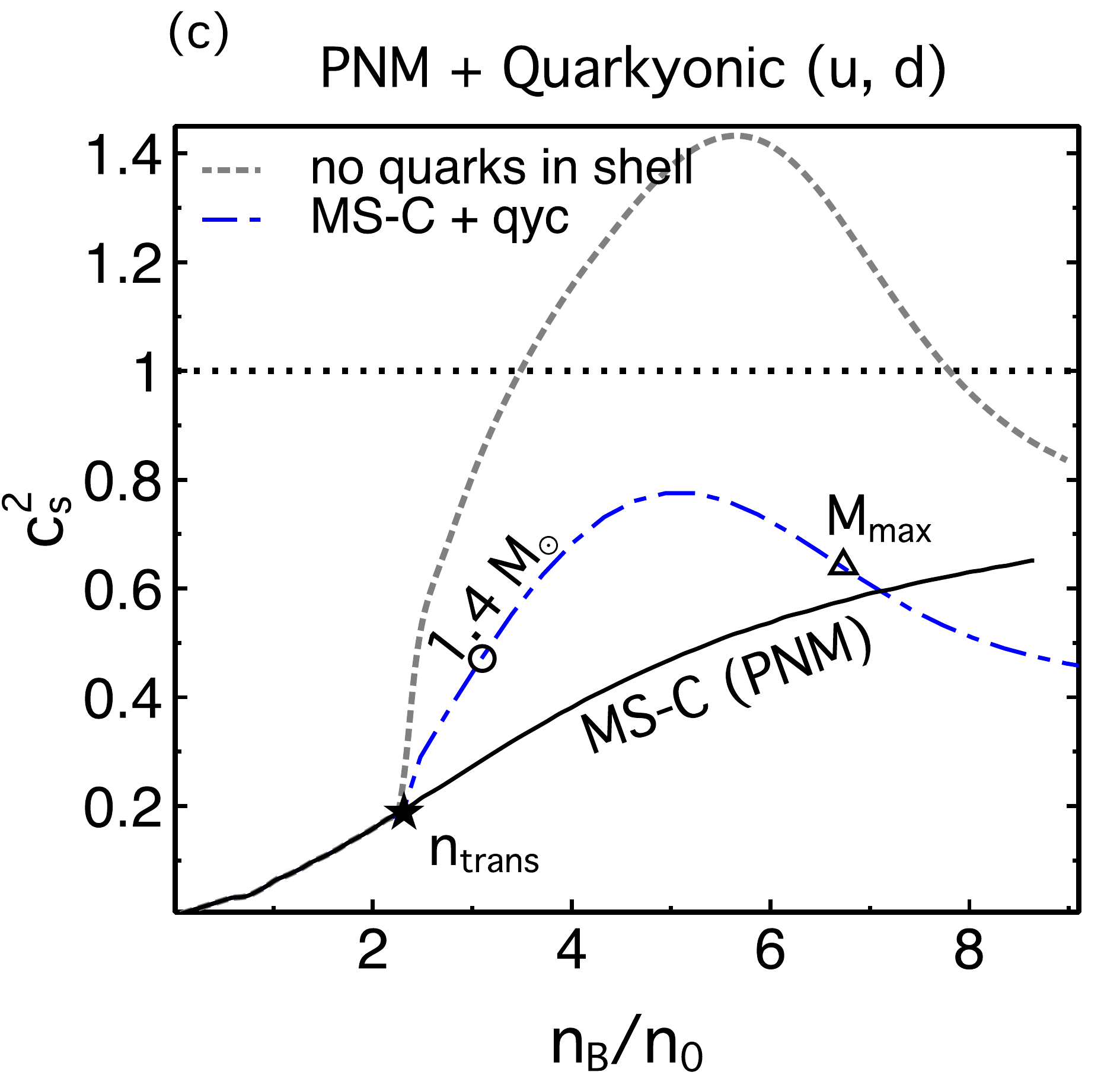}\\[-2ex]
}
\hspace{0.8cm}
\parbox{0.35\hsize}{
\includegraphics[width=\hsize]{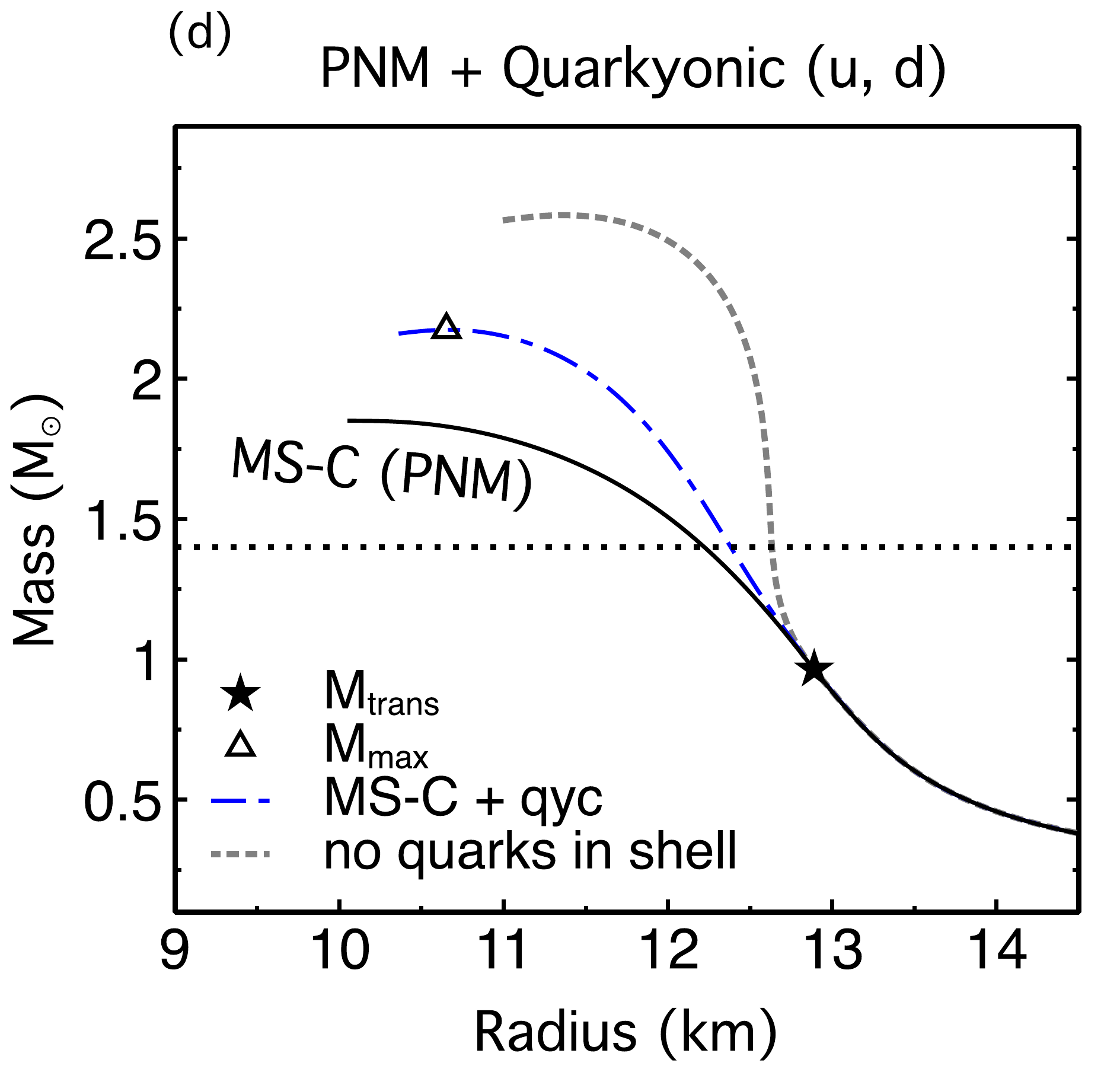}\\[-2ex]
}
\caption{(Color online) In panels (a) and (c), the squared speed of sound vs baryon density in PNM for the EoS models of Ref.~\cite{McLerran:2018hbz} and MS-C (PNM) of this work. The $1.4\,\Msolar$ and maximum-mass stars are marked with open circles and triangles respectively in $c_s^2 (\nb)$ plots. The right panels (b) and (d) show the corresponding $M$-$R$ curves. The different  curves illuminate the influence of the shell on the results.}
\label{fig:Cs2s}
\end{figure*}

An examination of the behavior of $c_s^2$ vs $\nb$ with and without quarks offers insights into the role played the presence of the  shell for $k_{Fn} > \De$  in the quarkyonic model. Fig.~\ref{fig:Cs2s} shows results of $c_s^2$ for the cases in which there is no shell (i.e. PNM throughout the star), neutrons only below and above $\De$ (i.e. with a shell but no quarks), and with the inclusion of quarks for $k_{Fn} > \De$. The results in this figure correspond to the neutron matter EoS used in Ref.~\cite{McLerran:2018hbz} and the MS-C+vNJL model of this work with $G_v/G_s=0.5$. For the former case, values of $\La_{Q}=420$ MeV and $\kappa = 1$ were used to calculate the shell width $\De$, and the onset of quarks occurs at $\ntrans = 0.37~{\rm fm}^{-3}$. This is to be compared with $\ntrans = 0.24~{\rm fm}^{-3}$ obtained with $\La_{Q}=380$ MeV and $\kappa = 0.3$ in the EoS of Ref.~\cite{McLerran:2018hbz}. For the two-flavor vNJL model used in this connection, values of the parameters used were $\La=631.4$ MeV and $G_s\La^2=1.835$ as in Ref.~\cite{Hatsuda:1994pi}. 

The main  differences between the models in Ref.~\cite{McLerran:2018hbz} and this work are: 

\noindent (i) For pure neutron matter (no quarks), the EoS of Ref.~\cite{McLerran:2018hbz} becomes acausal for $\nb/n_0 \simeq 6$ owing to the term proportional to $n_n^3$ in its interacting part. As the central density of the star is $\simeq 6.74\,n_0$, this feature may be of some concern. However, the MS-B/C+vNJL models - being relativistically covariant - remain causal for all densities, and \\
\noindent (ii) Interactions between quarks are not included in the EoS of Ref.~\cite{McLerran:2018hbz} except in the kinetic energy term with the use of $M_q=M_n/3$, whereas the MS-B/C+vNJL model uses density-dependent dynamically generated $u,d$ quark masses that steadily decrease with increasing density from their vacuum values of $\simeq M_n/3$. 
In addition, repulsive vector interactions between quarks were used in the vNJL models.  

The above differences notwithstanding, the inner workings of the quarkyonic model - particularly, the influence of quarks - are apparent from Fig.~\ref{fig:Cs2s} (a) and (c). Without the presence of quarks in the shell, the EoSs in both models are very stiff even to the point of being substantially acausal. The presence of quarks in the shell abates this undesirable behavior by softening the overall EoS (dash-dotted blue curves) relative to the case when only nucleons are present (dotted gray curves). With progressively increasing density, the density of nucleons is depleted within the shell whereas that of the quarks becomes predominant. As $c_s^2 \rightarrow 1/3$ for quarks at asymptotically high densities, it exhibits a maximum (as well as a minimum) at some intermediate density. Note however, that compared to the case of no shell, pure neutron matter everywhere (black solid curves), the overall EoS of the quarkyonic matter is still stiffer within the central densities of the corresponding stars. 

Insofar as $c_s^2$ is a measure of the stiffness of the EoS, the $M$-$R$ curves shown in Fig.~\ref{fig:Cs2s} (b) and (d) reflect the corresponding $c_s^2$ vs $\nb$ behavior. 
The presence of quarkyonic matter (dash-dotted blue curve) causes an increase in the $\Mmax$ for both models. If only the neutron content of quarkyonic matter is considered (dotted gray curve), then the increase in $\Mmax$ is more substantial. Similarly, the radii of both the maximum mass and $1.4\,\Msolar$ stars are significantly larger in quarkyonic matter. Quantitative differences between the two cases can be attributed to the presence of interactions between quarks in the MS-B+vNJL model.

\begin{figure*}[htb]
\parbox{0.34\hsize}{
\includegraphics[width=\hsize]{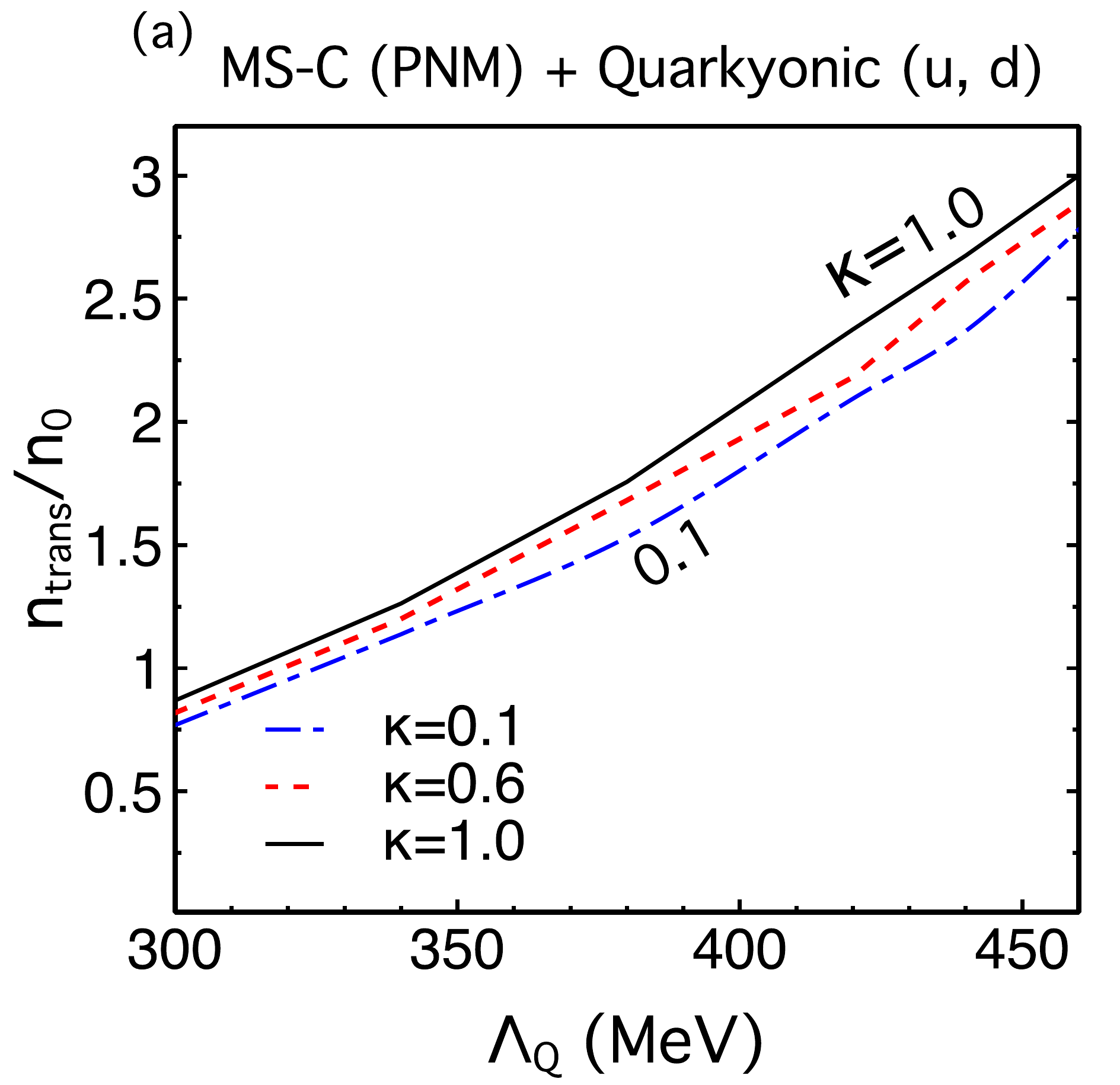}\\[-2ex]
}\parbox{0.34\hsize}{
\includegraphics[width=\hsize]{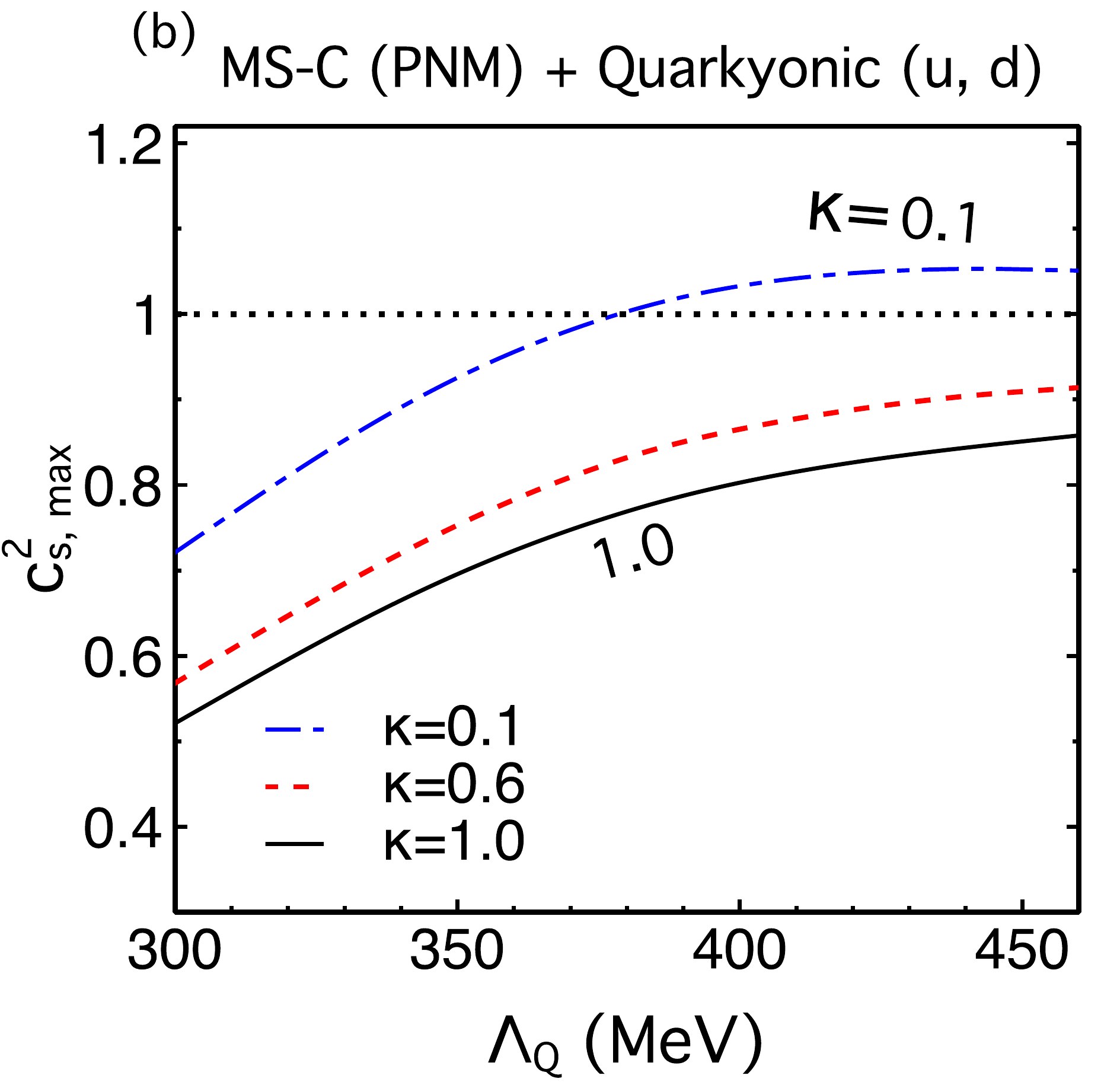}\\[-2ex]
}\parbox{0.34\hsize}{
\includegraphics[width=\hsize]{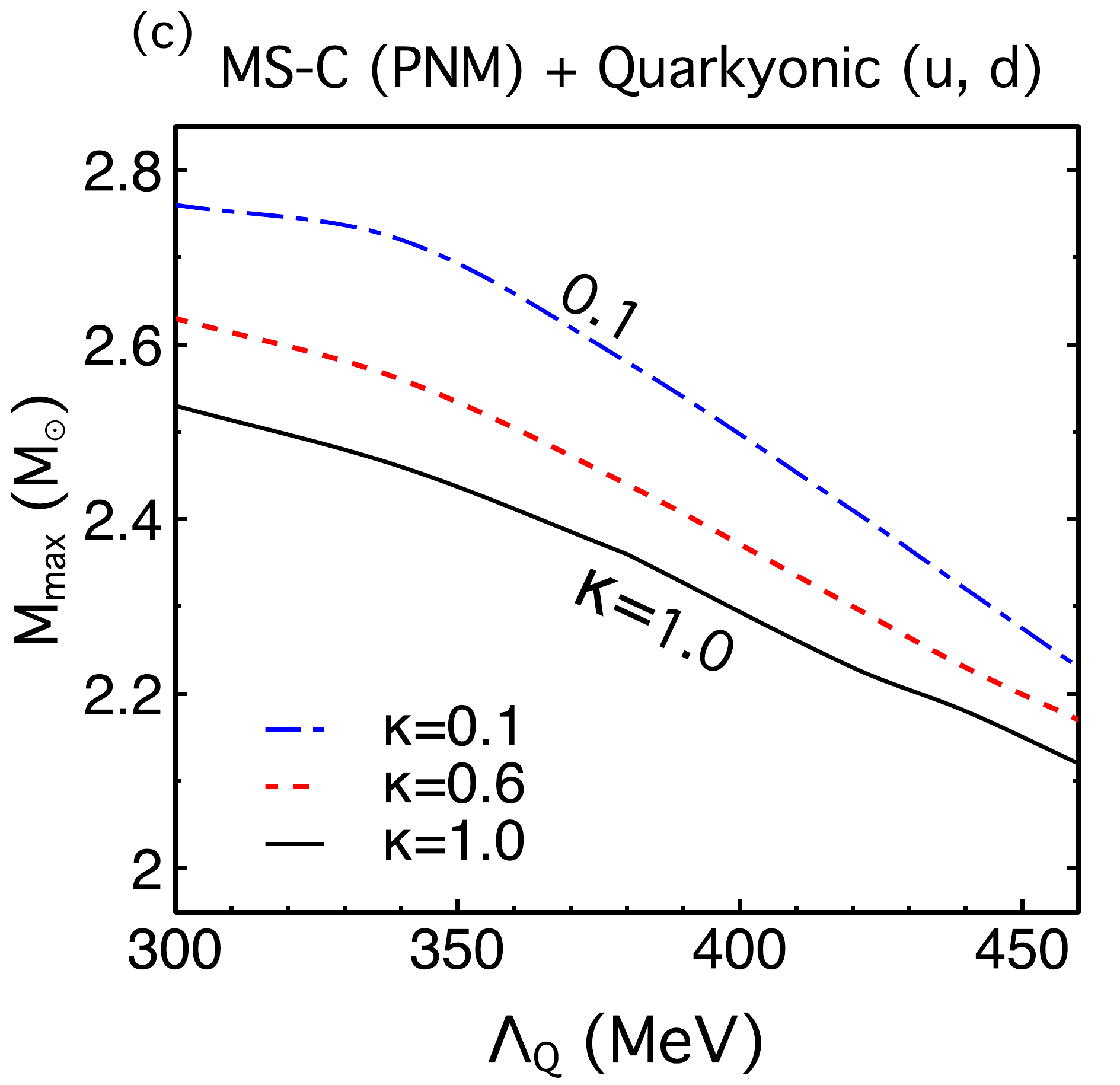}\\[-2ex]
}
\caption{(Color online) Variation of the hadron-to-quark transition density (in units of $n_0$), the squared speed of sound and the maximum mass as functions of the parameters $\La_Q$ and $\kappa$ that determine the shell width $\De$.
}
\label{fig:kappa-contour_qyc}
\end{figure*}

The hadron-to-quark transition density $\ntrans$, the peak value of the squared speed of sound $c_{s,\,{\rm max}}^2$, and the maximum mass $\Mmax$ all depend on the choice of $\La_{Q}$ and $\kappa$ used to calculate the shell width $\De$. Fig.~\ref{fig:kappa-contour_qyc} shows the variation of these quantities as a function of $\La_{Q}$ with $\kappa=0.1$, $0.6$ and 1 for the MS-C+vNJL model chosen here. Intermediate values of $\kappa$ lead to results that lie within the boundaries shown in this figure. Note that high values of both $\La_{Q}$ and $\kappa$ are required to ensure that $\ntrans \gtrsim 1.5\,n_0$ and $c_s^2 < 1$. This requirement decreases $\Mmax$, but masses above the current constraint of $\gtrsim 2\,\Msolar$ can be still obtained. In the absence of interactions between quarks (as in Ref.~\cite{McLerran:2018hbz}), the window of $\La_{Q}$ and $\kappa$ values that are usable is very small. We stress however, that the optimum choice of these parameters is model dependent in that if a different hadronic or quark EoS is used, the values of $\La_{Q}$ and $\kappa$ can change.

On a physical level, low values of $\La_{Q}$ and $\kappa$ lead to a substantial quark content in the star, but at the expense of $\ntrans\rightarrow n_0$ -- a disturbing trend.  Although quarks soften the overall EoS, the presence of the shell and the redistribution of baryon number between nucleons and quarks causes a substantial stiffening of the overall EoS, which in turn leads to very high values of $\Mmax$. Conversely, very high values of $\La_{Q}$ and $\kappa$ decrease the quark content which makes the overall EoS to be nearly that without quarks. This feature is generic to the quarkyonic model, which enables it to achieve maximum values consistent with the observational mass limit even when the EoS with hadrons only fails to meet this constraint. 

The low transition densities and the extreme stiffening of the EoS caused by the shell in quarkyonic matter bear further investigation. Although inspired by QCD and large $N_c$ physics, the width of the shell is independent of the EoSs in both hadronic and quark sectors, at least in the initial stage of the development of the model. The energy cost in creating such a shell in dense matter is another issue that warrants scrutiny.

\begin{table}[htb]
\begin{center}
\begin{tabular}{c|c|c|c}
\hline
&&&\\[-2ex]
Treatment  & $\ntrans/n_0$& $\Mtrans$  & Figure reference \\[0.5ex]
\hline  
&&&\\[-2ex]
Maxwell & 1.77 & $0.97\,\Msolar$ & Fig. 2 (b) \& (c) \\[0.5ex]
Gibbs & $\lesssim 1.5$ & $\lesssim 0.6\,\Msolar$  & Fig. 2 (e) \& (f) \\[0.5ex]
\hline  
&&&\\[-2ex]
Interpolation & 2.0, 1.5  & $0.81, 0.48 \,\Msolar$ & Fig. 5 (b) \& (c) \\[0.5ex]
\hline  
&&&\\[-2ex]
Quarkyonic & 2.31 & $0.97, 1.21\,\Msolar$  & Fig. 6 (b) \& (c)\\[0.5ex]
\hline
\end{tabular}
\end{center}
\caption{Summary of the minimum density and minimum neutron star mass when quarks start to appear in various treatments of phase transitions explored in our work; see also the indicated  figures for detailed information.
}
\label{tab:min-pt}
\end{table}  

\section{Conclusion and Outlook}
\label{sec:con}

In this work, we have performed a detailed comparison of  first-order phase transition and crossover treatments of the hadron-to-quark transition in neutron stars. For first-order transitions, results of both Maxwell and Gibbs constructions were examined. Also studied were interpolatory schemes and the second-order phase transition in quarkyonic matter, which fall in the class of crossover transitions. In both cases, sensitivity of the structural properties of neutron stars to variations in the EoSs in the hadronic as well as in the quark sectors were explored. The ensuing results were then tested for compatibility with the strict constraints imposed by the precise mass measurements of $2\,\Msolar$ neutron stars, the available limits on the tidal deformations of  neutron stars in the binary merger GW170817, and the radius estimates of $1.4\,\Msolar$ stars inferred from x-ray observations. 
These independent constraints from observations are significant in that the lower limit on the maximum mass reflects the behavior of the dense matter EoS for densities $\gtrsim 4-6\, n_0$, whereas bounds on binary tidal deformability $\tilde\La$ and estimates of $\Rtyp$ depend on the EoS for densities  $\gtrsim 2-3 \, n_0$, respectively.

Table \ref{tab:min-pt}  provides a summary of the transition density $\ntrans/n_0$ for the appearance of quarks and the associated neutron star mass, $\Mtrans$, for the EoSs and different treatments of the phase transition considered in this work. The entries in this table allow us to answer the first two of the three questions posed in the introduction:\\

\begin{table*}[htb]
\begin{center}
\begin{tabular}{c | c@{\quad} c  c }
\hline \\[-2ex]
HM $\to$ QM & First-order Transition (Maxwell) \footnote{See text for details if the Gibbs construction is applied. Gibbs construction satisfies many observational constraints such as $\Rtyp$ and $\Mmax$ due to the earlier onset of quarks. However, distinguishability from purely-hadronic stars is lost.} &  Crossover Transition\\[0.5ex]
\hline \\[-2ex]
\multirow{2}{*}{stiff to soft}     &   $\xmark$ vMIT: cannot support $\Mmax\geq 2\,\Msolar$  & \multirow{2}{*}{$\xmark$ unphysical decreasing function of $P(\nb)$} \\[0.5ex]
    & $\xmark$ vNJL: $\Mtrans\gtrsim 1.7 \,\Msolar$, $\tilde\La (\Mchirp=1.186\,\Msolar)>720$ &  \\[0.5ex]
\hline \\[-2ex]
\multirow{2}{*}{soft to stiff}  & \multirow{2}{*}{$\xmark$ no intersection for $P(\mu)$ \footnote{Limited by the specific quark models applied here; in a generic parametrization (e.g. CSS) the soft HM $\to$ stiff QM is possible.}} & \checkmark interpolation: $\Mtrans\lesssim 1.0 \,\Msolar$, $\Rtyp <13\, \rm{km}$ \\[0.5ex]
  &  & \checkmark quarkyonic:  $\Mmax \geq 2\,\Msolar$; $\Rtyp$ and $\Mtrans$ vary &  \\[0.5ex]
\hline \\[-2ex]
soft to soft       & $\xmark$ cannot support $\Mmax\geq 2\,\Msolar$   &  $\xmark$ cannot support $\Mmax\geq2\,\Msolar$ \\[0.5ex]
\hline \\[-2ex]
\multirow{2}{*}{stiff to stiff} & \checkmark vMIT: $\Mmax \geq 2\,\Msolar$; $\Rtyp$ and $\Mtrans$ vary   &   \multirow{2}{*}{$\xmark$ $\tilde\La (\Mchirp=1.186\,\Msolar)>720$, $\Rtyp > 13\,\rm{km}$}  \\[0.5ex]
&$\xmark$ vNJL: onset for quarks too high; immediately destabilize &  \\[0.5ex]
\hline
\end{tabular}
\end{center}
\caption{Summary of different treatments with the introduction of  quarks in the dense matter EoS. For a sharp first-order transition with Maxwell construction, the most readily compatible scenario is stiff hadronic matter undergoes phase transition into stiff ($\cQMsq\gtrsim 0.4$) quark matter. For a crossover transition, soft HM $\to$ stiff QM is necessary.
}
\label{tab:pt-summary}
\end{table*} 

(a) What is the minimum neutron star (NS) mass consistent with the observational lower limit of the maximum mass ($\Mmax$) that is likely to contain quarks?

The answer to this question depends on both the low-density hadronic and high-density quark EoSs, as well as the order and the method of implementing the phase transition.  Barring rare cases, such as in the Gibbs construction and interpolation (see why below), the minimum mass for phase transition is $\Mtrans \gtrsim 1\,\Msolar$.  \\

(b) What is the minimum physically reasonable density at which a hadron-quark transition of any sort can occur?

Our results indicate the minimum density $\ntrans$ to be $\gtrsim 2\,n_0$, again excluding the rare cases. The reasons for the exclusions are as follows. In the Gibbs construction, valid in the extreme case of the interface tension between the phases being zero, the onset density is generally lower than that of the corresponding Maxwell construction. Depending on the softness or stiffness of the hadronic and quark EoSs, $\ntrans$ can approach near-nuclear densities. Such cases should be discarded as being in conflict  with nuclear data near saturation. In the case of interpolation, the onset density is chosen a priori. In this approach, input values of $\ntrans \gtrsim 2\,n_0$ yield the minimum mass for phase transition $\Mtrans\gtrsim 1\,\Msolar$. 

The values of $\ntrans$ and $\Mtrans$ quoted in Table \ref{tab:min-pt} do not conflict with experimental data on nuclei that probe densities near and below $n_0$. A mild tension however, exists with theoretical interpretations of low-to-intermediate energy heavy-ion data \cite{Danielewicz:2002pu} which probe densities up to $3-4\,n_0$.  We wish to note, however, that analysis of such data using Boltzmann-type kinetic equations has not yet been performed with quark degrees of freedom and their subsequent hadronization as in RHIC and CERN experiments. \\

Table \ref{tab:pt-summary} provides a summary of the generic outcomes of our study. If the hadron-to-quark transition is strongly first-order, as is the case for standard quark models such as vMIT and vNJL that we used, then the hadronic part needs to be relatively stiff to guarantee a proper intersection in the $P$-$\mu$ plane. For a hadronic EoS as stiff as MS-A, this combination brings tension with $\La_{1.4}$ or $\Rtyp$ estimates. Concomitantly,  a too-high transition density that yields $\Mtrans\gtrsim 1.7\,\Msolar$ results in either very small quark cores or completely unstable stars that are indistinguishable from those resulting from the (stiff) purely-hadronic EoS. Thus, such hybrid EoSs can easily be ruled out. This is typical for NJL-type models; see e.g. Fig.~\ref{fig:MR-eos_vNJL}. In contrast, lower transition densities (that yield $\Mtrans\lesssim 1.0-1.6\, \Msolar$) are capable of decreasing radii, and if accompanied by a stiff quark matter EoS with sizable repulsive interactions, these hybrid EoSs produce $\Mmax \geq 2\,\Msolar$. Figs.~\ref{fig:MR-eos_vMIT} and \ref{fig:k2-Lam_vMIT} show examples in which the vMIT model was used with Maxwell/Gibbs constructions.   

Our analysis indicates that use of the Gibbs construction is beneficial in satisfying the current constraints from observation for many stiff hadronic EoSs, as it enlarges the parameter space of quark models. As similar  $M(R)$ and $\La (M)$ relations for hybrid and purely-hadronic stars can be obtained, the distinction between the two is, however, lost. This feature underscores the significance of dynamical properties such as neutron star cooling and spin-down, and the evolution of merger products.

To sum up the part about first-order phase transitions, current observational constraints disfavor weakly-interacting quarks at the densities reached in neutron star cores. Should a first-order transition into strongly-interacting quark matter (as described by the vMIT bag model or vNJL-type models) take place, the onset density is likely of relevance also to canonical neutron star masses in the range $1.0-1.6\,\Msolar$. 

One should keep in mind, however, that perturbative approaches to the quark matter EoS are not expected to hold in the density range $\approx 2-4 \, n_0$. This limitation brings the validity of first-order phase transitions caused by such EoSs into question. In this regard, model-independent parameterizations circumvent the issue and have the advantage of translating observational constraints more generically. For instance, specific QM models prohibit the transition into soft hadronic matter, but in the CSS parameterization this restriction disappears and a much larger parameter space can be explored including soft HM $\to$ stiff QM~\cite{Han:2018mtj}. However, such parameterizations lack a physical basis and beg for the invention of a non-perturbative approach.

If the hadron-to-quark transition is a smooth crossover, as in the case of interpolatory schemes and in quarkyonic matter, the pressure in the transition region is stiffened unlike the sudden softening of pressure caused by a first-order transition. This stiffening is also reflected in a local peak in the sound velocity before the pure quark phase is entered. This stiffening is responsible for supporting massive stars that are compatible with the current lower limit of 2\,$\Msolar$.  

It is also common that the onset density for quarks is somewhat low ($\Mtrans\lesssim 1.0-1.6 \,\Msolar$) in these crossover approaches. This feature implies that  all neutron stars we observe should contain some quarks admixed with hadrons. We find that at low densities soft hadronic EoSs are necessary, but above the transition changes in radii rely heavily on the methods of implementing the crossover in both  the interpolation approach and in quarkyonic matter. Consequently, it is difficult to obtain physical constraints on the crossover EoSs from a better determination of the radius, e.g. $\Rtyp$, or improved tidal deformability measurements. It is promising, however, in limiting parameters, e.g. the vector coupling strength $G_v/G_s$ in vNJL or $\kappa$ and $\La$ in the quarkyonic model,  pertinent to the required stiffening to satisfy the limits  imposed by mass measurements of heavy neutron stars.

Regardless of the phase transition being first-order or crossover, our results suggest that the presence of quarks in the pre-merger component neutron stars of GW170817 is a viable possibility. If quarks only appear after the merger (before the remnant collapsing into a black hole), there is a valid soft HM $\to$ stiff QM first-order transition that cannot be captured by the vMIT bag or vNJL models. There are exceptions when the onset occurs close to the $2\,\Msolar$ limit, so that quarks are precluded in cold beta-equilibrated NSs due to immediate collapse. While we rejected these solutions by default, these cases can however  be relevant for the dynamic products of mergers where quarks may emerge temporarily~\cite{Most:2018eaw,Chesler:2019osn}. Numerical simulations that involve quarks \cite{Most:2018eaw,Bauswein:2018bma} will assist in identifying such cases during post-merger gravitational-wave evolution. Better understanding and progress in theory, experiments, and observations are required to clarify the situation.
 
Although the presence of quarks in neutron stars is not ruled out by currently available constraints, it is nearly impossible to confirm it even with improved determinations of radii from x-ray observations and tidal deformabilities from gravitational wave detections. This conundrum arises because purely hadronic EoSs can also satisfy the current constraints of $\Mmax,~\Rtyp$ and $\tilde \La$; i.e., the ``masquerade problem'' \cite{Alford:2004pf} persists. Similarly, it will be difficult to identify the nature of the phase transition on the basis of $M$ and $R$ observations only, unless there is a sufficiently strong  first-order transition that gives rise to separate branches of twin stars with discontinuous $M$-$R$ and/or $\tilde\La$-$\Mchirp$ relations. 

We now turn to the third question in the introduction: (c) Which astronomical observations have the best potential to attest to the presence of quarks? Dynamical observables such as supernovae neutrino emission, thermal/spin evolution, global oscillation modes, continuous gravitational waves, dynamic collapse etc. that are sensitive to transport properties would potentially provide more distinct signatures of exotic matter in neutron stars~\cite{Fischer:2017lag, Ibanez:2018myp,Alford:2019oge}. In future work, it is worthwhile to achieve consistency with dynamical observables, particularly for the crossover scenarios of the transition to quark matter.

\section*{Acknowledgements} 
We thank Larry McLerran and Sanjay Reddy for helpful discussions relating to quarkyonic matter. This work was performed with research support from the National Science Foundation Grant PHY-1630782 and the Heising-Simons Foundation Grant 2017-228 for S.H., the U.S. Department of Energy Grant No. DE-FG02-93ER-40756 for M.A.A.M., S.L. and M.P. The research of C.C. is supported by the National Science Foundation Grant PHY-1748621.

\bibliographystyle{apsrev}
\bibliography{treat_qm_fin}

\begin{thebibliography}{140}
\expandafter\ifx\csname natexlab\endcsname\relax\def\natexlab#1{#1}\fi
\expandafter\ifx\csname bibnamefont\endcsname\relax
  \def\bibnamefont#1{#1}\fi
\expandafter\ifx\csname bibfnamefont\endcsname\relax
  \def\bibfnamefont#1{#1}\fi
\expandafter\ifx\csname citenamefont\endcsname\relax
  \def\citenamefont#1{#1}\fi
\expandafter\ifx\csname url\endcsname\relax
  \def\url#1{\texttt{#1}}\fi
\expandafter\ifx\csname urlprefix\endcsname\relax\def\urlprefix{URL }\fi
\providecommand{\bibinfo}[2]{#2}
\providecommand{\eprint}[2][]{\url{#2}}

\bibitem[{\citenamefont{Collins and Perry}(1975)}]{CP75}
\bibinfo{author}{\bibfnamefont{J.~C.} \bibnamefont{Collins}} \bibnamefont{and}
  \bibinfo{author}{\bibfnamefont{M.~J.} \bibnamefont{Perry}},
  \bibinfo{journal}{Phys. Rev. Lett.} \textbf{\bibinfo{volume}{34}},
  \bibinfo{pages}{1353} (\bibinfo{year}{1975}).

\bibitem[{\citenamefont{Meisel et~al.}(2018)\citenamefont{Meisel, Deibel, Keek,
  Shternin, and Elfritz}}]{Meisel:2018ufs}
\bibinfo{author}{\bibfnamefont{Z.}~\bibnamefont{Meisel}},
  \bibinfo{author}{\bibfnamefont{A.}~\bibnamefont{Deibel}},
  \bibinfo{author}{\bibfnamefont{L.}~\bibnamefont{Keek}},
  \bibinfo{author}{\bibfnamefont{P.}~\bibnamefont{Shternin}}, \bibnamefont{and}
  \bibinfo{author}{\bibfnamefont{J.}~\bibnamefont{Elfritz}},
  \bibinfo{journal}{J. Phys. G} \textbf{\bibinfo{volume}{45}},
  \bibinfo{pages}{093001} (\bibinfo{year}{2018}).

\bibitem[{\citenamefont{Baym et~al.}(1971)\citenamefont{Baym, Pethick, and
  Sutherland}}]{Baym71tg}
\bibinfo{author}{\bibfnamefont{G.}~\bibnamefont{Baym}},
  \bibinfo{author}{\bibfnamefont{C.}~\bibnamefont{Pethick}}, \bibnamefont{and}
  \bibinfo{author}{\bibfnamefont{P.}~\bibnamefont{Sutherland}},
  \bibinfo{journal}{Astrophys. J.} \textbf{\bibinfo{volume}{170}},
  \bibinfo{pages}{299} (\bibinfo{year}{1971}).

\bibitem[{\citenamefont{Negele and Vautherin}(1973)}]{Negele73ns}
\bibinfo{author}{\bibfnamefont{J.~W.} \bibnamefont{Negele}} \bibnamefont{and}
  \bibinfo{author}{\bibfnamefont{D.}~\bibnamefont{Vautherin}},
  \bibinfo{journal}{Nucl. Phys. A} \textbf{\bibinfo{volume}{207}},
  \bibinfo{pages}{298} (\bibinfo{year}{1973}).

\bibitem[{\citenamefont{Douchin and Haensel}(2001)}]{Douchin:2001sv}
\bibinfo{author}{\bibfnamefont{F.}~\bibnamefont{Douchin}} \bibnamefont{and}
  \bibinfo{author}{\bibfnamefont{P.}~\bibnamefont{Haensel}},
  \bibinfo{journal}{Astron. Astrophys.} \textbf{\bibinfo{volume}{380}},
  \bibinfo{pages}{151} (\bibinfo{year}{2001}).

\bibitem[{\citenamefont{Sharma et~al.}(2015)\citenamefont{Sharma, Centelles,
  Vi{\~n}as, Baldo, and Burgio}}]{Sharma:2015bna}
\bibinfo{author}{\bibfnamefont{B.~K.} \bibnamefont{Sharma}},
  \bibinfo{author}{\bibfnamefont{M.}~\bibnamefont{Centelles}},
  \bibinfo{author}{\bibfnamefont{X.}~\bibnamefont{Vi{\~n}as}},
  \bibinfo{author}{\bibfnamefont{M.}~\bibnamefont{Baldo}}, \bibnamefont{and}
  \bibinfo{author}{\bibfnamefont{G.~F.} \bibnamefont{Burgio}},
  \bibinfo{journal}{Astron. Astrophys.} \textbf{\bibinfo{volume}{584}},
  \bibinfo{pages}{A103} (\bibinfo{year}{2015}).

\bibitem[{\citenamefont{Hohenberg and Kohn}(1964)}]{HK64}
\bibinfo{author}{\bibfnamefont{P.}~\bibnamefont{Hohenberg}} \bibnamefont{and}
  \bibinfo{author}{\bibfnamefont{W.}~\bibnamefont{Kohn}},
  \bibinfo{journal}{Phys. Rev. B} \textbf{\bibinfo{volume}{76}},
  \bibinfo{pages}{6062} (\bibinfo{year}{1964}).

\bibitem[{\citenamefont{Kohn and Sham}(1965)}]{KS65}
\bibinfo{author}{\bibfnamefont{W.}~\bibnamefont{Kohn}} \bibnamefont{and}
  \bibinfo{author}{\bibfnamefont{L.~J.} \bibnamefont{Sham}},
  \bibinfo{journal}{Phys. Rev. A} \textbf{\bibinfo{volume}{140}},
  \bibinfo{pages}{1133} (\bibinfo{year}{1965}).

\bibitem[{\citenamefont{Myers and Swiatecki}(1966)}]{Myers:1966zz}
\bibinfo{author}{\bibfnamefont{W.~D.} \bibnamefont{Myers}} \bibnamefont{and}
  \bibinfo{author}{\bibfnamefont{W.~J.} \bibnamefont{Swiatecki}},
  \bibinfo{journal}{Nucl. Phys.} \textbf{\bibinfo{volume}{81}},
  \bibinfo{pages}{1} (\bibinfo{year}{1966}).

\bibitem[{\citenamefont{Myers and Swiatecki}(1996)}]{Myers:1995wx}
\bibinfo{author}{\bibfnamefont{W.~D.} \bibnamefont{Myers}} \bibnamefont{and}
  \bibinfo{author}{\bibfnamefont{W.~J.} \bibnamefont{Swiatecki}},
  \bibinfo{journal}{Nucl. Phys. A} \textbf{\bibinfo{volume}{601}},
  \bibinfo{pages}{141} (\bibinfo{year}{1996}).

\bibitem[{\citenamefont{Day}(1978)}]{Day:1978zz}
\bibinfo{author}{\bibfnamefont{B.~D.} \bibnamefont{Day}},
  \bibinfo{journal}{Rev. Mod. Phys.} \textbf{\bibinfo{volume}{50}},
  \bibinfo{pages}{495} (\bibinfo{year}{1978}).

\bibitem[{\citenamefont{Garg}(2004{\natexlab{a}})}]{Garg:2004fsg}
\bibinfo{author}{\bibfnamefont{U.}~\bibnamefont{Garg}}, \bibinfo{journal}{Nucl.
  Phys. A} \textbf{\bibinfo{volume}{731}}, \bibinfo{pages}{3}
  (\bibinfo{year}{2004}{\natexlab{a}}).

\bibitem[{\citenamefont{Col{\`o} et~al.}(2004)\citenamefont{Col{\`o}, Van~Giai,
  Meyer, Bennaceur, and Bonche}}]{Colo:2004mj}
\bibinfo{author}{\bibfnamefont{G.}~\bibnamefont{Col{\`o}}},
  \bibinfo{author}{\bibfnamefont{N.}~\bibnamefont{Van~Giai}},
  \bibinfo{author}{\bibfnamefont{J.}~\bibnamefont{Meyer}},
  \bibinfo{author}{\bibfnamefont{K.}~\bibnamefont{Bennaceur}},
  \bibnamefont{and} \bibinfo{author}{\bibfnamefont{P.}~\bibnamefont{Bonche}},
  \bibinfo{journal}{Phys. Rev. C} \textbf{\bibinfo{volume}{70}},
  \bibinfo{pages}{024307} (\bibinfo{year}{2004}).

\bibitem[{\citenamefont{Shlomo et~al.}(2006)\citenamefont{Shlomo, Kolomietz,
  and Col{\`o}}}]{Shlomo06}
\bibinfo{author}{\bibfnamefont{S.}~\bibnamefont{Shlomo}},
  \bibinfo{author}{\bibfnamefont{V.~M.} \bibnamefont{Kolomietz}},
  \bibnamefont{and} \bibinfo{author}{\bibfnamefont{G.}~\bibnamefont{Col{\`o}}},
  \bibinfo{journal}{The European Physical Journal A - Hadrons and Nuclei}
  \textbf{\bibinfo{volume}{30}}, \bibinfo{pages}{23} (\bibinfo{year}{2006}).

\bibitem[{\citenamefont{Bohigas et~al.}(1979)\citenamefont{Bohigas, Lane, and
  Martorell}}]{Bohigas:1978qu}
\bibinfo{author}{\bibfnamefont{O.}~\bibnamefont{Bohigas}},
  \bibinfo{author}{\bibfnamefont{A.~M.} \bibnamefont{Lane}}, \bibnamefont{and}
  \bibinfo{author}{\bibfnamefont{J.}~\bibnamefont{Martorell}},
  \bibinfo{journal}{Phys. Rept.} \textbf{\bibinfo{volume}{51}},
  \bibinfo{pages}{267} (\bibinfo{year}{1979}).

\bibitem[{\citenamefont{Krivine et~al.}(1980)\citenamefont{Krivine, Treiner,
  and Bohigas}}]{Krivine:1980kzz}
\bibinfo{author}{\bibfnamefont{H.}~\bibnamefont{Krivine}},
  \bibinfo{author}{\bibfnamefont{J.}~\bibnamefont{Treiner}}, \bibnamefont{and}
  \bibinfo{author}{\bibfnamefont{O.}~\bibnamefont{Bohigas}},
  \bibinfo{journal}{Nucl. Phys. A} \textbf{\bibinfo{volume}{336}},
  \bibinfo{pages}{155} (\bibinfo{year}{1980}).

\bibitem[{\citenamefont{Margueron et~al.}(2018)\citenamefont{Margueron,
  Hoffmann~Casali, and Gulminelli}}]{Margueron:2017eqc}
\bibinfo{author}{\bibfnamefont{J.}~\bibnamefont{Margueron}},
  \bibinfo{author}{\bibfnamefont{R.}~\bibnamefont{Hoffmann~Casali}},
  \bibnamefont{and}
  \bibinfo{author}{\bibfnamefont{F.}~\bibnamefont{Gulminelli}},
  \bibinfo{journal}{Phys. Rev. C} \textbf{\bibinfo{volume}{97}},
  \bibinfo{pages}{025805} (\bibinfo{year}{2018}).

\bibitem[{\citenamefont{Lattimer and Lim}(2013)}]{Lattimer:2012xj}
\bibinfo{author}{\bibfnamefont{J.~M.} \bibnamefont{Lattimer}} \bibnamefont{and}
  \bibinfo{author}{\bibfnamefont{Y.}~\bibnamefont{Lim}},
  \bibinfo{journal}{Astrophys. J.} \textbf{\bibinfo{volume}{771}},
  \bibinfo{pages}{51} (\bibinfo{year}{2013}).

\bibitem[{\citenamefont{Tsang et~al.}(2012)}]{Tsang:2012se}
\bibinfo{author}{\bibfnamefont{M.~B.} \bibnamefont{Tsang}}
  \bibnamefont{et~al.}, \bibinfo{journal}{Phys. Rev. C}
  \textbf{\bibinfo{volume}{86}}, \bibinfo{pages}{015803}
  (\bibinfo{year}{2012}).

\bibitem[{\citenamefont{Gale et~al.}(1987)\citenamefont{Gale, Bertsch, and
  Das~Gupta}}]{Gale87}
\bibinfo{author}{\bibfnamefont{C.}~\bibnamefont{Gale}},
  \bibinfo{author}{\bibfnamefont{G.}~\bibnamefont{Bertsch}}, \bibnamefont{and}
  \bibinfo{author}{\bibfnamefont{S.}~\bibnamefont{Das~Gupta}},
  \bibinfo{journal}{Phys. Rev. C} \textbf{\bibinfo{volume}{35}},
  \bibinfo{pages}{1666} (\bibinfo{year}{1987}).

\bibitem[{\citenamefont{Prakash
  et~al.}(1988{\natexlab{a}})\citenamefont{Prakash, Kuo, and
  Das~Gupta}}]{Prakash88b}
\bibinfo{author}{\bibfnamefont{M.}~\bibnamefont{Prakash}},
  \bibinfo{author}{\bibfnamefont{T.~T.~S.} \bibnamefont{Kuo}},
  \bibnamefont{and}
  \bibinfo{author}{\bibfnamefont{S.}~\bibnamefont{Das~Gupta}},
  \bibinfo{journal}{Phys. Rev. C} \textbf{\bibinfo{volume}{37}},
  \bibinfo{pages}{2253} (\bibinfo{year}{1988}{\natexlab{a}}).

\bibitem[{\citenamefont{Welke et~al.}(1988)\citenamefont{Welke, Prakash, Kuo,
  Das~Gupta, and Gale}}]{Welke88}
\bibinfo{author}{\bibfnamefont{G.~M.} \bibnamefont{Welke}},
  \bibinfo{author}{\bibfnamefont{M.}~\bibnamefont{Prakash}},
  \bibinfo{author}{\bibfnamefont{T.~T.~S.} \bibnamefont{Kuo}},
  \bibinfo{author}{\bibfnamefont{S.}~\bibnamefont{Das~Gupta}},
  \bibnamefont{and} \bibinfo{author}{\bibfnamefont{C.}~\bibnamefont{Gale}},
  \bibinfo{journal}{Phys. Rev. C} \textbf{\bibinfo{volume}{38}},
  \bibinfo{pages}{2101} (\bibinfo{year}{1988}).

\bibitem[{\citenamefont{Gale et~al.}(1990)\citenamefont{Gale, Welke, Prakash,
  Lee, and Das~Gupta}}]{Gale90}
\bibinfo{author}{\bibfnamefont{C.}~\bibnamefont{Gale}},
  \bibinfo{author}{\bibfnamefont{G.~M.} \bibnamefont{Welke}},
  \bibinfo{author}{\bibfnamefont{M.}~\bibnamefont{Prakash}},
  \bibinfo{author}{\bibfnamefont{S.~J.} \bibnamefont{Lee}}, \bibnamefont{and}
  \bibinfo{author}{\bibfnamefont{S.}~\bibnamefont{Das~Gupta}},
  \bibinfo{journal}{Phys. Rev. C} \textbf{\bibinfo{volume}{41}},
  \bibinfo{pages}{1545} (\bibinfo{year}{1990}).

\bibitem[{\citenamefont{Danielewicz}(2000)}]{Danielewicz00}
\bibinfo{author}{\bibfnamefont{P.}~\bibnamefont{Danielewicz}},
  \bibinfo{journal}{Nucl. Phys. A} \textbf{\bibinfo{volume}{673}},
  \bibinfo{pages}{375} (\bibinfo{year}{2000}).

\bibitem[{\citenamefont{Danielewicz et~al.}(2002)\citenamefont{Danielewicz,
  Lacey, and Lynch}}]{Danielewicz:2002pu}
\bibinfo{author}{\bibfnamefont{P.}~\bibnamefont{Danielewicz}},
  \bibinfo{author}{\bibfnamefont{R.}~\bibnamefont{Lacey}}, \bibnamefont{and}
  \bibinfo{author}{\bibfnamefont{W.~G.} \bibnamefont{Lynch}},
  \bibinfo{journal}{Science} \textbf{\bibinfo{volume}{298}},
  \bibinfo{pages}{1592} (\bibinfo{year}{2002}).

\bibitem[{\citenamefont{Youngblood et~al.}(1999)\citenamefont{Youngblood,
  Clark, and Lui}}]{Youngblood99}
\bibinfo{author}{\bibfnamefont{D.~H.} \bibnamefont{Youngblood}},
  \bibinfo{author}{\bibfnamefont{H.~L.} \bibnamefont{Clark}}, \bibnamefont{and}
  \bibinfo{author}{\bibfnamefont{Y.~W.} \bibnamefont{Lui}},
  \bibinfo{journal}{Phys. Rev. Lett.} \textbf{\bibinfo{volume}{82}},
  \bibinfo{pages}{691} (\bibinfo{year}{1999}).

\bibitem[{\citenamefont{Garg}(2004{\natexlab{b}})}]{Garg04}
\bibinfo{author}{\bibfnamefont{U.}~\bibnamefont{Garg}}, \bibinfo{journal}{Nucl.
  Phys. A} \textbf{\bibinfo{volume}{731}}, \bibinfo{pages}{3}
  (\bibinfo{year}{2004}{\natexlab{b}}).

\bibitem[{\citenamefont{Colo et~al.}(2004)\citenamefont{Colo, Van~Giai, Meyer,
  Bennaceur, and Bonche}}]{Colo04}
\bibinfo{author}{\bibfnamefont{G.}~\bibnamefont{Colo}},
  \bibinfo{author}{\bibfnamefont{N.}~\bibnamefont{Van~Giai}},
  \bibinfo{author}{\bibfnamefont{J.}~\bibnamefont{Meyer}},
  \bibinfo{author}{\bibfnamefont{K.}~\bibnamefont{Bennaceur}},
  \bibnamefont{and} \bibinfo{author}{\bibfnamefont{P.}~\bibnamefont{Bonche}},
  \bibinfo{journal}{Phys. Rev. C} \textbf{\bibinfo{volume}{70}},
  \bibinfo{pages}{024307} (\bibinfo{year}{2004}).

\bibitem[{\citenamefont{Bludman and Dover}(1980)}]{Bludman:1980xn}
\bibinfo{author}{\bibfnamefont{S.~A.} \bibnamefont{Bludman}} \bibnamefont{and}
  \bibinfo{author}{\bibfnamefont{C.~B.} \bibnamefont{Dover}},
  \bibinfo{journal}{Phys. Rev. D} \textbf{\bibinfo{volume}{22}},
  \bibinfo{pages}{1333} (\bibinfo{year}{1980}).

\bibitem[{\citenamefont{Prakash
  et~al.}(1988{\natexlab{b}})\citenamefont{Prakash, Ainsworth, and
  Lattimer}}]{Prakash:1988md}
\bibinfo{author}{\bibfnamefont{M.}~\bibnamefont{Prakash}},
  \bibinfo{author}{\bibfnamefont{T.~L.} \bibnamefont{Ainsworth}},
  \bibnamefont{and} \bibinfo{author}{\bibfnamefont{J.~M.}
  \bibnamefont{Lattimer}}, \bibinfo{journal}{Phys. Rev. Lett.}
  \textbf{\bibinfo{volume}{61}}, \bibinfo{pages}{2518}
  (\bibinfo{year}{1988}{\natexlab{b}}).

\bibitem[{\citenamefont{Nauenberg and Chapline}(1973)}]{Nauenberg73}
\bibinfo{author}{\bibfnamefont{M.}~\bibnamefont{Nauenberg}} \bibnamefont{and}
  \bibinfo{author}{\bibfnamefont{G.}~\bibnamefont{Chapline}},
  \bibinfo{journal}{Astrophys. J.} \textbf{\bibinfo{volume}{179}},
  \bibinfo{pages}{277} (\bibinfo{year}{1973}).

\bibitem[{\citenamefont{Lattimer et~al.}(1990)\citenamefont{Lattimer, Prakash,
  Masak, and Yahil}}]{Lattimer:1990zz}
\bibinfo{author}{\bibfnamefont{J.~M.} \bibnamefont{Lattimer}},
  \bibinfo{author}{\bibfnamefont{M.}~\bibnamefont{Prakash}},
  \bibinfo{author}{\bibfnamefont{D.}~\bibnamefont{Masak}}, \bibnamefont{and}
  \bibinfo{author}{\bibfnamefont{A.}~\bibnamefont{Yahil}},
  \bibinfo{journal}{Astrophys. J.} \textbf{\bibinfo{volume}{355}},
  \bibinfo{pages}{241} (\bibinfo{year}{1990}).

\bibitem[{\citenamefont{Hebeler et~al.}(2010)\citenamefont{Hebeler, Lattimer,
  Pethick, and Schwenk}}]{Hebeler:2010jx}
\bibinfo{author}{\bibfnamefont{K.}~\bibnamefont{Hebeler}},
  \bibinfo{author}{\bibfnamefont{J.~M.} \bibnamefont{Lattimer}},
  \bibinfo{author}{\bibfnamefont{C.~J.} \bibnamefont{Pethick}},
  \bibnamefont{and} \bibinfo{author}{\bibfnamefont{A.}~\bibnamefont{Schwenk}},
  \bibinfo{journal}{Phys. Rev. Lett.} \textbf{\bibinfo{volume}{105}},
  \bibinfo{pages}{161102} (\bibinfo{year}{2010}).

\bibitem[{\citenamefont{Tews et~al.}(2013)\citenamefont{Tews, Kr{\"u}ger,
  Hebeler, and Schwenk}}]{Tews:2012fj}
\bibinfo{author}{\bibfnamefont{I.}~\bibnamefont{Tews}},
  \bibinfo{author}{\bibfnamefont{T.}~\bibnamefont{Kr{\"u}ger}},
  \bibinfo{author}{\bibfnamefont{K.}~\bibnamefont{Hebeler}}, \bibnamefont{and}
  \bibinfo{author}{\bibfnamefont{A.}~\bibnamefont{Schwenk}},
  \bibinfo{journal}{Phys. Rev. Lett.} \textbf{\bibinfo{volume}{110}},
  \bibinfo{pages}{032504} (\bibinfo{year}{2013}).

\bibitem[{\citenamefont{Lattimer and Prakash}(2016)}]{Lattimer:2015nhk}
\bibinfo{author}{\bibfnamefont{J.~M.} \bibnamefont{Lattimer}} \bibnamefont{and}
  \bibinfo{author}{\bibfnamefont{M.}~\bibnamefont{Prakash}},
  \bibinfo{journal}{Phys. Rept.} \textbf{\bibinfo{volume}{621}},
  \bibinfo{pages}{127} (\bibinfo{year}{2016}).

\bibitem[{\citenamefont{Steiner et~al.}(2016)\citenamefont{Steiner, Lattimer,
  and Brown}}]{Steiner:2015aea}
\bibinfo{author}{\bibfnamefont{A.~W.} \bibnamefont{Steiner}},
  \bibinfo{author}{\bibfnamefont{J.~M.} \bibnamefont{Lattimer}},
  \bibnamefont{and} \bibinfo{author}{\bibfnamefont{E.~F.} \bibnamefont{Brown}},
  \bibinfo{journal}{Eur. Phys. J. A} \textbf{\bibinfo{volume}{52}},
  \bibinfo{pages}{18} (\bibinfo{year}{2016}).

\bibitem[{\citenamefont{Tews et~al.}(2018)\citenamefont{Tews, Carlson,
  Gandolfi, and Reddy}}]{Tews:2018kmu}
\bibinfo{author}{\bibfnamefont{I.}~\bibnamefont{Tews}},
  \bibinfo{author}{\bibfnamefont{J.}~\bibnamefont{Carlson}},
  \bibinfo{author}{\bibfnamefont{S.}~\bibnamefont{Gandolfi}}, \bibnamefont{and}
  \bibinfo{author}{\bibfnamefont{S.}~\bibnamefont{Reddy}},
  \bibinfo{journal}{Astrophys. J.} \textbf{\bibinfo{volume}{860}},
  \bibinfo{pages}{149} (\bibinfo{year}{2018}).

\bibitem[{\citenamefont{Zhao and Lattimer}(2018)}]{Zhao:2018nyf}
\bibinfo{author}{\bibfnamefont{T.}~\bibnamefont{Zhao}} \bibnamefont{and}
  \bibinfo{author}{\bibfnamefont{J.~M.} \bibnamefont{Lattimer}},
  \bibinfo{journal}{Phys. Rev. D} \textbf{\bibinfo{volume}{98}},
  \bibinfo{pages}{063020} (\bibinfo{year}{2018}).

\bibitem[{\citenamefont{Chatterjee and Vida{\~n}a}(2016)}]{Chatterjee:2015pua}
\bibinfo{author}{\bibfnamefont{D.}~\bibnamefont{Chatterjee}} \bibnamefont{and}
  \bibinfo{author}{\bibfnamefont{I.}~\bibnamefont{Vida{\~n}a}},
  \bibinfo{journal}{Eur. Phys. J. A} \textbf{\bibinfo{volume}{52}},
  \bibinfo{pages}{29} (\bibinfo{year}{2016}).

\bibitem[{\citenamefont{Vida{\~n}a}(2018)}]{Vidana:2018bdi}
\bibinfo{author}{\bibfnamefont{I.}~\bibnamefont{Vida{\~n}a}},
  \bibinfo{journal}{Proc. Roy. Soc. Lond. A} \textbf{\bibinfo{volume}{474}},
  \bibinfo{pages}{0145} (\bibinfo{year}{2018}).

\bibitem[{\citenamefont{Baym and Chin}(1976{\natexlab{a}})}]{Baym:1976yu}
\bibinfo{author}{\bibfnamefont{G.}~\bibnamefont{Baym}} \bibnamefont{and}
  \bibinfo{author}{\bibfnamefont{S.~A.} \bibnamefont{Chin}},
  \bibinfo{journal}{Phys. Lett. B} \textbf{\bibinfo{volume}{62}},
  \bibinfo{pages}{241} (\bibinfo{year}{1976}{\natexlab{a}}).

\bibitem[{\citenamefont{Baym and Chin}(1976{\natexlab{b}})}]{Baym:1975va}
\bibinfo{author}{\bibfnamefont{G.}~\bibnamefont{Baym}} \bibnamefont{and}
  \bibinfo{author}{\bibfnamefont{S.~A.} \bibnamefont{Chin}},
  \bibinfo{journal}{Nucl. Phys. A} \textbf{\bibinfo{volume}{262}},
  \bibinfo{pages}{527} (\bibinfo{year}{1976}{\natexlab{b}}).

\bibitem[{\citenamefont{Freedman and McLerran}(1978)}]{Freedman:1977gz}
\bibinfo{author}{\bibfnamefont{B.}~\bibnamefont{Freedman}} \bibnamefont{and}
  \bibinfo{author}{\bibfnamefont{L.~D.} \bibnamefont{McLerran}},
  \bibinfo{journal}{Phys. Rev. D} \textbf{\bibinfo{volume}{17}},
  \bibinfo{pages}{1109} (\bibinfo{year}{1978}).

\bibitem[{\citenamefont{Farhi and Jaffe}(1984)}]{Farhi:1984qu}
\bibinfo{author}{\bibfnamefont{E.}~\bibnamefont{Farhi}} \bibnamefont{and}
  \bibinfo{author}{\bibfnamefont{R.~L.} \bibnamefont{Jaffe}},
  \bibinfo{journal}{Phys. Rev. D} \textbf{\bibinfo{volume}{30}},
  \bibinfo{pages}{2379} (\bibinfo{year}{1984}).

\bibitem[{\citenamefont{Kurkela et~al.}(2010)\citenamefont{Kurkela, Romatschke,
  and Vuorinen}}]{Kurkela:2009gj}
\bibinfo{author}{\bibfnamefont{A.}~\bibnamefont{Kurkela}},
  \bibinfo{author}{\bibfnamefont{P.}~\bibnamefont{Romatschke}},
  \bibnamefont{and} \bibinfo{author}{\bibfnamefont{A.}~\bibnamefont{Vuorinen}},
  \bibinfo{journal}{Phys. Rev. D} \textbf{\bibinfo{volume}{81}},
  \bibinfo{pages}{105021} (\bibinfo{year}{2010}).

\bibitem[{\citenamefont{Kurkela and Vuorinen}(2016)}]{Kurkela:2016was}
\bibinfo{author}{\bibfnamefont{A.}~\bibnamefont{Kurkela}} \bibnamefont{and}
  \bibinfo{author}{\bibfnamefont{A.}~\bibnamefont{Vuorinen}},
  \bibinfo{journal}{Phys. Rev. Lett.} \textbf{\bibinfo{volume}{117}},
  \bibinfo{pages}{042501} (\bibinfo{year}{2016}).

\bibitem[{\citenamefont{Gorda et~al.}(2018)\citenamefont{Gorda, Kurkela,
  Romatschke, S{\"a}ppi, and Vuorinen}}]{Gorda:2018gpy}
\bibinfo{author}{\bibfnamefont{T.}~\bibnamefont{Gorda}},
  \bibinfo{author}{\bibfnamefont{A.}~\bibnamefont{Kurkela}},
  \bibinfo{author}{\bibfnamefont{P.}~\bibnamefont{Romatschke}},
  \bibinfo{author}{\bibfnamefont{M.}~\bibnamefont{S{\"a}ppi}},
  \bibnamefont{and} \bibinfo{author}{\bibfnamefont{A.}~\bibnamefont{Vuorinen}},
  \bibinfo{journal}{Phys. Rev. Lett.} \textbf{\bibinfo{volume}{121}},
  \bibinfo{pages}{202701} (\bibinfo{year}{2018}).

\bibitem[{\citenamefont{Alford et~al.}(2008)\citenamefont{Alford, Schmitt,
  Rajagopal, and Sch{\"a}fer}}]{Alford:2007xm}
\bibinfo{author}{\bibfnamefont{M.~G.} \bibnamefont{Alford}},
  \bibinfo{author}{\bibfnamefont{A.}~\bibnamefont{Schmitt}},
  \bibinfo{author}{\bibfnamefont{K.}~\bibnamefont{Rajagopal}},
  \bibnamefont{and}
  \bibinfo{author}{\bibfnamefont{T.}~\bibnamefont{Sch{\"a}fer}},
  \bibinfo{journal}{Rev. Mod. Phys.} \textbf{\bibinfo{volume}{80}},
  \bibinfo{pages}{1455} (\bibinfo{year}{2008}).

\bibitem[{\citenamefont{Nambu and Jona-Lasinio}(1961)}]{Nambu:1961tp}
\bibinfo{author}{\bibfnamefont{Y.}~\bibnamefont{Nambu}} \bibnamefont{and}
  \bibinfo{author}{\bibfnamefont{G.}~\bibnamefont{Jona-Lasinio}},
  \bibinfo{journal}{Phys. Rev.} \textbf{\bibinfo{volume}{122}},
  \bibinfo{pages}{345} (\bibinfo{year}{1961}).

\bibitem[{\citenamefont{Kunihiro}(1989)}]{Kunihiro:1989my}
\bibinfo{author}{\bibfnamefont{T.}~\bibnamefont{Kunihiro}},
  \bibinfo{journal}{Phys. Lett. B} \textbf{\bibinfo{volume}{219}},
  \bibinfo{pages}{363} (\bibinfo{year}{1989}).

\bibitem[{\citenamefont{Buballa and Oertel}(1999)}]{Buballa:1998pr}
\bibinfo{author}{\bibfnamefont{M.}~\bibnamefont{Buballa}} \bibnamefont{and}
  \bibinfo{author}{\bibfnamefont{M.}~\bibnamefont{Oertel}},
  \bibinfo{journal}{Phys. Lett. B} \textbf{\bibinfo{volume}{457}},
  \bibinfo{pages}{261} (\bibinfo{year}{1999}).

\bibitem[{\citenamefont{Buballa}(2005)}]{Buballa:2003qv}
\bibinfo{author}{\bibfnamefont{M.}~\bibnamefont{Buballa}},
  \bibinfo{journal}{Phys. Rept.} \textbf{\bibinfo{volume}{407}},
  \bibinfo{pages}{205} (\bibinfo{year}{2005}).

\bibitem[{\citenamefont{Kl{\"a}hn and Fischer}(2015)}]{Klahn:2015mfa}
\bibinfo{author}{\bibfnamefont{T.}~\bibnamefont{Kl{\"a}hn}} \bibnamefont{and}
  \bibinfo{author}{\bibfnamefont{T.}~\bibnamefont{Fischer}},
  \bibinfo{journal}{Astrophys. J.} \textbf{\bibinfo{volume}{810}},
  \bibinfo{pages}{134} (\bibinfo{year}{2015}).

\bibitem[{\citenamefont{Gomes et~al.}(2019)\citenamefont{Gomes, Char, and
  Schramm}}]{Gomes:2018eiv}
\bibinfo{author}{\bibfnamefont{R.~O.} \bibnamefont{Gomes}},
  \bibinfo{author}{\bibfnamefont{P.}~\bibnamefont{Char}}, \bibnamefont{and}
  \bibinfo{author}{\bibfnamefont{S.}~\bibnamefont{Schramm}},
  \bibinfo{journal}{Astrophys. J.} \textbf{\bibinfo{volume}{877}},
  \bibinfo{pages}{139} (\bibinfo{year}{2019}).

\bibitem[{\citenamefont{Nandi and Char}(2018)}]{Nandi:2017rhy}
\bibinfo{author}{\bibfnamefont{R.}~\bibnamefont{Nandi}} \bibnamefont{and}
  \bibinfo{author}{\bibfnamefont{P.}~\bibnamefont{Char}},
  \bibinfo{journal}{Astrophys. J.} \textbf{\bibinfo{volume}{857}},
  \bibinfo{pages}{12} (\bibinfo{year}{2018}).

\bibitem[{\citenamefont{Paschalidis et~al.}(2018)\citenamefont{Paschalidis,
  Yagi, Alvarez-Castillo, Blaschke, and Sedrakian}}]{Paschalidis:2017qmb}
\bibinfo{author}{\bibfnamefont{V.}~\bibnamefont{Paschalidis}},
  \bibinfo{author}{\bibfnamefont{K.}~\bibnamefont{Yagi}},
  \bibinfo{author}{\bibfnamefont{D.}~\bibnamefont{Alvarez-Castillo}},
  \bibinfo{author}{\bibfnamefont{D.~B.} \bibnamefont{Blaschke}},
  \bibnamefont{and}
  \bibinfo{author}{\bibfnamefont{A.}~\bibnamefont{Sedrakian}},
  \bibinfo{journal}{Phys. Rev. D} \textbf{\bibinfo{volume}{97}},
  \bibinfo{pages}{084038} (\bibinfo{year}{2018}).

\bibitem[{\citenamefont{Alvarez-Castillo
  et~al.}(2019)\citenamefont{Alvarez-Castillo, Blaschke, Grunfeld, and
  Pagura}}]{Alvarez-Castillo:2018pve}
\bibinfo{author}{\bibfnamefont{D.~E.} \bibnamefont{Alvarez-Castillo}},
  \bibinfo{author}{\bibfnamefont{D.~B.} \bibnamefont{Blaschke}},
  \bibinfo{author}{\bibfnamefont{A.~G.} \bibnamefont{Grunfeld}},
  \bibnamefont{and} \bibinfo{author}{\bibfnamefont{V.~P.}
  \bibnamefont{Pagura}}, \bibinfo{journal}{Phys. Rev. D}
  \textbf{\bibinfo{volume}{99}}, \bibinfo{pages}{063010}
  (\bibinfo{year}{2019}).

\bibitem[{\citenamefont{Wei et~al.}(2018)\citenamefont{Wei, Irving, Kl{\"a}hn,
  and Jaikumar}}]{Wei:2018mxy}
\bibinfo{author}{\bibfnamefont{W.}~\bibnamefont{Wei}},
  \bibinfo{author}{\bibfnamefont{B.}~\bibnamefont{Irving}},
  \bibinfo{author}{\bibfnamefont{T.}~\bibnamefont{Kl{\"a}hn}},
  \bibnamefont{and} \bibinfo{author}{\bibfnamefont{P.}~\bibnamefont{Jaikumar}},
  \bibinfo{journal}{arXiv e-prints}  (\bibinfo{year}{2018}),
  \eprint{1811.09441}.

\bibitem[{\citenamefont{Alford et~al.}(2001)\citenamefont{Alford, Rajagopal,
  Reddy, and Wilczek}}]{Alford:2001zr}
\bibinfo{author}{\bibfnamefont{M.~G.} \bibnamefont{Alford}},
  \bibinfo{author}{\bibfnamefont{K.}~\bibnamefont{Rajagopal}},
  \bibinfo{author}{\bibfnamefont{S.}~\bibnamefont{Reddy}}, \bibnamefont{and}
  \bibinfo{author}{\bibfnamefont{F.}~\bibnamefont{Wilczek}},
  \bibinfo{journal}{Phys. Rev. D} \textbf{\bibinfo{volume}{64}},
  \bibinfo{pages}{074017} (\bibinfo{year}{2001}).

\bibitem[{\citenamefont{Mintz et~al.}(2010)\citenamefont{Mintz, Fraga,
  Pagliara, and Schaffner-Bielich}}]{Mintz:2009ay}
\bibinfo{author}{\bibfnamefont{B.~W.} \bibnamefont{Mintz}},
  \bibinfo{author}{\bibfnamefont{E.~S.} \bibnamefont{Fraga}},
  \bibinfo{author}{\bibfnamefont{G.}~\bibnamefont{Pagliara}}, \bibnamefont{and}
  \bibinfo{author}{\bibfnamefont{J.}~\bibnamefont{Schaffner-Bielich}},
  \bibinfo{journal}{Phys. Rev. D} \textbf{\bibinfo{volume}{81}},
  \bibinfo{pages}{123012} (\bibinfo{year}{2010}).

\bibitem[{\citenamefont{Lugones et~al.}(2013)\citenamefont{Lugones, Grunfeld,
  and Ajmi}}]{Lugones:2013ema}
\bibinfo{author}{\bibfnamefont{G.}~\bibnamefont{Lugones}},
  \bibinfo{author}{\bibfnamefont{A.~G.} \bibnamefont{Grunfeld}},
  \bibnamefont{and} \bibinfo{author}{\bibfnamefont{M.~A.} \bibnamefont{Ajmi}},
  \bibinfo{journal}{Phys. Rev. C} \textbf{\bibinfo{volume}{88}},
  \bibinfo{pages}{045803} (\bibinfo{year}{2013}).

\bibitem[{\citenamefont{Fraga et~al.}(2019)\citenamefont{Fraga, Hippert, and
  Schmitt}}]{Fraga:2018cvr}
\bibinfo{author}{\bibfnamefont{E.~S.} \bibnamefont{Fraga}},
  \bibinfo{author}{\bibfnamefont{M.}~\bibnamefont{Hippert}}, \bibnamefont{and}
  \bibinfo{author}{\bibfnamefont{A.}~\bibnamefont{Schmitt}},
  \bibinfo{journal}{Phys. Rev. D} \textbf{\bibinfo{volume}{99}},
  \bibinfo{pages}{014046} (\bibinfo{year}{2019}).

\bibitem[{\citenamefont{Glendenning}(1992)}]{Glendenning:1992vb}
\bibinfo{author}{\bibfnamefont{N.~K.} \bibnamefont{Glendenning}},
  \bibinfo{journal}{Phys. Rev. D} \textbf{\bibinfo{volume}{46}},
  \bibinfo{pages}{1274} (\bibinfo{year}{1992}).

\bibitem[{\citenamefont{Glendenning}(2001)}]{Glendenning:2001pe}
\bibinfo{author}{\bibfnamefont{N.~K.} \bibnamefont{Glendenning}},
  \bibinfo{journal}{Phys. Rept.} \textbf{\bibinfo{volume}{342}},
  \bibinfo{pages}{393} (\bibinfo{year}{2001}).

\bibitem[{\citenamefont{Baym et~al.}(2018)\citenamefont{Baym, Hatsuda, Kojo,
  Powell, Song, and Takatsuka}}]{Baym:2017whm}
\bibinfo{author}{\bibfnamefont{G.}~\bibnamefont{Baym}},
  \bibinfo{author}{\bibfnamefont{T.}~\bibnamefont{Hatsuda}},
  \bibinfo{author}{\bibfnamefont{T.}~\bibnamefont{Kojo}},
  \bibinfo{author}{\bibfnamefont{P.~D.} \bibnamefont{Powell}},
  \bibinfo{author}{\bibfnamefont{Y.}~\bibnamefont{Song}}, \bibnamefont{and}
  \bibinfo{author}{\bibfnamefont{T.}~\bibnamefont{Takatsuka}},
  \bibinfo{journal}{Rept. Prog. Phys.} \textbf{\bibinfo{volume}{81}},
  \bibinfo{pages}{056902} (\bibinfo{year}{2018}).

\bibitem[{\citenamefont{Masuda et~al.}(2013)\citenamefont{Masuda, Hatsuda, and
  Takatsuka}}]{Masuda:2012ed}
\bibinfo{author}{\bibfnamefont{K.}~\bibnamefont{Masuda}},
  \bibinfo{author}{\bibfnamefont{T.}~\bibnamefont{Hatsuda}}, \bibnamefont{and}
  \bibinfo{author}{\bibfnamefont{T.}~\bibnamefont{Takatsuka}},
  \bibinfo{journal}{PTEP} \textbf{\bibinfo{volume}{2013}},
  \bibinfo{pages}{073D01} (\bibinfo{year}{2013}).

\bibitem[{\citenamefont{Fukushima and Kojo}(2016)}]{Fukushima:2015bda}
\bibinfo{author}{\bibfnamefont{K.}~\bibnamefont{Fukushima}} \bibnamefont{and}
  \bibinfo{author}{\bibfnamefont{T.}~\bibnamefont{Kojo}},
  \bibinfo{journal}{Astrophys. J.} \textbf{\bibinfo{volume}{817}},
  \bibinfo{pages}{180} (\bibinfo{year}{2016}).

\bibitem[{\citenamefont{Kojo et~al.}(2015)\citenamefont{Kojo, Powell, Song, and
  Baym}}]{Kojo:2014rca}
\bibinfo{author}{\bibfnamefont{T.}~\bibnamefont{Kojo}},
  \bibinfo{author}{\bibfnamefont{P.~D.} \bibnamefont{Powell}},
  \bibinfo{author}{\bibfnamefont{Y.}~\bibnamefont{Song}}, \bibnamefont{and}
  \bibinfo{author}{\bibfnamefont{G.}~\bibnamefont{Baym}},
  \bibinfo{journal}{Phys. Rev. D} \textbf{\bibinfo{volume}{91}},
  \bibinfo{pages}{045003} (\bibinfo{year}{2015}).

\bibitem[{\citenamefont{Dexheimer et~al.}(2015)\citenamefont{Dexheimer,
  Negreiros, and Schramm}}]{Dexheimer:2014pea}
\bibinfo{author}{\bibfnamefont{V.}~\bibnamefont{Dexheimer}},
  \bibinfo{author}{\bibfnamefont{R.}~\bibnamefont{Negreiros}},
  \bibnamefont{and} \bibinfo{author}{\bibfnamefont{S.}~\bibnamefont{Schramm}},
  \bibinfo{journal}{Phys. Rev. C} \textbf{\bibinfo{volume}{91}},
  \bibinfo{pages}{055808} (\bibinfo{year}{2015}).

\bibitem[{\citenamefont{McLerran and Reddy}(2019)}]{McLerran:2018hbz}
\bibinfo{author}{\bibfnamefont{L.}~\bibnamefont{McLerran}} \bibnamefont{and}
  \bibinfo{author}{\bibfnamefont{S.}~\bibnamefont{Reddy}},
  \bibinfo{journal}{Phys. Rev. Lett.} \textbf{\bibinfo{volume}{122}},
  \bibinfo{pages}{122701} (\bibinfo{year}{2019}).

\bibitem[{\citenamefont{Prakash et~al.}(1997)\citenamefont{Prakash, Bombaci,
  Prakash, Ellis, Lattimer, and Knorren}}]{Prakash:1996xs}
\bibinfo{author}{\bibfnamefont{M.}~\bibnamefont{Prakash}},
  \bibinfo{author}{\bibfnamefont{I.}~\bibnamefont{Bombaci}},
  \bibinfo{author}{\bibfnamefont{M.}~\bibnamefont{Prakash}},
  \bibinfo{author}{\bibfnamefont{P.~J.} \bibnamefont{Ellis}},
  \bibinfo{author}{\bibfnamefont{J.~M.} \bibnamefont{Lattimer}},
  \bibnamefont{and} \bibinfo{author}{\bibfnamefont{R.}~\bibnamefont{Knorren}},
  \bibinfo{journal}{Phys. Rept.} \textbf{\bibinfo{volume}{280}},
  \bibinfo{pages}{1} (\bibinfo{year}{1997}).

\bibitem[{\citenamefont{Alford et~al.}(2005)\citenamefont{Alford, Braby, Paris,
  and Reddy}}]{Alford:2004pf}
\bibinfo{author}{\bibfnamefont{M.}~\bibnamefont{Alford}},
  \bibinfo{author}{\bibfnamefont{M.}~\bibnamefont{Braby}},
  \bibinfo{author}{\bibfnamefont{M.}~\bibnamefont{Paris}}, \bibnamefont{and}
  \bibinfo{author}{\bibfnamefont{S.}~\bibnamefont{Reddy}},
  \bibinfo{journal}{Astrophys. J.} \textbf{\bibinfo{volume}{629}},
  \bibinfo{pages}{969} (\bibinfo{year}{2005}).

\bibitem[{\citenamefont{Demorest et~al.}(2010)\citenamefont{Demorest, Pennucci,
  Ransom, Roberts, and Hessels}}]{Demorest:2010bx}
\bibinfo{author}{\bibfnamefont{P.}~\bibnamefont{Demorest}},
  \bibinfo{author}{\bibfnamefont{T.}~\bibnamefont{Pennucci}},
  \bibinfo{author}{\bibfnamefont{S.}~\bibnamefont{Ransom}},
  \bibinfo{author}{\bibfnamefont{M.}~\bibnamefont{Roberts}}, \bibnamefont{and}
  \bibinfo{author}{\bibfnamefont{J.}~\bibnamefont{Hessels}},
  \bibinfo{journal}{Nature} \textbf{\bibinfo{volume}{467}},
  \bibinfo{pages}{1081} (\bibinfo{year}{2010}).

\bibitem[{\citenamefont{Antoniadis et~al.}(2013)\citenamefont{Antoniadis,
  Freire, Wex, Tauris, Lynch et~al.}}]{Antoniadis:2013pzd}
\bibinfo{author}{\bibfnamefont{J.}~\bibnamefont{Antoniadis}},
  \bibinfo{author}{\bibfnamefont{P.~C.} \bibnamefont{Freire}},
  \bibinfo{author}{\bibfnamefont{N.}~\bibnamefont{Wex}},
  \bibinfo{author}{\bibfnamefont{T.~M.} \bibnamefont{Tauris}},
  \bibinfo{author}{\bibfnamefont{R.~S.} \bibnamefont{Lynch}},
  \bibnamefont{et~al.}, \bibinfo{journal}{Science}
  \textbf{\bibinfo{volume}{340}}, \bibinfo{pages}{6131} (\bibinfo{year}{2013}).

\bibitem[{\citenamefont{Fonseca et~al.}(2016)}]{Fonseca:2016tux}
\bibinfo{author}{\bibfnamefont{E.}~\bibnamefont{Fonseca}} \bibnamefont{et~al.},
  \bibinfo{journal}{Astrophys. J.} \textbf{\bibinfo{volume}{832}},
  \bibinfo{pages}{167} (\bibinfo{year}{2016}).

\bibitem[{\citenamefont{Cromartie et~al.}(2019)}]{Cromartie:2019kug}
\bibinfo{author}{\bibfnamefont{H.~T.} \bibnamefont{Cromartie}}
  \bibnamefont{et~al.}, \bibinfo{journal}{arXiv e-prints}
  (\bibinfo{year}{2019}), \eprint{1904.06759}.

\bibitem[{\citenamefont{Abbott et~al.}(2017{\natexlab{a}})}]{LIGO:2017qsa}
\bibinfo{author}{\bibfnamefont{B.~P.} \bibnamefont{Abbott}}
  \bibnamefont{et~al.} (\bibinfo{collaboration}{Virgo, LIGO Scientific}),
  \bibinfo{journal}{Phys. Rev. Lett.} \textbf{\bibinfo{volume}{119}},
  \bibinfo{pages}{161101} (\bibinfo{year}{2017}{\natexlab{a}}).

\bibitem[{\citenamefont{Abbott et~al.}(2018)}]{LIGO:2018exr}
\bibinfo{author}{\bibfnamefont{B.~P.} \bibnamefont{Abbott}}
  \bibnamefont{et~al.} (\bibinfo{collaboration}{Virgo, LIGO Scientific}),
  \bibinfo{journal}{Phys. Rev. Lett.} \textbf{\bibinfo{volume}{121}},
  \bibinfo{pages}{161101} (\bibinfo{year}{2018}).

\bibitem[{\citenamefont{De et~al.}(2018)\citenamefont{De, Finstad, Lattimer,
  Brown, Berger, and Biwer}}]{De:2018uhw}
\bibinfo{author}{\bibfnamefont{S.}~\bibnamefont{De}},
  \bibinfo{author}{\bibfnamefont{D.}~\bibnamefont{Finstad}},
  \bibinfo{author}{\bibfnamefont{J.~M.} \bibnamefont{Lattimer}},
  \bibinfo{author}{\bibfnamefont{D.~A.} \bibnamefont{Brown}},
  \bibinfo{author}{\bibfnamefont{E.}~\bibnamefont{Berger}}, \bibnamefont{and}
  \bibinfo{author}{\bibfnamefont{C.~M.} \bibnamefont{Biwer}},
  \bibinfo{journal}{Phys. Rev. Lett.} \textbf{\bibinfo{volume}{121}},
  \bibinfo{pages}{091102} (\bibinfo{year}{2018}).

\bibitem[{\citenamefont{Abbott et~al.}(2019)}]{LIGO:2018wiz}
\bibinfo{author}{\bibfnamefont{B.~P.} \bibnamefont{Abbott}}
  \bibnamefont{et~al.} (\bibinfo{collaboration}{LIGO Scientific, Virgo}),
  \bibinfo{journal}{Phys. Rev. X} \textbf{\bibinfo{volume}{9}},
  \bibinfo{pages}{011001} (\bibinfo{year}{2019}).

\bibitem[{\citenamefont{{\"O}zel and Freire}(2016)}]{Ozel:2016oaf}
\bibinfo{author}{\bibfnamefont{F.}~\bibnamefont{{\"O}zel}} \bibnamefont{and}
  \bibinfo{author}{\bibfnamefont{P.}~\bibnamefont{Freire}},
  \bibinfo{journal}{Ann. Rev. Astron. Astrophys.}
  \textbf{\bibinfo{volume}{54}}, \bibinfo{pages}{401} (\bibinfo{year}{2016}).

\bibitem[{\citenamefont{Steiner et~al.}(2000)\citenamefont{Steiner, Prakash,
  and Lattimer}}]{Steiner:2000bi}
\bibinfo{author}{\bibfnamefont{A.}~\bibnamefont{Steiner}},
  \bibinfo{author}{\bibfnamefont{M.}~\bibnamefont{Prakash}}, \bibnamefont{and}
  \bibinfo{author}{\bibfnamefont{J.~M.} \bibnamefont{Lattimer}},
  \bibinfo{journal}{Phys. Lett. B} \textbf{\bibinfo{volume}{486}},
  \bibinfo{pages}{239} (\bibinfo{year}{2000}).

\bibitem[{\citenamefont{Hanauske et~al.}(2001)\citenamefont{Hanauske, Satarov,
  Mishustin, Stocker, and Greiner}}]{Hanauske:2001nc}
\bibinfo{author}{\bibfnamefont{M.}~\bibnamefont{Hanauske}},
  \bibinfo{author}{\bibfnamefont{L.~M.} \bibnamefont{Satarov}},
  \bibinfo{author}{\bibfnamefont{I.~N.} \bibnamefont{Mishustin}},
  \bibinfo{author}{\bibfnamefont{H.}~\bibnamefont{Stocker}}, \bibnamefont{and}
  \bibinfo{author}{\bibfnamefont{W.}~\bibnamefont{Greiner}},
  \bibinfo{journal}{Phys. Rev. D} \textbf{\bibinfo{volume}{64}},
  \bibinfo{pages}{043005} (\bibinfo{year}{2001}).

\bibitem[{\citenamefont{Bhattacharyya et~al.}(2010)\citenamefont{Bhattacharyya,
  Mishustin, and Greiner}}]{Bhattacharyya:2009fg}
\bibinfo{author}{\bibfnamefont{A.}~\bibnamefont{Bhattacharyya}},
  \bibinfo{author}{\bibfnamefont{I.~N.} \bibnamefont{Mishustin}},
  \bibnamefont{and} \bibinfo{author}{\bibfnamefont{W.}~\bibnamefont{Greiner}},
  \bibinfo{journal}{J. Phys. G} \textbf{\bibinfo{volume}{37}},
  \bibinfo{pages}{025201} (\bibinfo{year}{2010}).

\bibitem[{\citenamefont{Magalhaes et~al.}(2012)\citenamefont{Magalhaes,
  Miranda, and Frajuca}}]{Magalhaes:2012za}
\bibinfo{author}{\bibfnamefont{N.~S.} \bibnamefont{Magalhaes}},
  \bibinfo{author}{\bibfnamefont{T.~A.} \bibnamefont{Miranda}},
  \bibnamefont{and} \bibinfo{author}{\bibfnamefont{C.}~\bibnamefont{Frajuca}},
  \bibinfo{journal}{Astrophys. J.} \textbf{\bibinfo{volume}{755}},
  \bibinfo{pages}{54} (\bibinfo{year}{2012}).

\bibitem[{\citenamefont{Hamil et~al.}(2015)\citenamefont{Hamil, Stone, Urbanec,
  and Urbancov\'a}}]{Hamil:2015hqa}
\bibinfo{author}{\bibfnamefont{O.}~\bibnamefont{Hamil}},
  \bibinfo{author}{\bibfnamefont{J.~R.} \bibnamefont{Stone}},
  \bibinfo{author}{\bibfnamefont{M.}~\bibnamefont{Urbanec}}, \bibnamefont{and}
  \bibinfo{author}{\bibfnamefont{G.}~\bibnamefont{Urbancov\'a}},
  \bibinfo{journal}{Phys. Rev. D} \textbf{\bibinfo{volume}{91}},
  \bibinfo{pages}{063007} (\bibinfo{year}{2015}).

\bibitem[{\citenamefont{Johnston and Karastergiou}(2017)}]{Johnston:2017wgm}
\bibinfo{author}{\bibfnamefont{S.}~\bibnamefont{Johnston}} \bibnamefont{and}
  \bibinfo{author}{\bibfnamefont{A.}~\bibnamefont{Karastergiou}},
  \bibinfo{journal}{Mon. Not. Roy. Astron. Soc.}
  \textbf{\bibinfo{volume}{467}}, \bibinfo{pages}{3493} (\bibinfo{year}{2017}).

\bibitem[{\citenamefont{Akmal et~al.}(1998)\citenamefont{Akmal, Pandharipande,
  and Ravenhall}}]{APR98}
\bibinfo{author}{\bibfnamefont{A.}~\bibnamefont{Akmal}},
  \bibinfo{author}{\bibfnamefont{V.~R.} \bibnamefont{Pandharipande}},
  \bibnamefont{and} \bibinfo{author}{\bibfnamefont{D.~G.}
  \bibnamefont{Ravenhall}}, \bibinfo{journal}{Phys. Rev. C}
  \textbf{\bibinfo{volume}{58}}, \bibinfo{pages}{1804} (\bibinfo{year}{1998}).

\bibitem[{\citenamefont{Akmal and Pandharipande}(1997)}]{Akmal:1997ft}
\bibinfo{author}{\bibfnamefont{A.}~\bibnamefont{Akmal}} \bibnamefont{and}
  \bibinfo{author}{\bibfnamefont{V.~R.} \bibnamefont{Pandharipande}},
  \bibinfo{journal}{Phys. Rev. C} \textbf{\bibinfo{volume}{56}},
  \bibinfo{pages}{2261} (\bibinfo{year}{1997}).

\bibitem[{\citenamefont{Constantinou et~al.}(2014)\citenamefont{Constantinou,
  Muccioli, Prakash, and Lattimer}}]{Constantinou:2014hha}
\bibinfo{author}{\bibfnamefont{C.}~\bibnamefont{Constantinou}},
  \bibinfo{author}{\bibfnamefont{B.}~\bibnamefont{Muccioli}},
  \bibinfo{author}{\bibfnamefont{M.}~\bibnamefont{Prakash}}, \bibnamefont{and}
  \bibinfo{author}{\bibfnamefont{J.~M.} \bibnamefont{Lattimer}},
  \bibinfo{journal}{Phys. Rev. C} \textbf{\bibinfo{volume}{89}},
  \bibinfo{pages}{065802} (\bibinfo{year}{2014}).

\bibitem[{\citenamefont{Steiner et~al.}(2005)\citenamefont{Steiner, Prakash,
  Lattimer, and Ellis}}]{Steiner:2004fi}
\bibinfo{author}{\bibfnamefont{A.~W.} \bibnamefont{Steiner}},
  \bibinfo{author}{\bibfnamefont{M.}~\bibnamefont{Prakash}},
  \bibinfo{author}{\bibfnamefont{J.~M.} \bibnamefont{Lattimer}},
  \bibnamefont{and} \bibinfo{author}{\bibfnamefont{P.~J.} \bibnamefont{Ellis}},
  \bibinfo{journal}{Phys. Rept.} \textbf{\bibinfo{volume}{411}},
  \bibinfo{pages}{325} (\bibinfo{year}{2005}).

\bibitem[{\citenamefont{Schneider et~al.}(2019)\citenamefont{Schneider,
  Constantinou, Muccioli, and Prakash}}]{Schneider:2019vdm}
\bibinfo{author}{\bibfnamefont{A.~S.} \bibnamefont{Schneider}},
  \bibinfo{author}{\bibfnamefont{C.}~\bibnamefont{Constantinou}},
  \bibinfo{author}{\bibfnamefont{B.}~\bibnamefont{Muccioli}}, \bibnamefont{and}
  \bibinfo{author}{\bibfnamefont{M.}~\bibnamefont{Prakash}},
  \bibinfo{journal}{Phys. Rev. C} \textbf{\bibinfo{volume}{100}},
  \bibinfo{pages}{025803} (\bibinfo{year}{2019}).

\bibitem[{\citenamefont{Mueller and Serot}(1996)}]{Mueller:1996pm}
\bibinfo{author}{\bibfnamefont{H.}~\bibnamefont{Mueller}} \bibnamefont{and}
  \bibinfo{author}{\bibfnamefont{B.~D.} \bibnamefont{Serot}},
  \bibinfo{journal}{Nucl. Phys. A} \textbf{\bibinfo{volume}{606}},
  \bibinfo{pages}{508} (\bibinfo{year}{1996}).

\bibitem[{\citenamefont{Horowitz and
  Piekarewicz}(2001{\natexlab{a}})}]{Horowitz:2001ya}
\bibinfo{author}{\bibfnamefont{C.~J.} \bibnamefont{Horowitz}} \bibnamefont{and}
  \bibinfo{author}{\bibfnamefont{J.}~\bibnamefont{Piekarewicz}},
  \bibinfo{journal}{Phys. Rev. C} \textbf{\bibinfo{volume}{64}},
  \bibinfo{pages}{062802(R)} (\bibinfo{year}{2001}{\natexlab{a}}).

\bibitem[{\citenamefont{Horowitz and
  Piekarewicz}(2001{\natexlab{b}})}]{Horowitz:2000xj}
\bibinfo{author}{\bibfnamefont{C.~J.} \bibnamefont{Horowitz}} \bibnamefont{and}
  \bibinfo{author}{\bibfnamefont{J.}~\bibnamefont{Piekarewicz}},
  \bibinfo{journal}{Phys. Rev. Lett.} \textbf{\bibinfo{volume}{86}},
  \bibinfo{pages}{5647} (\bibinfo{year}{2001}{\natexlab{b}}).

\bibitem[{\citenamefont{Chin}(1977)}]{Chin:1977iz}
\bibinfo{author}{\bibfnamefont{S.~A.} \bibnamefont{Chin}},
  \bibinfo{journal}{Annals Phys.} \textbf{\bibinfo{volume}{108}},
  \bibinfo{pages}{301} (\bibinfo{year}{1977}).

\bibitem[{\citenamefont{Zhang and Prakash}(2016)}]{Zhang:2016bem}
\bibinfo{author}{\bibfnamefont{X.}~\bibnamefont{Zhang}} \bibnamefont{and}
  \bibinfo{author}{\bibfnamefont{M.}~\bibnamefont{Prakash}},
  \bibinfo{journal}{Phys. Rev. C} \textbf{\bibinfo{volume}{93}},
  \bibinfo{pages}{055805} (\bibinfo{year}{2016}).

\bibitem[{\citenamefont{Constantinou et~al.}(2017)\citenamefont{Constantinou,
  Lalit, and Prakash}}]{Constantinou:2016hvf}
\bibinfo{author}{\bibfnamefont{C.}~\bibnamefont{Constantinou}},
  \bibinfo{author}{\bibfnamefont{S.}~\bibnamefont{Lalit}}, \bibnamefont{and}
  \bibinfo{author}{\bibfnamefont{M.}~\bibnamefont{Prakash}},
  \bibinfo{journal}{Int. J. Mod. Phys. E} \textbf{\bibinfo{volume}{26}},
  \bibinfo{pages}{1740005} (\bibinfo{year}{2017}).

\bibitem[{\citenamefont{Constantinou et~al.}(2015)\citenamefont{Constantinou,
  Muccioli, Prakash, and Lattimer}}]{Constantinou:2015mna}
\bibinfo{author}{\bibfnamefont{C.}~\bibnamefont{Constantinou}},
  \bibinfo{author}{\bibfnamefont{B.}~\bibnamefont{Muccioli}},
  \bibinfo{author}{\bibfnamefont{M.}~\bibnamefont{Prakash}}, \bibnamefont{and}
  \bibinfo{author}{\bibfnamefont{J.~M.} \bibnamefont{Lattimer}},
  \bibinfo{journal}{Phys. Rev. C} \textbf{\bibinfo{volume}{92}},
  \bibinfo{pages}{025801} (\bibinfo{year}{2015}).

\bibitem[{\citenamefont{Oertel et~al.}(2017)\citenamefont{Oertel, Hempel,
  Kl{\"a}hn, and Typel}}]{Oertel:2016bki}
\bibinfo{author}{\bibfnamefont{M.}~\bibnamefont{Oertel}},
  \bibinfo{author}{\bibfnamefont{M.}~\bibnamefont{Hempel}},
  \bibinfo{author}{\bibfnamefont{T.}~\bibnamefont{Kl{\"a}hn}},
  \bibnamefont{and} \bibinfo{author}{\bibfnamefont{S.}~\bibnamefont{Typel}},
  \bibinfo{journal}{Rev. Mod. Phys.} \textbf{\bibinfo{volume}{89}},
  \bibinfo{pages}{015007} (\bibinfo{year}{2017}).

\bibitem[{\citenamefont{Chen and Piekarewicz}(2014)}]{Chen:2014sca}
\bibinfo{author}{\bibfnamefont{W.-C.} \bibnamefont{Chen}} \bibnamefont{and}
  \bibinfo{author}{\bibfnamefont{J.}~\bibnamefont{Piekarewicz}},
  \bibinfo{journal}{Phys. Rev. C} \textbf{\bibinfo{volume}{90}},
  \bibinfo{pages}{044305} (\bibinfo{year}{2014}).

\bibitem[{\citenamefont{Fattoyev et~al.}(2018)\citenamefont{Fattoyev,
  Piekarewicz, and Horowitz}}]{Fattoyev:2017jql}
\bibinfo{author}{\bibfnamefont{F.~J.} \bibnamefont{Fattoyev}},
  \bibinfo{author}{\bibfnamefont{J.}~\bibnamefont{Piekarewicz}},
  \bibnamefont{and} \bibinfo{author}{\bibfnamefont{C.~J.}
  \bibnamefont{Horowitz}}, \bibinfo{journal}{Phys. Rev. Lett.}
  \textbf{\bibinfo{volume}{120}}, \bibinfo{pages}{172702}
  (\bibinfo{year}{2018}).

\bibitem[{\citenamefont{Lattimer et~al.}(1991)\citenamefont{Lattimer, Pethick,
  Prakash, and Haensel}}]{Lattimer:1991ib}
\bibinfo{author}{\bibfnamefont{J.~M.} \bibnamefont{Lattimer}},
  \bibinfo{author}{\bibfnamefont{C.~J.} \bibnamefont{Pethick}},
  \bibinfo{author}{\bibfnamefont{M.}~\bibnamefont{Prakash}}, \bibnamefont{and}
  \bibinfo{author}{\bibfnamefont{P.}~\bibnamefont{Haensel}},
  \bibinfo{journal}{Phys. Rev. Lett.} \textbf{\bibinfo{volume}{66}},
  \bibinfo{pages}{2701} (\bibinfo{year}{1991}).

\bibitem[{\citenamefont{Hornick et~al.}(2018)\citenamefont{Hornick, Tolos,
  Zacchi, Christian, and Schaffner-Bielich}}]{Hornick:2018kfi}
\bibinfo{author}{\bibfnamefont{N.}~\bibnamefont{Hornick}},
  \bibinfo{author}{\bibfnamefont{L.}~\bibnamefont{Tolos}},
  \bibinfo{author}{\bibfnamefont{A.}~\bibnamefont{Zacchi}},
  \bibinfo{author}{\bibfnamefont{J.-E.} \bibnamefont{Christian}},
  \bibnamefont{and}
  \bibinfo{author}{\bibfnamefont{J.}~\bibnamefont{Schaffner-Bielich}},
  \bibinfo{journal}{Phys. Rev. C} \textbf{\bibinfo{volume}{98}},
  \bibinfo{pages}{065804} (\bibinfo{year}{2018}).

\bibitem[{\citenamefont{Prakash et~al.}(1990)\citenamefont{Prakash, Baron, and
  Prakash}}]{Prakash:1990at}
\bibinfo{author}{\bibfnamefont{M.}~\bibnamefont{Prakash}},
  \bibinfo{author}{\bibfnamefont{E.}~\bibnamefont{Baron}}, \bibnamefont{and}
  \bibinfo{author}{\bibfnamefont{M.}~\bibnamefont{Prakash}},
  \bibinfo{journal}{Phys. Lett. B} \textbf{\bibinfo{volume}{243}},
  \bibinfo{pages}{175} (\bibinfo{year}{1990}).

\bibitem[{\citenamefont{Xia et~al.}(2019)\citenamefont{Xia, Zhu, Zhou, and
  Li}}]{Xia:2019xax}
\bibinfo{author}{\bibfnamefont{C.}~\bibnamefont{Xia}},
  \bibinfo{author}{\bibfnamefont{Z.}~\bibnamefont{Zhu}},
  \bibinfo{author}{\bibfnamefont{X.}~\bibnamefont{Zhou}}, \bibnamefont{and}
  \bibinfo{author}{\bibfnamefont{A.}~\bibnamefont{Li}}, \bibinfo{journal}{arXiv
  e-prints}  (\bibinfo{year}{2019}), \eprint{1906.00826}.

\bibitem[{\citenamefont{'t~Hooft}(1986)}]{tHooft:1986ooh}
\bibinfo{author}{\bibfnamefont{G.}~\bibnamefont{'t~Hooft}},
  \bibinfo{journal}{Phys. Rept.} \textbf{\bibinfo{volume}{142}},
  \bibinfo{pages}{357} (\bibinfo{year}{1986}).

\bibitem[{\citenamefont{Hatsuda and Kunihiro}(1994)}]{Hatsuda:1994pi}
\bibinfo{author}{\bibfnamefont{T.}~\bibnamefont{Hatsuda}} \bibnamefont{and}
  \bibinfo{author}{\bibfnamefont{T.}~\bibnamefont{Kunihiro}},
  \bibinfo{journal}{Phys. Rept.} \textbf{\bibinfo{volume}{247}},
  \bibinfo{pages}{221} (\bibinfo{year}{1994}).

\bibitem[{\citenamefont{Lamb et~al.}(1983)\citenamefont{Lamb, Lattimer,
  Pethick, and Ravenhall}}]{Lamb:1983djd}
\bibinfo{author}{\bibfnamefont{D.~Q.} \bibnamefont{Lamb}},
  \bibinfo{author}{\bibfnamefont{J.~M.} \bibnamefont{Lattimer}},
  \bibinfo{author}{\bibfnamefont{C.~J.} \bibnamefont{Pethick}},
  \bibnamefont{and} \bibinfo{author}{\bibfnamefont{D.~G.}
  \bibnamefont{Ravenhall}}, \bibinfo{journal}{Nucl. Phys. A}
  \textbf{\bibinfo{volume}{411}}, \bibinfo{pages}{449} (\bibinfo{year}{1983}).

\bibitem[{\citenamefont{Li et~al.}(2018)\citenamefont{Li, Yan, Geng, Huang, and
  Zong}}]{Li:2018ayl}
\bibinfo{author}{\bibfnamefont{C.-M.} \bibnamefont{Li}},
  \bibinfo{author}{\bibfnamefont{Y.}~\bibnamefont{Yan}},
  \bibinfo{author}{\bibfnamefont{J.-J.} \bibnamefont{Geng}},
  \bibinfo{author}{\bibfnamefont{Y.-F.} \bibnamefont{Huang}}, \bibnamefont{and}
  \bibinfo{author}{\bibfnamefont{H.-S.} \bibnamefont{Zong}},
  \bibinfo{journal}{Phys. Rev. D} \textbf{\bibinfo{volume}{98}},
  \bibinfo{pages}{083013} (\bibinfo{year}{2018}).

\bibitem[{\citenamefont{Dexheimer and Schramm}(2010)}]{Dexheimer:2009hi}
\bibinfo{author}{\bibfnamefont{V.~A.} \bibnamefont{Dexheimer}}
  \bibnamefont{and} \bibinfo{author}{\bibfnamefont{S.}~\bibnamefont{Schramm}},
  \bibinfo{journal}{Phys. Rev. C} \textbf{\bibinfo{volume}{81}},
  \bibinfo{pages}{045201} (\bibinfo{year}{2010}).

\bibitem[{\citenamefont{Negreiros et~al.}(2010)\citenamefont{Negreiros,
  Dexheimer, and Schramm}}]{Negreiros:2010hk}
\bibinfo{author}{\bibfnamefont{R.}~\bibnamefont{Negreiros}},
  \bibinfo{author}{\bibfnamefont{V.~A.} \bibnamefont{Dexheimer}},
  \bibnamefont{and} \bibinfo{author}{\bibfnamefont{S.}~\bibnamefont{Schramm}},
  \bibinfo{journal}{Phys. Rev. C} \textbf{\bibinfo{volume}{82}},
  \bibinfo{pages}{035803} (\bibinfo{year}{2010}).

\bibitem[{\citenamefont{McLerran and Pisarski}(2007)}]{McLerran:2007qj}
\bibinfo{author}{\bibfnamefont{L.}~\bibnamefont{McLerran}} \bibnamefont{and}
  \bibinfo{author}{\bibfnamefont{R.~D.} \bibnamefont{Pisarski}},
  \bibinfo{journal}{Nucl. Phys. A} \textbf{\bibinfo{volume}{796}},
  \bibinfo{pages}{83} (\bibinfo{year}{2007}).

\bibitem[{\citenamefont{Flanagan and Hinderer}(2008)}]{Flanagan:2007ix}
\bibinfo{author}{\bibfnamefont{E.~E.} \bibnamefont{Flanagan}} \bibnamefont{and}
  \bibinfo{author}{\bibfnamefont{T.}~\bibnamefont{Hinderer}},
  \bibinfo{journal}{Phys. Rev. D} \textbf{\bibinfo{volume}{77}},
  \bibinfo{pages}{021502(R)} (\bibinfo{year}{2008}).

\bibitem[{\citenamefont{Favata}(2014)}]{Favata:2013rwa}
\bibinfo{author}{\bibfnamefont{M.}~\bibnamefont{Favata}},
  \bibinfo{journal}{Phys. Rev. Lett.} \textbf{\bibinfo{volume}{112}},
  \bibinfo{pages}{101101} (\bibinfo{year}{2014}).

\bibitem[{\citenamefont{Love}(1909)}]{L1909}
\bibinfo{author}{\bibfnamefont{A.~E.~H.} \bibnamefont{Love}},
  \bibinfo{journal}{Proc. R. Soc. A} \textbf{\bibinfo{volume}{82}},
  \bibinfo{pages}{73} (\bibinfo{year}{1909}).

\bibitem[{\citenamefont{Thorne and Campolattaro}(1967)}]{TC67}
\bibinfo{author}{\bibfnamefont{K.}~\bibnamefont{Thorne}} \bibnamefont{and}
  \bibinfo{author}{\bibfnamefont{A.}~\bibnamefont{Campolattaro}},
  \bibinfo{journal}{Astrophys. J.} \textbf{\bibinfo{volume}{149}},
  \bibinfo{pages}{591} (\bibinfo{year}{1967}).

\bibitem[{\citenamefont{Hinderer}(2008)}]{Hinderer:2007mb}
\bibinfo{author}{\bibfnamefont{T.}~\bibnamefont{Hinderer}},
  \bibinfo{journal}{Astrophys. J.} \textbf{\bibinfo{volume}{677}},
  \bibinfo{pages}{1216} (\bibinfo{year}{2008}).

\bibitem[{\citenamefont{Damour and Nagar}(2009)}]{Damour:2009vw}
\bibinfo{author}{\bibfnamefont{T.}~\bibnamefont{Damour}} \bibnamefont{and}
  \bibinfo{author}{\bibfnamefont{A.}~\bibnamefont{Nagar}},
  \bibinfo{journal}{Phys. Rev. D} \textbf{\bibinfo{volume}{80}},
  \bibinfo{pages}{084035} (\bibinfo{year}{2009}).

\bibitem[{\citenamefont{Hinderer et~al.}(2010)\citenamefont{Hinderer, Lackey,
  Lang, and Read}}]{Hinderer:2009ca}
\bibinfo{author}{\bibfnamefont{T.}~\bibnamefont{Hinderer}},
  \bibinfo{author}{\bibfnamefont{B.~D.} \bibnamefont{Lackey}},
  \bibinfo{author}{\bibfnamefont{R.~N.} \bibnamefont{Lang}}, \bibnamefont{and}
  \bibinfo{author}{\bibfnamefont{J.~S.} \bibnamefont{Read}},
  \bibinfo{journal}{Phys. Rev. D} \textbf{\bibinfo{volume}{81}},
  \bibinfo{pages}{123016} (\bibinfo{year}{2010}).

\bibitem[{\citenamefont{Postnikov et~al.}(2010)\citenamefont{Postnikov,
  Prakash, and Lattimer}}]{Postnikov:2010yn}
\bibinfo{author}{\bibfnamefont{S.}~\bibnamefont{Postnikov}},
  \bibinfo{author}{\bibfnamefont{M.}~\bibnamefont{Prakash}}, \bibnamefont{and}
  \bibinfo{author}{\bibfnamefont{J.~M.} \bibnamefont{Lattimer}},
  \bibinfo{journal}{Phys. Rev. D} \textbf{\bibinfo{volume}{82}},
  \bibinfo{pages}{024016} (\bibinfo{year}{2010}).

\bibitem[{\citenamefont{Abbott et~al.}(2017{\natexlab{b}})}]{GBM:2017lvd}
\bibinfo{author}{\bibfnamefont{B.~P.} \bibnamefont{Abbott}}
  \bibnamefont{et~al.}, \bibinfo{journal}{Astrophys. J.}
  \textbf{\bibinfo{volume}{848}}, \bibinfo{pages}{L12}
  (\bibinfo{year}{2017}{\natexlab{b}}).

\bibitem[{\citenamefont{Margalit and Metzger}(2017)}]{Margalit:2017dij}
\bibinfo{author}{\bibfnamefont{B.}~\bibnamefont{Margalit}} \bibnamefont{and}
  \bibinfo{author}{\bibfnamefont{B.~D.} \bibnamefont{Metzger}},
  \bibinfo{journal}{Astrophys. J.} \textbf{\bibinfo{volume}{850}},
  \bibinfo{pages}{L19} (\bibinfo{year}{2017}).

\bibitem[{\citenamefont{Shibata et~al.}(2019)\citenamefont{Shibata, Zhou,
  Kiuchi, and Fujibayashi}}]{Shibata:2019ctb}
\bibinfo{author}{\bibfnamefont{M.}~\bibnamefont{Shibata}},
  \bibinfo{author}{\bibfnamefont{E.}~\bibnamefont{Zhou}},
  \bibinfo{author}{\bibfnamefont{K.}~\bibnamefont{Kiuchi}}, \bibnamefont{and}
  \bibinfo{author}{\bibfnamefont{S.}~\bibnamefont{Fujibayashi}},
  \bibinfo{journal}{Phys. Rev. D} \textbf{\bibinfo{volume}{100}},
  \bibinfo{pages}{023015} (\bibinfo{year}{2019}).

\bibitem[{\citenamefont{Rezzolla et~al.}(2018)\citenamefont{Rezzolla, Most, and
  Weih}}]{Rezzolla:2017aly}
\bibinfo{author}{\bibfnamefont{L.}~\bibnamefont{Rezzolla}},
  \bibinfo{author}{\bibfnamefont{E.~R.} \bibnamefont{Most}}, \bibnamefont{and}
  \bibinfo{author}{\bibfnamefont{L.~R.} \bibnamefont{Weih}},
  \bibinfo{journal}{Astrophys. J.} \textbf{\bibinfo{volume}{852}},
  \bibinfo{pages}{L25} (\bibinfo{year}{2018}), \bibinfo{note}{[Astrophys. J.
  Lett.852,L25(2018)]}.

\bibitem[{\citenamefont{Alford et~al.}(2013)\citenamefont{Alford, Han, and
  Prakash}}]{Alford:2013aca}
\bibinfo{author}{\bibfnamefont{M.~G.} \bibnamefont{Alford}},
  \bibinfo{author}{\bibfnamefont{S.}~\bibnamefont{Han}}, \bibnamefont{and}
  \bibinfo{author}{\bibfnamefont{M.}~\bibnamefont{Prakash}},
  \bibinfo{journal}{Phys. Rev. D} \textbf{\bibinfo{volume}{88}},
  \bibinfo{pages}{083013} (\bibinfo{year}{2013}).

\bibitem[{\citenamefont{Burgio et~al.}(2018)\citenamefont{Burgio, Drago,
  Pagliara, Schulze, and Wei}}]{Burgio:2018yix}
\bibinfo{author}{\bibfnamefont{G.~F.} \bibnamefont{Burgio}},
  \bibinfo{author}{\bibfnamefont{A.}~\bibnamefont{Drago}},
  \bibinfo{author}{\bibfnamefont{G.}~\bibnamefont{Pagliara}},
  \bibinfo{author}{\bibfnamefont{H.~J.} \bibnamefont{Schulze}},
  \bibnamefont{and} \bibinfo{author}{\bibfnamefont{J.~B.} \bibnamefont{Wei}},
  \bibinfo{journal}{Astrophys. J.} \textbf{\bibinfo{volume}{860}},
  \bibinfo{pages}{139} (\bibinfo{year}{2018}).

\bibitem[{\citenamefont{Christian et~al.}(2019)\citenamefont{Christian, Zacchi,
  and Schaffner-Bielich}}]{Christian:2018jyd}
\bibinfo{author}{\bibfnamefont{J.-E.} \bibnamefont{Christian}},
  \bibinfo{author}{\bibfnamefont{A.}~\bibnamefont{Zacchi}}, \bibnamefont{and}
  \bibinfo{author}{\bibfnamefont{J.}~\bibnamefont{Schaffner-Bielich}},
  \bibinfo{journal}{Phys. Rev. D} \textbf{\bibinfo{volume}{99}},
  \bibinfo{pages}{023009} (\bibinfo{year}{2019}).

\bibitem[{\citenamefont{Montana et~al.}(2019)\citenamefont{Montana, Tolos,
  Hanauske, and Rezzolla}}]{Montana:2018bkb}
\bibinfo{author}{\bibfnamefont{G.}~\bibnamefont{Montana}},
  \bibinfo{author}{\bibfnamefont{L.}~\bibnamefont{Tolos}},
  \bibinfo{author}{\bibfnamefont{M.}~\bibnamefont{Hanauske}}, \bibnamefont{and}
  \bibinfo{author}{\bibfnamefont{L.}~\bibnamefont{Rezzolla}},
  \bibinfo{journal}{Phys. Rev. D} \textbf{\bibinfo{volume}{99}},
  \bibinfo{pages}{103009} (\bibinfo{year}{2019}).

\bibitem[{\citenamefont{{Seidov}}(1971)}]{Seidov:1971}
\bibinfo{author}{\bibfnamefont{Z.~F.} \bibnamefont{{Seidov}}},
  \bibinfo{journal}{Sov. Astron.} \textbf{\bibinfo{volume}{15}},
  \bibinfo{pages}{347} (\bibinfo{year}{1971}).

\bibitem[{\citenamefont{Schaeffer et~al.}(1983)\citenamefont{Schaeffer, Zdunik,
  and Haensel}}]{Schaeffer:1983}
\bibinfo{author}{\bibfnamefont{R.}~\bibnamefont{Schaeffer}},
  \bibinfo{author}{\bibfnamefont{L.}~\bibnamefont{Zdunik}}, \bibnamefont{and}
  \bibinfo{author}{\bibfnamefont{P.}~\bibnamefont{Haensel}},
  \bibinfo{journal}{Astron. Astrophys.} \textbf{\bibinfo{volume}{126}},
  \bibinfo{pages}{121} (\bibinfo{year}{1983}).

\bibitem[{\citenamefont{Lindblom}(1998)}]{Lindblom:1998dp}
\bibinfo{author}{\bibfnamefont{L.}~\bibnamefont{Lindblom}},
  \bibinfo{journal}{Phys. Rev. D} \textbf{\bibinfo{volume}{58}},
  \bibinfo{pages}{024008} (\bibinfo{year}{1998}).

\bibitem[{\citenamefont{Sieniawska et~al.}(2019)\citenamefont{Sieniawska,
  Turczanski, Bejger, and Zdunik}}]{Sieniawska:2018zzj}
\bibinfo{author}{\bibfnamefont{M.}~\bibnamefont{Sieniawska}},
  \bibinfo{author}{\bibfnamefont{W.}~\bibnamefont{Turczanski}},
  \bibinfo{author}{\bibfnamefont{M.}~\bibnamefont{Bejger}}, \bibnamefont{and}
  \bibinfo{author}{\bibfnamefont{J.~L.} \bibnamefont{Zdunik}},
  \bibinfo{journal}{Astron. Astrophys.} \textbf{\bibinfo{volume}{622}},
  \bibinfo{pages}{A174} (\bibinfo{year}{2019}).

\bibitem[{\citenamefont{Han and Steiner}(2019)}]{Han:2018mtj}
\bibinfo{author}{\bibfnamefont{S.}~\bibnamefont{Han}} \bibnamefont{and}
  \bibinfo{author}{\bibfnamefont{A.~W.} \bibnamefont{Steiner}},
  \bibinfo{journal}{Phys. Rev. D} \textbf{\bibinfo{volume}{99}},
  \bibinfo{pages}{083014} (\bibinfo{year}{2019}).

\bibitem[{\citenamefont{Most et~al.}(2019)\citenamefont{Most, Papenfort,
  Dexheimer, Hanauske, Schramm, St{\"o}cker, and Rezzolla}}]{Most:2018eaw}
\bibinfo{author}{\bibfnamefont{E.~R.} \bibnamefont{Most}},
  \bibinfo{author}{\bibfnamefont{L.~J.} \bibnamefont{Papenfort}},
  \bibinfo{author}{\bibfnamefont{V.}~\bibnamefont{Dexheimer}},
  \bibinfo{author}{\bibfnamefont{M.}~\bibnamefont{Hanauske}},
  \bibinfo{author}{\bibfnamefont{S.}~\bibnamefont{Schramm}},
  \bibinfo{author}{\bibfnamefont{H.}~\bibnamefont{St{\"o}cker}},
  \bibnamefont{and} \bibinfo{author}{\bibfnamefont{L.}~\bibnamefont{Rezzolla}},
  \bibinfo{journal}{Phys. Rev. Lett.} \textbf{\bibinfo{volume}{122}},
  \bibinfo{pages}{061101} (\bibinfo{year}{2019}).

\bibitem[{\citenamefont{Chesler et~al.}(2019)\citenamefont{Chesler, Jokela,
  Loeb, and Vuorinen}}]{Chesler:2019osn}
\bibinfo{author}{\bibfnamefont{P.~M.} \bibnamefont{Chesler}},
  \bibinfo{author}{\bibfnamefont{N.}~\bibnamefont{Jokela}},
  \bibinfo{author}{\bibfnamefont{A.}~\bibnamefont{Loeb}}, \bibnamefont{and}
  \bibinfo{author}{\bibfnamefont{A.}~\bibnamefont{Vuorinen}},
  \bibinfo{journal}{Phys. Rev. D} \textbf{\bibinfo{volume}{100}},
  \bibinfo{pages}{066027} (\bibinfo{year}{2019}).

\bibitem[{\citenamefont{Bauswein et~al.}(2019)\citenamefont{Bauswein, Bastian,
  Blaschke, Chatziioannou, Clark, Fischer, and Oertel}}]{Bauswein:2018bma}
\bibinfo{author}{\bibfnamefont{A.}~\bibnamefont{Bauswein}},
  \bibinfo{author}{\bibfnamefont{N.-U.~F.} \bibnamefont{Bastian}},
  \bibinfo{author}{\bibfnamefont{D.~B.} \bibnamefont{Blaschke}},
  \bibinfo{author}{\bibfnamefont{K.}~\bibnamefont{Chatziioannou}},
  \bibinfo{author}{\bibfnamefont{J.~A.} \bibnamefont{Clark}},
  \bibinfo{author}{\bibfnamefont{T.}~\bibnamefont{Fischer}}, \bibnamefont{and}
  \bibinfo{author}{\bibfnamefont{M.}~\bibnamefont{Oertel}},
  \bibinfo{journal}{Phys. Rev. Lett.} \textbf{\bibinfo{volume}{122}},
  \bibinfo{pages}{061102} (\bibinfo{year}{2019}).

\bibitem[{\citenamefont{Fischer et~al.}(2018)\citenamefont{Fischer, Bastian,
  Wu, Baklanov, Sorokina, Blinnikov, Typel, Kl{\"a}hn, and
  Blaschke}}]{Fischer:2017lag}
\bibinfo{author}{\bibfnamefont{T.}~\bibnamefont{Fischer}},
  \bibinfo{author}{\bibfnamefont{N.-U.~F.} \bibnamefont{Bastian}},
  \bibinfo{author}{\bibfnamefont{M.-R.} \bibnamefont{Wu}},
  \bibinfo{author}{\bibfnamefont{P.}~\bibnamefont{Baklanov}},
  \bibinfo{author}{\bibfnamefont{E.}~\bibnamefont{Sorokina}},
  \bibinfo{author}{\bibfnamefont{S.}~\bibnamefont{Blinnikov}},
  \bibinfo{author}{\bibfnamefont{S.}~\bibnamefont{Typel}},
  \bibinfo{author}{\bibfnamefont{T.}~\bibnamefont{Kl{\"a}hn}},
  \bibnamefont{and} \bibinfo{author}{\bibfnamefont{D.~B.}
  \bibnamefont{Blaschke}}, \bibinfo{journal}{Nat. Astron.}
  \textbf{\bibinfo{volume}{2}}, \bibinfo{pages}{980} (\bibinfo{year}{2018}).

\bibitem[{\citenamefont{Aloy et~al.}(2019)\citenamefont{Aloy, Ib{\'a}{\~n}ez,
  Sanchis-Gual, Obergaulinger, Font, Serna, and Marquina}}]{Ibanez:2018myp}
\bibinfo{author}{\bibfnamefont{M.~A.} \bibnamefont{Aloy}},
  \bibinfo{author}{\bibfnamefont{J.~M.} \bibnamefont{Ib{\'a}{\~n}ez}},
  \bibinfo{author}{\bibfnamefont{N.}~\bibnamefont{Sanchis-Gual}},
  \bibinfo{author}{\bibfnamefont{M.}~\bibnamefont{Obergaulinger}},
  \bibinfo{author}{\bibfnamefont{J.~A.} \bibnamefont{Font}},
  \bibinfo{author}{\bibfnamefont{S.}~\bibnamefont{Serna}}, \bibnamefont{and}
  \bibinfo{author}{\bibfnamefont{A.}~\bibnamefont{Marquina}},
  \bibinfo{journal}{Mon. Not. Roy. Astron. Soc.}
  \textbf{\bibinfo{volume}{484}}, \bibinfo{pages}{4980} (\bibinfo{year}{2019}).

\bibitem[{\citenamefont{Alford et~al.}(2019)\citenamefont{Alford, Han, and
  Schwenzer}}]{Alford:2019oge}
\bibinfo{author}{\bibfnamefont{M.~G.} \bibnamefont{Alford}},
  \bibinfo{author}{\bibfnamefont{S.}~\bibnamefont{Han}}, \bibnamefont{and}
  \bibinfo{author}{\bibfnamefont{K.}~\bibnamefont{Schwenzer}},
  \bibinfo{journal}{J. Phys. G} \textbf{\bibinfo{volume}{46}},
  \bibinfo{pages}{114001} (\bibinfo{year}{2019}).

\end{thebibliography}

\appendix
\section{Thermodynamics of nucleons in the shell of quarkyonic matter}
\label{sec:TI}
In this Appendix, we provide some details of the evaluation of the kinetic parts of the energy density, chemical potential and energy density and pressure for nucleons in the shell. The expressions we obtain will then be used to establishing the thermodynamic identity (TI) in the presence of a shell. For the evaluation of these quantities the relation 
\ba
\frac {d}{d\alpha} \int_{\phi_1(\alpha)}^{\phi_2(\alpha)} F(x,\alpha) \, dx &=&  
\int_{\phi_1(\alpha)}^{\phi_2(\alpha)} \frac {\partial F(x,\alpha)}{\partial \alpha} \, dx \nonumber \\
&+& F(\phi_2,\alpha)  \frac {\partial \phi_2}{\partial \alpha}
-  F(\phi_1,\alpha)  \frac {\partial \phi_1}{\partial \alpha} \,, \nonumber \\
\label{eqn:Leibnitz}
\ea
where $\alpha$ is a parameter in the functions $\phi_1,\phi_2$ and $F$ will be useful. 

\subsection*{Energy density}

The kinetic energy density of nucleons, neutrons to be specific, in the shell is 
\ba
\ep_n^{\rm (kin)} &=& \frac {1}{\pi^2} \left ( \int_0^{k_{Fn}} - \int_0^{k_{Fn} - \De}\right) dk \,k^2 {\sqrt{k^2+M_n^2}} \,, \nonumber \\
&=& F_1(k_{Fn}) -  F_2(k_{Fn}-\De) \,,
\ea
where $\Delta= \La_Q^3/k_{Fn}^2 + \kappa \La_Q /N_c^2$.  In analytical form, 
\ba
\ep_n^{({\rm kin})} = \frac {1}{4\pi^2} \left[k^3e_k + \frac {M_n^2}{2} k e_k - \frac{M_n^4}{2} \ln (k+e_k) \right]_L^U \,.
\label{epsnkin}
\ea
For $k_{Fn} < \De$, the upper limit $U=k_{Fn}$, the lower limit $L=0$, and $e_k={\sqrt{k^2+M_n^2}}$, which leads to the familiar expression for spin-$\frac 12$ relativistic particles of mass $M_n$. 
For neutrons in the shell with $k_{Fn} > \De$, however, $U=k_{Fn}$ and  $L=(k_{Fn}-\De)$ with $e_k={\sqrt{(k-\De)^2+M_n^2}}$.

\subsection*{Chemical potential}

The associated chemical potential ensues from 
\ba
\mu_n^{\rm (kin)} &=& \frac {d \ep_n^{\rm (kin)}}{dk_{Fn}} \frac {dk_{Fn}}{dn_n} \nonumber \\
&=& \left( \frac{dF_1}{dk_{Fn}} -    \frac{dF_2}{dk_{Fn}}\right)   \frac {dk_{Fn}}{dn_n} \,.
\ea
For the neutrons in the shell, 
\ba
n_n &=& \frac {1}{3\pi^2} \left[ k_{Fn}^3 - (k_{Fn} - \De)^3 \right] \,, \nonumber \\
\frac {dn_n}{dk_{Fn}} &=& \frac {1}{\pi^2} \left[ k_{Fn}^2 - (k_{Fn} - \De)^2 \left(1 - \frac {\partial \De}{\partial k_{Fn}}\right) \right] 
\nonumber \\
&=& \frac {1}{\pi^2} \left[ k_{Fn}^2 - (k_{Fn} - \De)^2 \left(1 + \frac {2\La_{\rm Q}^3}{k_{Fn}^3}\right) \right] \,.
\ea
For evaluating  $dF_1/dk_{Fn}$, use of the relations
\ba
\phi_1(k_{Fn}) &=& 0, ~~ \phi_2(k_{Fn}) = k_{Fn} \nonumber \\
 \frac {\partial \phi_1}{\partial k_{Fn}} &=& 0\,, ~~ \frac {\partial \phi_2}{\partial k_{Fn}} = 1 
\ea
in Eq. (\ref{eqn:Leibnitz}) yields 
\ba
\frac {dF_1}{dk_{Fn}} = \frac {1}{\pi^2} k_{Fn}^2 {\sqrt{k_{Fn}^2 + M_n^2}} \,.
\ea
The evaluation of $dF_2/dk_{Fn}$ proceeds along similar lines, but with
\ba
\phi_1(k_{Fn}) &=& 0, ~~ \phi_2(k_{Fn}) = k_{Fn} - \De \nonumber \\
 \frac {\partial \phi_1}{\partial k_{Fn}} &=& 0\,, ~~ \frac {\partial \phi_2}{\partial k_{Fn}} = 1 - \frac {\partial \De}{\partial k_{Fn}}
\ea
with the result 
\ba
\frac {dF_2}{dk_{Fn}} = \frac {1}{\pi^2} (k_{Fn}-\De) ^2 
\sqrt{(k_{Fn}-\De)^2 + M_n^2}  \left( 1 + \frac {2\La_{\rm Q}^3}{k_{Fn}^3}\right) \nonumber \\
\ea
Putting these results together, we obtain after some simplification 
\ba
\mu_n^{({\rm kin})} &=& \frac { {\sqrt{k_{Fn}^2+M_n^2}} - {\cal R} \sqrt{(k_{Fn} - \De)^2+M_n^2 }} {1 - {\cal R}} \nonumber \\
{\cal R} &=& 
\left(1 - \frac{\De}{k_{Fn}} \right)^2 \left( 1 + \frac {2\La_{\rm Q}^3}{k_{Fn}^3} \right) \,.
\label{munkin}
\ea

\subsection*{Pressure}

The kinetic theory expression for a single species of spin-$\frac 12$ fermions is
\ba
P = 2\, T  \int_L^U \frac {d^3k}{(2\pi)^3}   \ln [1 + e^{(\mu-e)\beta}] \,,  
\ea
where $\beta=1/T$, $e$ is the single particle spectrum and $\mu$ the chemical potential. A partial integration on the right-hand side  yields
\ba
P &=& T  \left[ \frac {1}{3\pi^2}  k^3 \ln  [1 + e^{(\mu-e)\beta}]  \right]_L^U \nonumber \\
&+& \frac {1}{3} \, 2  \int_L^U \frac {d^3k}{(2\pi)^3} \, k \frac {de_k}{dk} \frac {1}{1+  e^{(\mu-e)\beta}} \,.
\ea
At finite $T$, the first term vanishes when $U=\infty$ and $L=0$, leaving the second term as the kinetic pressure. For $T\rightarrow 0$, and finite $L$ and $U$, however, we have 
\ba
P = \frac {1}{3\pi^2} \left[ k^3 (\mu-e) \right]_L^U + \frac {1}{3} \, 2  \int_L^U \frac {d^3k}{(2\pi)^3} \, k \frac {de_k}{dk} \,.
\ea

The expressions for $P_n^{{\rm (kin)}}$, thus take different forms in the regions $k_{Fn} < \De$ and  $k_{Fn} > \De$. For $k_{Fn} < \De$, the limits $U=k_{Fn}$ and $L=0$ yield the familiar kinetic theory expression
 \ba
 P_n^{{\rm (kin)}} &=& \frac 13\,2 \int_L^U \frac {d^3k}{(2\pi)^3} \,   k \frac {de_k}{dk}  \nonumber \\
 &=& \frac{1}{12\pi^2} \left[ k^3e_k -\frac 32 M_n^2 ke_k + \frac 32 M_n^4 \ln (k+e_k) \right]_L^U \,, \nonumber \\
 \label{Pnkin}
 \ea
where $e_k=\sqrt{k^2+M_n^2}$. The last two terms in Eqs. (\ref{epsnkin}) and (\ref{Pnkin}) cancel, and thus in this region,  
$\ep_n^{\rm (kin)} +  P_n^{{\rm (kin)}} = n_n \mu_n^{{\rm kin}}$ (the TI) with $n_n=k_{Fn}^3/(3\pi^2)$ and $\mu_n^{\rm (kin)}= \sqrt{k_{Fn}^2+M_n^2}$.  

In the region $k_{Fn} > \De$ with $U=k_{Fn}$ and   $L=(k_{Fn}-\De)$, the kinetic theory pressure becomes 
\ba
 P_n^{{\rm (kin)}} &=& \frac {1}{3\pi^2} \left[ k^3 (\mu_n^{\rm kin} - e_k) \right]_L^U 
 +   \frac {1}{3}\, 2 \int_L^U \frac {d^3k}{(2\pi)^3} \,   k \frac {de_k}{dk} \,. \nonumber \\
 \label{PnTI1}
\ea
The first term above gives the contribution from the boundaries of the shell. Inserting the appropriate limits  for the shell, this term reads as
\ba
\label{PnTI2}
\frac {1}{3\pi^2} \bigg[ \mu_n \left( k_{Fn}^3 - (k_{Fn} - \De)^3 \right)  \nonumber \\
-  \left(k_{Fn}^3 \sqrt{k_{Fn}^2+M_n^2} - \left(k_{Fn} - \De \right)^3  \sqrt{(k_{Fn}-\De)^2+M_n^2} \right) \bigg] \,.\nonumber \\
\ea

\subsection*{Thermodynamic identity}

Collecting the results in Eqs. (\ref{PnTI1}) and (\ref{PnTI2}) and evaluating  $\ep_n^{\rm (kin)}$ in \Eqn{epsnkin} for the shell, we obtain the TI (after many cancellations)
\ba
\ep_n^{\rm (kin)} +  P_n^{{\rm (kin)}} &=& n_n \mu_n^{{\rm kin}} \quad {\rm with} \nonumber \\
n_n &=& \frac {1}{3\pi^2} \left[ k_{Fn}^3 - (k_{Fn} - \De)^3 \right] \,,
\ea
the last relation giving the number density of neutrons in the shell.  The neutron chemical $\mu_n^{\rm kin}$ here is independently calculated from \Eqn{munkin}. We have verified that, numerical calculations of $\mu_n^{\rm kin}= d\ep_n^{\rm (kin)}/dn_n$ and $P_n^{\rm (kin)} = n_n^2 d(\ep_n^{\rm (kin)}/n_n)/dn_n$ in the shell yield the same results as evaluations from the analytical formulas above, which provide additional physical insight concerning the role played by the shell.   

All expressions in this Appendix apply also to protons so that a generalization to a 2-component system consisting of neutrons and protons is straightforward.  

\end{document}